\documentclass[twoside]{article}
\usepackage{xcolor}
\usepackage{colordef}
\usepackage[utf8]{inputenc}
\usepackage[T1]{fontenc}
\usepackage{lmodern} 
\usepackage{graphicx}

\usepackage[sc]{mathpazo} 
\usepackage[T1]{fontenc} 
\usepackage{microtype} 

\usepackage[hmarginratio=1:1,top=32mm,columnsep=20pt]{geometry} 
\usepackage{multicol} 
\usepackage[hang, small,labelfont=bf,up,textfont=it,up]{caption} 
\usepackage{booktabs} 
\usepackage{float} 
\usepackage{hyperref} 

\usepackage{paralist} 

\usepackage{abstract} 

\usepackage{titlesec} 

\usepackage{fancyhdr} 
\usepackage{colordef}
\usepackage{float}
\usepackage{subfig}
\newcommand{\quantnet}{\hspace*{\fill} \raisebox{-1pt}{\includegraphics[scale=0.05]{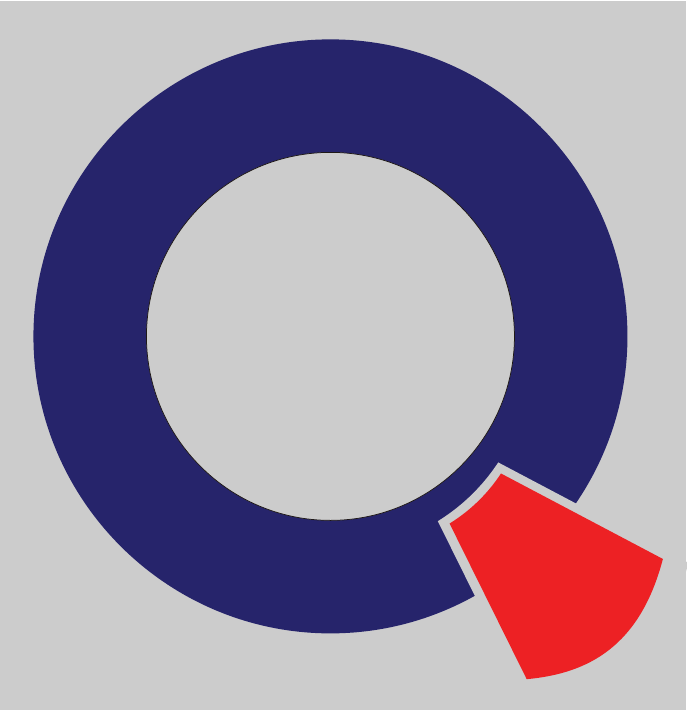}}\,}

\usepackage[authoryear]{natbib}

\title{\vspace{-15mm}\fontsize{24pt}{10pt}\selectfont\textbf{A Mortality Model for Multi-populations:\\ A Semi-Parametric Approach}\thanks{The authors gratefully acknowledge financial support from the Deutsche Forschungsgemeinschaft through the International Research Training Group IRTG 1792 "High Dimensional Non Stationary Time Series" and the Collaborative Research Center CRC 649 "Economic Risk".}} 

\author{
Lei Fang$^1$, Wolfgang K. H\"ardle$^1$, Juhyun Park$^2$\thanks{Corresponding author: Department of Mathematics and Statistics, Lancaster University, Lancaster LA1 4YF, UK., email: juhyun.park@lancaster.ac.uk} \\
$^1$Humboldt-Universit\"at zu Berlin, $^2$Lancaster University \\
}

\date{}

\begin{document}

\maketitle 

\thispagestyle{fancy} 


\begin{abstract}
\noindent Mortality is different across countries, states and regions. Several empirical research works however reveal that mortality trends exhibit a common pattern and show similar structures across populations. The key element in analyzing mortality rate is a time-varying indicator curve. Our main interest lies in validating the existence of the common trends among these curves, the similar gender differences and their variability in location among the curves at the national level. Motivated by the empirical findings, we make the study of estimating and forecasting mortality rates based on a semi-parametric approach, which is applied to multiple curves with the shape-related nonlinear variation. This approach allows us to capture the common features contained in the curve functions and meanwhile provides the possibility to characterize the nonlinear variation via a few deviation parameters. These parameters carry an instructive summary of the time-varying curve functions and can be further used to make a suggestive forecast analysis for countries with barren data sets. In this research the model is illustrated with mortality rates of Japan and China, and extended to incorporate more countries.\\
\end{abstract}



\newpage

\section{Introduction} \label{sec:intro}

In recent years, global population trend has received widespread attention with regard to the growth of the aging populations on the globe, which raises demographic risk in most developed and even some developing countries.
Demographic risk is understood to be an imbalance of the age distribution of a society with the obvious implications in economic growth, social stability, political decisions and resource allocation. The factor demography is particularly important in understanding the challenges of developing countries and their global impacts, yet, due to limited access and available resources, most studies have focused on the cases from developed countries. This motivated us to develop a novel approach to forecasting a population trend with limited data, using the example from China. 

\begin{figure}[htp]
\begin{multicols}{3}
\includegraphics[width=0.32\textwidth]{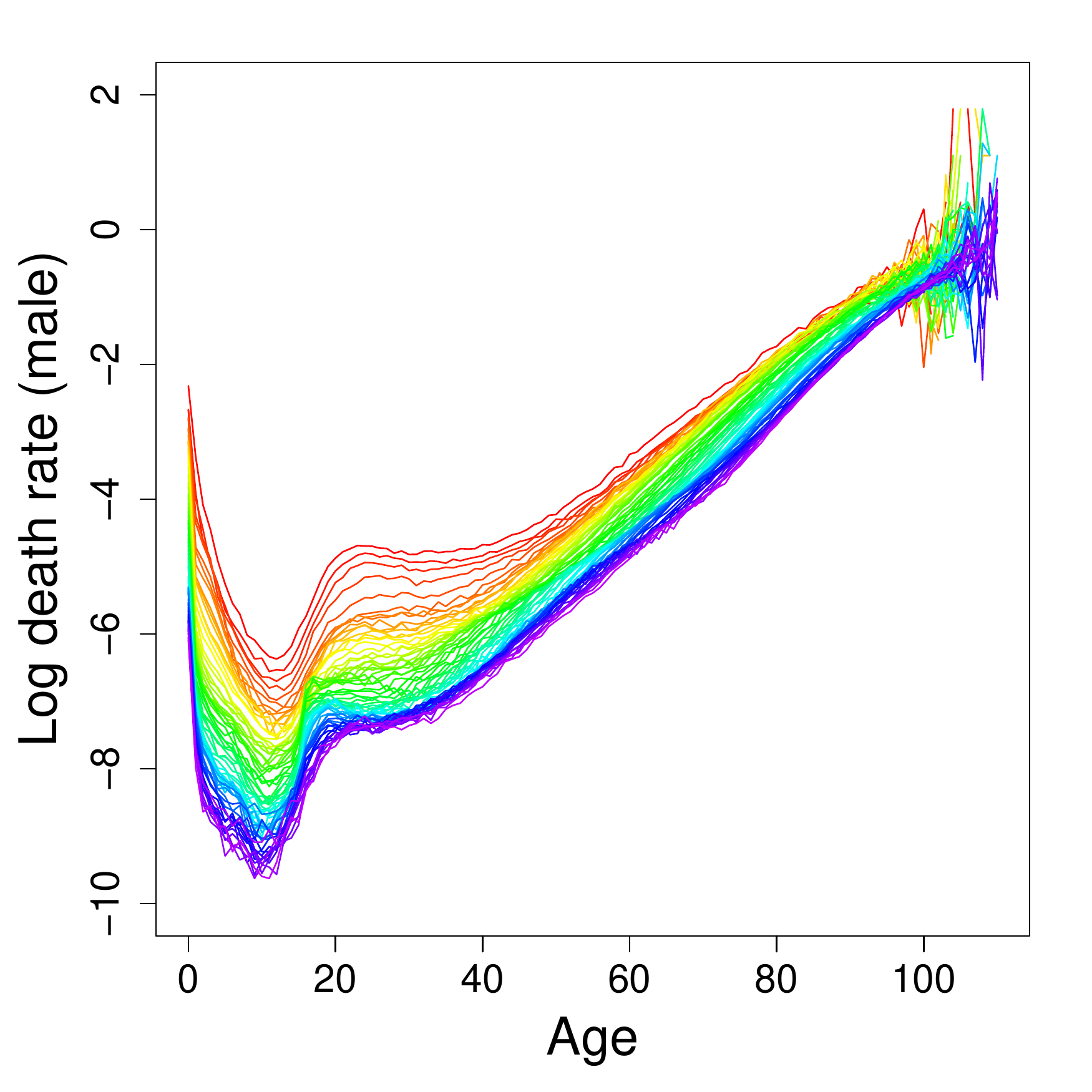}
\includegraphics[width=0.32\textwidth]{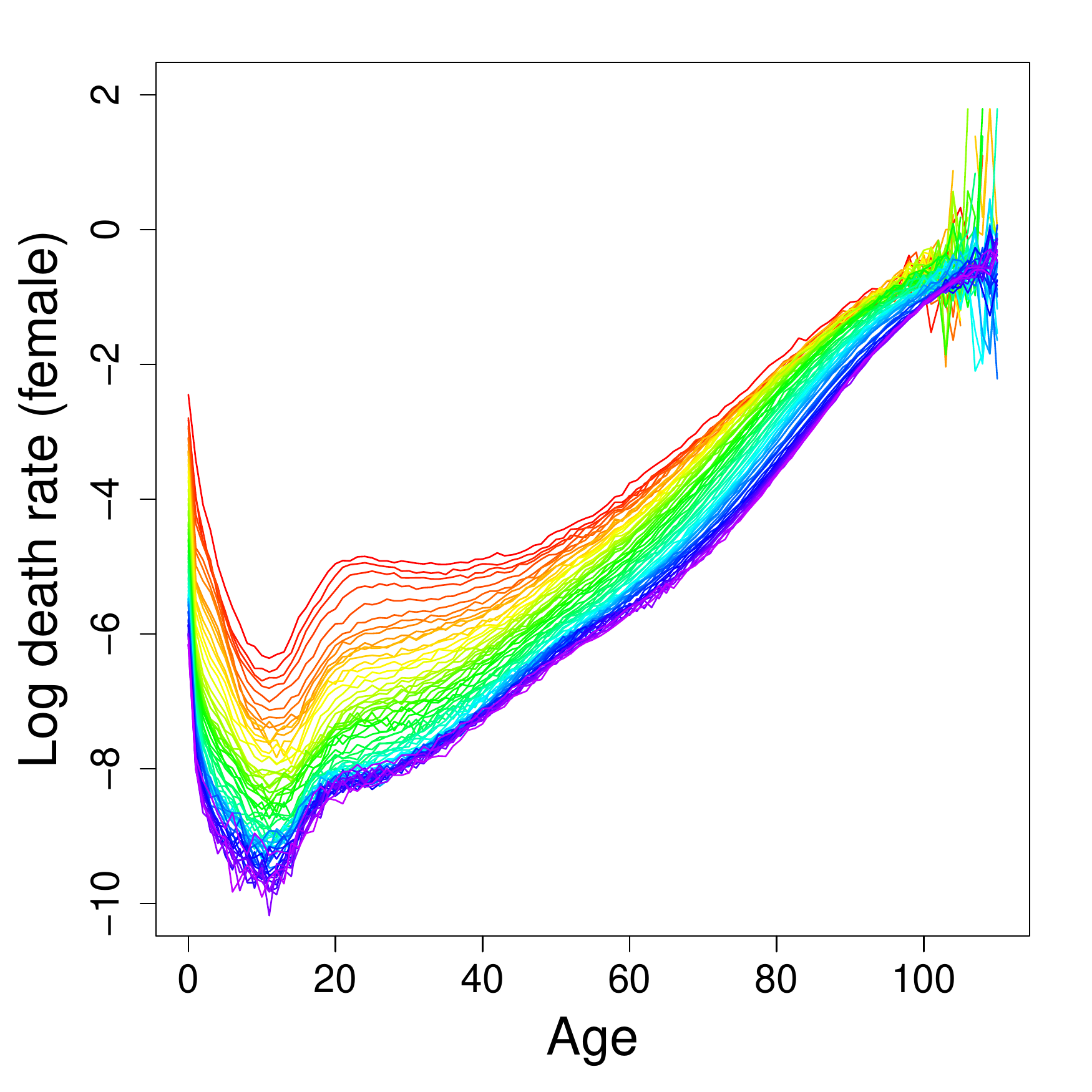}
\includegraphics[width=0.32\textwidth]{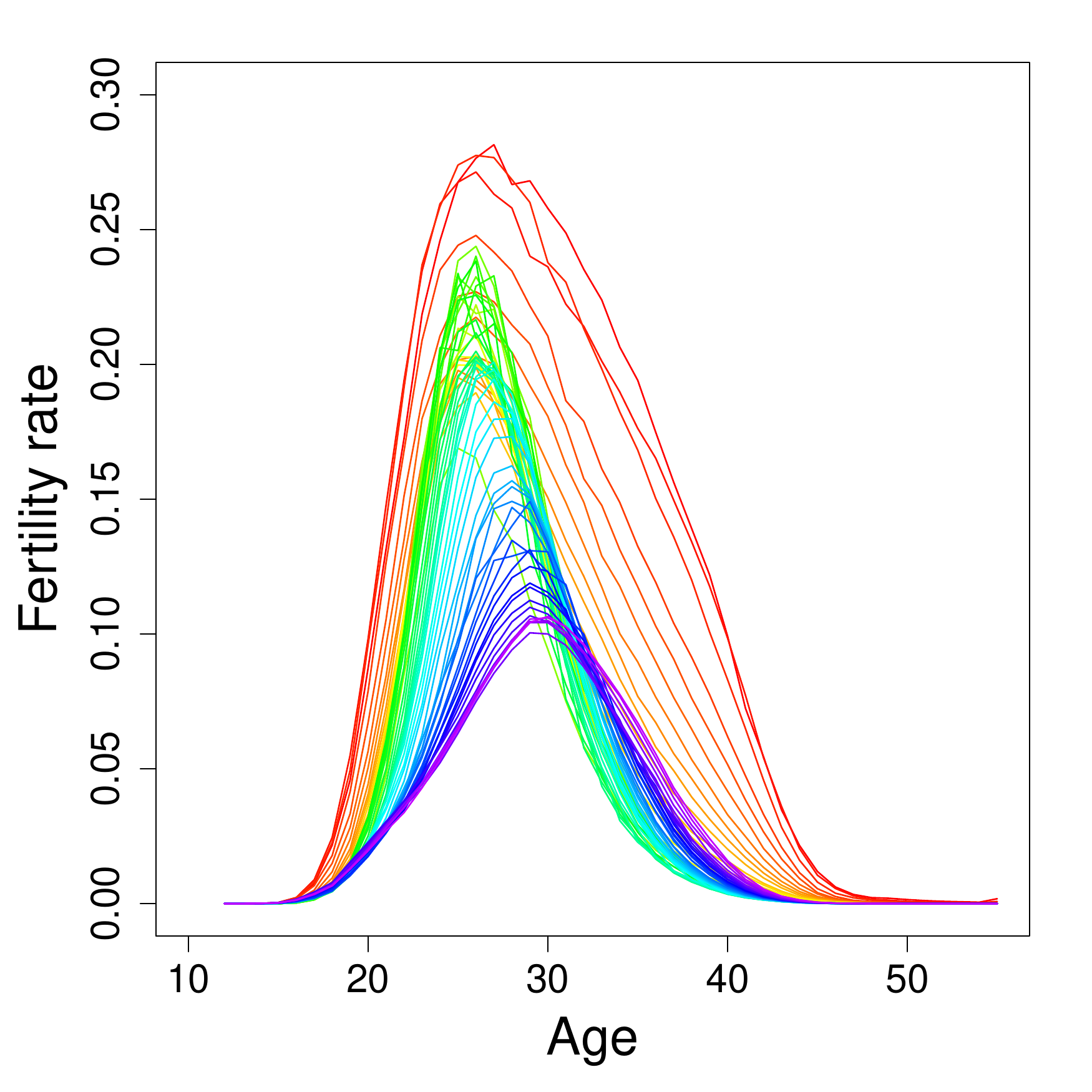}
\end{multicols}
\caption{Japan's descriptive demography - male (left) and female (middle) mortality dynamics from 1947 to 2012, fertility (right) dynamics from 1947 to 2009, rates in different years are plotted in rainbow palette order. \label{fig:japdes}}
\href{https://github.com/QuantLet/MuPoMo}{\quantnet MuPoMo}
\end{figure}

China, as a large developing Asian country, is experiencing the transformation to an aging society at an even faster rate, and is therefore a good example with which to study demographic risk. However, due to political reasons and delays in construction of a national system for systematically collecting statistics, the problem of insufficient and unsatisfying demographic data sets remains unsolved, and this situation poses a unique challenge in statistical analysis. 
Japan, China's neighbour, has undergone a dramatic demographic change during the last several decades. In addition, the Japanese government has set up a complete national statistical system in the middle of the last century, which provides fine demographic data sets in longer time horizons to help researchers explore Japan's demographic transition. A typical example of demographic data in terms of mortality and fertility is shown in Figure~\ref{fig:japdes} from Japan. 

As we shall demonstrate later with our new formulation of the problem, those Japanese data sets could offer a good reference point to studying the pattern in China. However, due to the time-lag in the phase of the development of each country, a direct comparison is not feasible. In this work, we propose to incorporate a nonlinear transformation across different periods by formulating a semi-parametric regression model. The advantages and shortcomings of our approach will be discussed in detail. To overcome some of the shortcomings, our approach will be extended by taking into account a trend in multi-populations using global~\citet{HMD}. 

Our approach could be understood as an extension of the \emph{coherent forecasting} advocated by \citep{Li:Lee:2005} with the idea that a local trend estimation could be improved by taking into account the patterns in a larger group it belongs to. This requires a serious consideration of the notion of similarity between the members. Until now, the idea of coherent forecasting was mostly discussed in the context of concurrent comparison with a small well-defined neighboring countries. In this article we attempt to demonstrate its generality in a wider context, with a new formulation under the shape invariant models \citep{Lawton:Sylvestre:Maggio:1972}. We first introduce the datasets at our disposal to demonstrate our approach before presenting a brief literature review.

\subsection{Data Sets}
The demographic data sets are collected from different sources: China data sets are extracted from the China Statistical Year Book, the other 35 countries are obtained from the Human Mortality Database (HMD). For data sets on China the mortality rates are gender-age specific starting from 0 to 90+ years old, while the mortality rates of other countries are gender-age specific from 0 to 110+ since the Human Mortality Database makes estimates and adjustments on the raw data to extend into wider age group.
As for the sample size, they are all different as well: China mortality data spans from 1994 to 2010 with missing data of years 1996, 1997, 2001 and 2006, while the sample size from other countries range from 14 years (Chile) to 261 years (Sweden). Referring to the missing values, we use moving average of neighboring five years to compute these.

Recall that the definition of the mortality rate is the number of deaths per 1000 living individuals per single calender year. To fit the mortality trend more precisely and for visual convenience, we present the log mortality.

Data and sources codes are available online (\href{https://github.com/QuantLet/MuPoMo}{https://github.com/QuantLet/MuPoMo}) and linked to the figures.

\subsection{Literature Review}
Since 1980, one of the challenges in demography has been to analyze and forecast mortality in a purely statistical way without involving the subjective opinions of experts. \citet{Lee:Carter:1992} (LC) firstly proposed a stochastic method based on a Singular Value Decomposition technique to explore the unobserved demographic information. This proved insightful and gained a good reputation. Since then, several methods based on stochastic population modelling and forecasting have been developed, see e.g. \citet{Cairns:Blake:Dowd:2008}, \citet{Tickle:2008} and \citet{Booth:2006} for review.
Among all the stochastic models, the most popular one is the LC model, which was used to analyze the U.S. mortality rates from 1933 to 1987. Based on the idea of the LC model, comparisons of different methods and some variants or extensions have been developed. \citet{Lee:Miller:2001} compared the forecasts of LC model with the U.S. social security system forecasts. \citet{Li:Lee:Tuljapurkar:2004} proposed another method when there are fewer observations at uneven intervals, and applied it to China and South Korea. \citet{Hyndman:Ullah:2007} developed a more general method by treating the underlying demographic process as functional data, employing the functional principal components analysis to extract more than one explanatory components and providing robust estimation and forecast. This idea is further extended to multilevel functional data in~\citet{Shang:2016, Gao:Shang:2017}. These linear extensions are successful in capturing complex variation at the aggregate level at the expanse of ease of interpretability of the LC model. 

In light of limited data access combined with fragile quality, less technically refined methods are available for Asian countries compared to, for example, developed western countries. An exception is the stochastic population approach on Asian data by \citet{Li:Lee:Tuljapurkar:2004}, who implemented the LC model to sparse data. In their work, they generated central forecast with just the first and last observations along the time horizon, improved the estimates by additional observations and evaluated its performance with other existing methods. \citet{Raftery:Li:2012} proposed a Bayesian method for probabilistic population projections for all countries, where the Bayesian hierarchical models, estimated via Markov chain Monte Carlo, are applied to the United Nations population data. In cases of limited data and similar demographic trends between two populations (regional or national level), the Bayesian stochastic modelling for two populations is proposed by \citet{Cairns:Blake:Dowd:2011}. This motivates us to analyze Chinese demography via taking Japan into reference.

One might wonder why Japan, not Taiwan or any other neighboring countries. An interesting finding from \citet{Fang:Haerdle:2015} is that China has a demographic trend closer to Japan than to Taiwan, particularly visible in the mortality trend. On the economic level, China and Japan have been both important economies in last several decades and the development pattern is also quite similar. \citet{Hanewald:2011} found that the LC mortality index $k_t$ correlates significantly with macroeconomic fluctuations in some periods, which provides a good reference with which to connect the mortality trends between China and Japan. Nevertheless, due to their differences in the developmental phase, it is much desirable to incorporate nonlinear transformation of the trend to capture the effect of time trend in a flexible manner.

\subsection{Goal and Outline}

Due to the sensitivity of fertility to social policies and induced unpredictability, our research is restricted to mortality analysis. The purpose of our analysis is to build a modeling framework that takes into account similarities in trend across different countries. We focus on the popular time-varying mortality indicator $k_t$ extracted from the standard LC method. Following the usual interpretation of the mortality indicator as a time trend, we seek to compare the trends in terms of their shape variation in time or in phase to measure similarities among different countries. Most linear approaches such as those based on principal component analysis are inefficient in capturing these nonlinear variations and thus difficult to interpret the results. Instead we directly model the shape variation of the curves. To increase flexibility and interpretability of the shape of the trend function, we adopt the framework of the semiparametric comparison approach~\citep{Haerdle:Marron:1990, Kneip:Engel:1995} and accordingly demonstrate potential improvements in mortality forecasting. We analyse similarities in mortality between China and Japan, and then extend our approach to a global common mortality trend and study a sub-group pattern.

A typical setting for the semiparametric approach assumes that (i) measurements are available on a common interval, often at common and dense grid points, (ii) the population has a well defined trend in terms of identifiable features and (iii) the errors are independent,  all of which are violated in our demographic application. We demonstrate its ramifications and additional considerations throughout the analysis.   

The general methodology is presented in Section~\ref{sec:method}.  Section~\ref{sec:two-pop} focuses on mortality forecast with two country comparison, using the example of China and Japan. This serves a motivation for further development in Section~\ref{sec:multi-pop} with multiple populations using the Human Mortality Database.  More discussions on economic insights, global aging trend influences and suggestions are given in the last section.


\section{Methodology}\label{sec:method}

In this section, we firstly introduce the parameters of interests and then outline the LC method, semi-parametric comparison of nonlinear curves and common trend modeling in details.

\subsection{Notations and Parameters of Interest}

The parameters of interests are age-specific mortality rates from multiple countries. We use the symbols $m$ to denote the mortality rate. All the parameters are indexed by a one-year age group, denoted by $x$, and in addition indexed by time, denoted by $t$. For instance, $m(x,t)$ is the mortality rate for age $x$ in year $t$.
Based on the LC model, $m(x,t)$ is assumed to be decomposed into the average age pattern and the time-varying index $k_t$ (for single country, state or region). Characterizing $k_t$ is essential in understanding the mortality trend, which is the main goal of this paper. 
When it comes to multiple countries, we use $k_i(t)$ to denote the derived time-varying mortality indicator for country $i$, with $i\in \{1,...,N\}$.

\subsection{Lee-Carter (LC) Method}

The benchmark LC model employs the Singular Value Decomposition (SVD) to analyse the time series on the log of the age-specific mortality. The method relies on the standard statistical analysis of the time series. Nonetheless, the LC model does not fit well in some cases where missing data is common or the horizon of time series is not sufficient, the reason being the assumption of long-term stationarity.

The basic idea for demography dynamics analysis is to regress mortality $m(x,t)$ on non-observable regressors for prediction. The regressors are obtained via SVD of the demographic indicators. It separates the age pattern from the time-dependent components, takes time series analysis on the time-dependent components only and hence forecasts the future trend.

The mortality rate $m(x,t)$ is hence calibrated via the following model:
\begin{eqnarray}
         \log\{m(x,t)\} & = & a_x + b_x k_t + \varepsilon_{x,t}, \label{lc}
\end{eqnarray}
or
\begin{eqnarray*}
         m(x,t) & = & \mbox{exp}(a_x + b_x k_t + \varepsilon_{x,t}),
\end{eqnarray*}
where 
$a_x$ is the derived age pattern averaged across years and $k_t$ represents the only time-varying index of mortality level. Thus $b_x$ measures the sensitivity of the mortality rates to the change of $k_t$, reflecting how fast the mortality rate changes over ages and $\varepsilon_{x,t}$ is the residual term at age $x$ in year $t$ with $\mathop{\mbox{\sf E}}(\varepsilon_{x,t})=0$ and $\mathop{\mbox{\sf Var}}(\varepsilon_{x,t})=\sigma_\varepsilon^2$.

Three unobserved parameters $a_x$, $b_x$ and $k_t$ in the single equation (\ref{lc}) mean that the LC model is over-parameterized and therefore two normalisation constraints are imposed:
\begin{eqnarray*}
         \sum k_t = 0 , \  \sum b_x = 1 \,.
\end{eqnarray*}
By SVD, one obtains $k_t$ and $b_x$. The evolution of $k_t$ can be further fitted by standard time series models such as using ARIMA techniques. \citet{Lee:Carter:1992} found that a random walk with drift describes $k_t$ quite well:
\begin{eqnarray*}
         k_t = k_{t-1} + d + e_t
\end{eqnarray*}
where $d$ is the drift parameter reflecting the average annual change and $e_t$ is an uncorrelated error. Others such as \citet{Chan:Li:Cheung:2008} pointed out that the mortality index $k_t$ may be better fitted with a trend-stationary model for Canada, England and United States when accounting for a break in mortality rate decline during the 1970s.

Given an $h$-step ahead forecasting $k_{t+h}$ from the time series models, the forecast of the mortality rates in future period $t+h$ can be made via the following formula:
\begin{eqnarray*}
         m(x,t+h) & = & \mbox{exp}(a_x + b_x k_{t+h}).
\end{eqnarray*}

\subsection{Semi-Parametric Comparison of Nonlinear Curves}

\begin{figure}[H]
	  \centering
	  \includegraphics[width=0.7\textwidth]{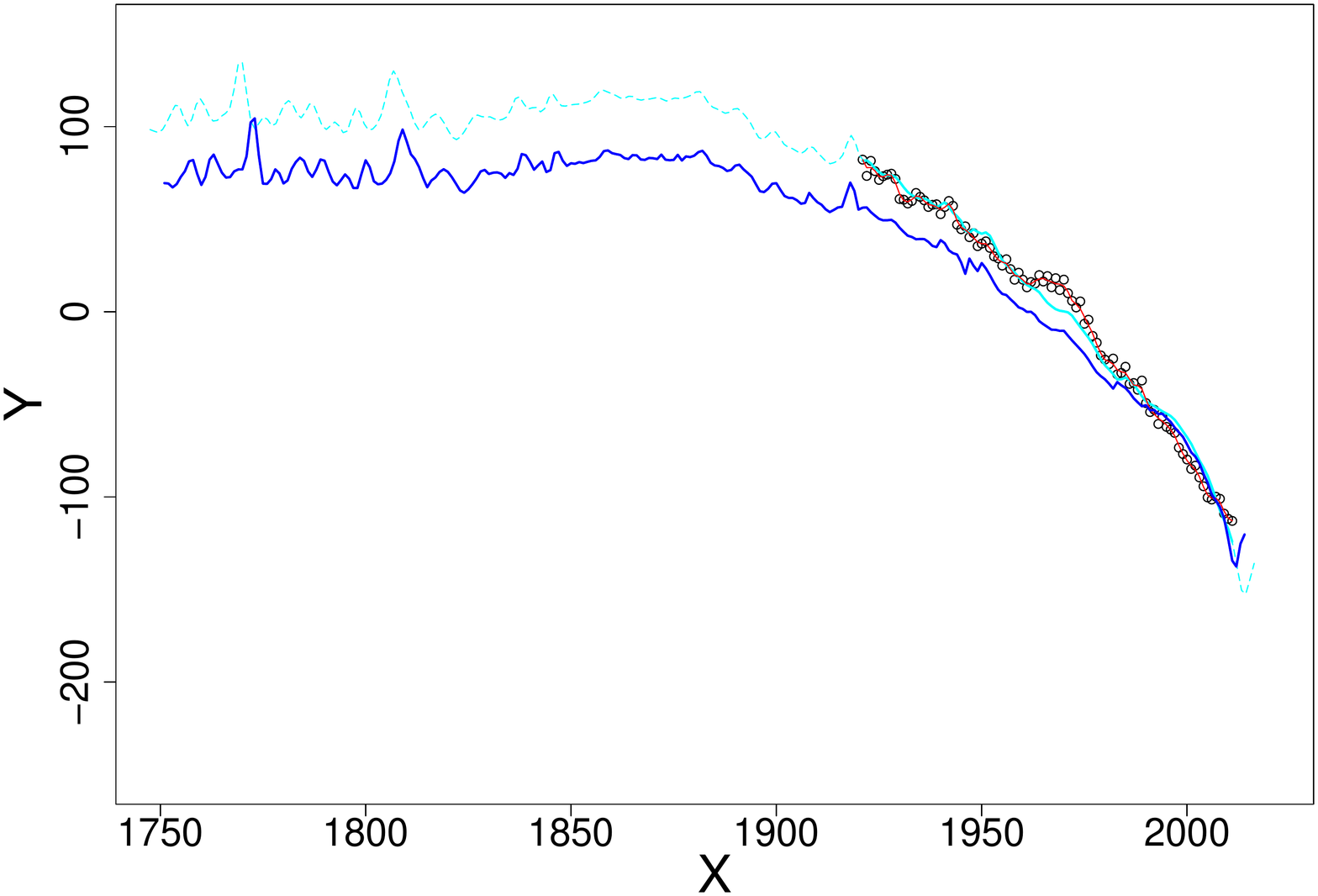}
\caption{Semi-parametric Comparison of Nonlinear Curves: the dark blue curve and the red curve have similar pattern, while the light blue curve is semi-parametrically shifted to represent the red curve via the dark blue curve. \label{semiexample}}
\href{https://github.com/QuantLet/MuPoMo}{\quantnet MuPoMo}
\end{figure}

When the observable curves are noisy versions of similar regression curves, comparison of regression curves from related samples is not trivial. The problem aggravates when the data are sparse, partially available for some and the domain of the curves do not match, as demonstrated in Figure~\ref{semiexample} where we wish to compare the short segment of black dots on the right and the long blue curve below. The black dots on the right with the red solid curve underneath represent the mortality rates from China, while the elongated blue curve below represents the mortality rates from Sweden. The curve above in cyan is a transformed Swedish curve that gives a {\it best} match to the partial observations of China, which will be defined precisely later.

It is not unreasonable to assume that developing countries will catch up with the developed countries, and  due to globalization the trend in mortality rates is expected to follow a similar pattern. 
Then the main differences among those curves could be explained by shifted time axis and vertical re-scaling, which could further be parametrized for parsimonious representation and ease of interpretation.
This is our motivation to take a view of semiparametric comparison in taking into account mortality variations of other countries.

To handle the noisy data, nonparametric smoothing techniques could be incorporated to estimate the underlying curves when the solid theory is unavailable in modeling them. Hence, we seek general semi-parametric models that allow for nonlinear transformation on the domain as well as the image of the curves. 

Simply denoting the underlying curves by $f_1$ and $f_2$, the semi-parametric comparison of the nonlinear curves can be expressed as
\begin{eqnarray}
          f_2(t)=\theta_{1} f_1\left( \frac{t-\theta_{2}}{\theta_{3}}\right) +\theta_{4}, \label{semi}
\end{eqnarray}
where we assume that $f_2$ has a similar pattern to $f_1$ and $\theta=(\theta_{1},\theta_{2},\theta_{3},\theta_{4})^{\top}$ are shape deviation parameters. Since normally only noisy measurements of the curves are available, $f_1$ and $f_2$ are not directly available, there will be measurements models given by
\[
Y_i(t_{ij}) \equiv Y_{ij} = f_i(t_{ij}) + \varepsilon_{ij} \qquad i=1,2; j=1,\ldots,n_i\,.
\] 
More detailed discussions on this method can be referred to \citet{Haerdle:Marron:1990} on semiparametric comparison of regression curves.

\subsection{Common Trend Modeling} \label{sec:commontrend}

When more than two regression curves share similar pattern or trend, a common trend model can be built based on the technique of semi-parametric comparison of nonlinear curves we discussed previously. Suppose we are given $N$ noisy curves $\mbox{Y}_i, i=1,\ldots,N$ that exhibit some similar patterns. A general regression model can be expressed as,
\begin{equation}\label{e:additive}
          \mbox{Y}_i = f_i + \varepsilon_i\,,
\end{equation}
where $f_i$ denote unknown smoothing regression functions while $\varepsilon_i$ represent independent errors with mean 0 and variance $\sigma_i^{2}$. 

The relationship among these similar curves can be described as
\begin{eqnarray}
       f_i(t) = \theta_{i1} g\left( \frac{t-\theta_{i2}} {\theta_{i3}} \right)  + \theta_{i4}. \label{common}
\end{eqnarray}
Here $\theta_i = \left( \theta_{i1}, \theta_{i2}, \theta_{i3}, \theta_{i4} \right)$ are unknown parameters describing shape deviations, and $g$ is a unknown function specifying the common shape of these curves, which can be interpreted as a reference curve. The model in~(\ref{common}) is commonly known as shape invariant model (SIM), firstly proposed by \citet{Lawton:Sylvestre:Maggio:1972} and further studied by \citet{Kneip:Engel:1995} and provides an extension of the model in~(\ref{semi}) to multiple curves. A detailed investigation of this model to estimate the mortality trend will be given in the following section. 


\section{Mortality forecasting: two-country case}\label{sec:two-pop}

\begin{figure}[htb]
\captionsetup[subfigure]{labelformat=empty}
\subfloat[]
  {\includegraphics[width=0.5\linewidth]{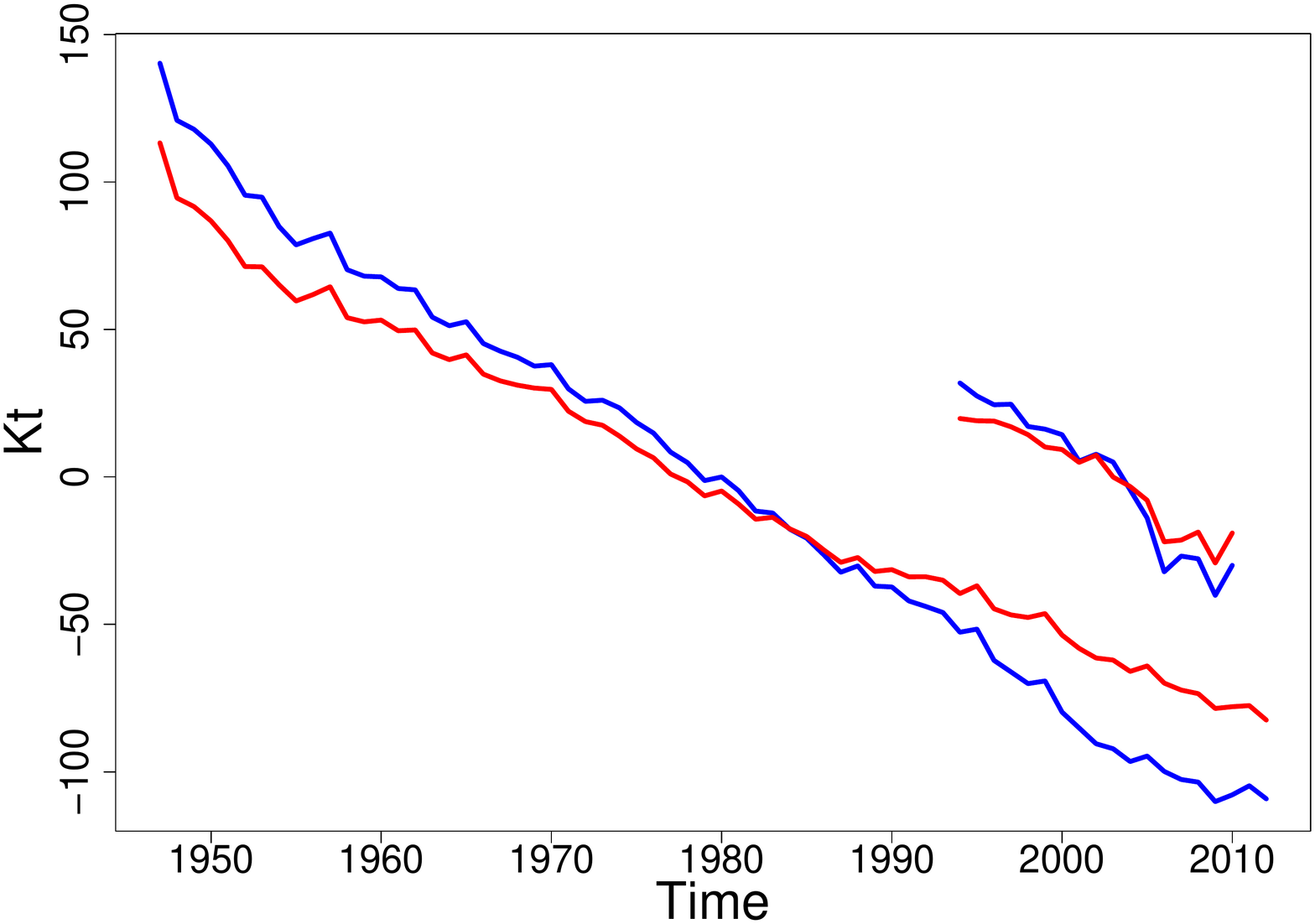}}\hfill
\subfloat[]
  {\includegraphics[width=0.5\linewidth]{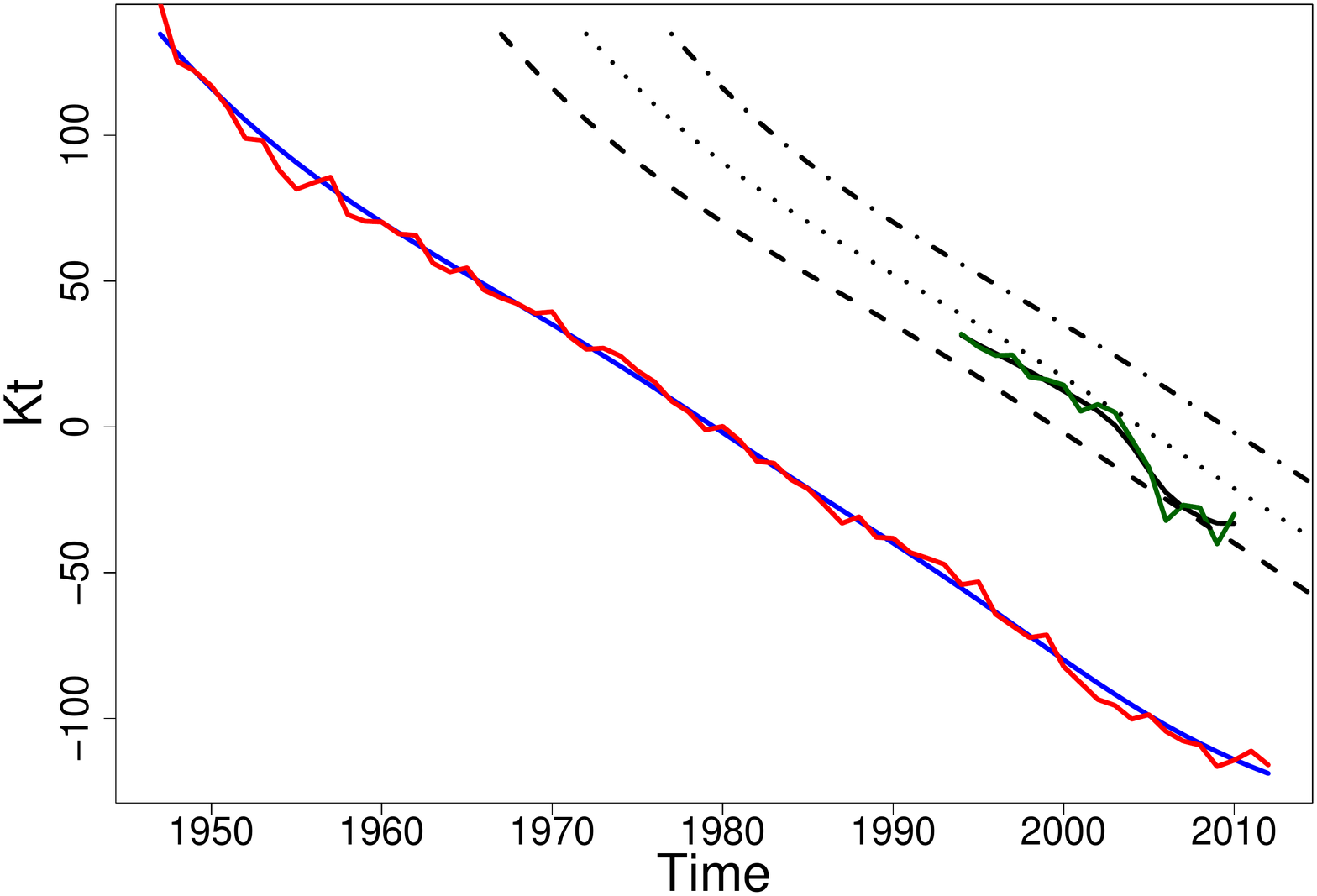}}
  \vspace{-0.5cm}
\caption{China mortality trend vs. Japan mortality trend: \color{blue}{female}, \color{red}{male} \color{black} {(left) and} 
\color{red}{Japan trend}, \color{blue}{Japan smoothed trend}, \color{darkgreen}{China trend} \color{black} and {China smoothed trends} (right).The dotted lines (right) from left to right are shifted Japan's smoothed trends of 20-, 23- and 25- years forward respectively. \label{ktcom}}
\vspace{0.1cm}
\href{https://github.com/QuantLet/MuPoMo}{\quantnet MuPoMo}
\end{figure}

With reasons mentioned in Section~\ref{sec:intro}, we will first focus on analysing mortality similarities between China and Japan.
Before presenting our formulation, we first graphically show the empirical trends in the left of Figure \ref{ktcom}, which suggests that the mortality trends from both gender groups of China correlate with those of Japan respectively. However, due to limited sample they all seemingly reflect linear mortality trend over time, which would contradict to the theoretical and conceptual views on mortality trends. 
To provide intuitive comparison between these two countries, the horizontal shift of Japan's mortality curve over time axis is plotted in the right part of Figure \ref{ktcom}. The dotted lines from left to right are shifted Japan's smoothed trends of 20-, 23- and 25- years forward respectively. Graphically we see that Japan's mortality trend is 23 years earlier than China.

\subsection{Model formulation}
To parameterize the potential relationship between China and Japan mortality trend, we specify the model as following, using $k_t$ derived from LC model in (\ref{lc}):
\begin{eqnarray*}
         \log\{m(x,t)\} & = & a_x + b_x k_t + \varepsilon_{x,t}\,.
      \end{eqnarray*}
Then we infer China's mortality trend via Japan's trend through the technique of semi-parametric comparison of regression curves defined in (\ref{semi})
         \begin{eqnarray}
          k_c(t)=\theta_{1}k_j\left( \frac{t-\theta_{2}}{\theta_{3}}\right) +\theta_{4} , \label{equtheta}
         \end{eqnarray}
where $k_c(t)$ is the time-varying indicator for China, $k_j(t)$ is the time-varying indicator for Japan, and $\theta=(\theta_{1},\theta_{2},\theta_{3},\theta_{4})^{\top}$ are shape deviation parameters.


\paragraph{Understanding $\theta$}
It is probably easiest to interpret the parameters (\ref{equtheta}) by starting with $\theta=(\theta_{1},\theta_{2},\theta_{3},\theta_{4})^{\top}=(1,\theta_{2},1,\theta_{4})^{\top}$.
           \begin{compactitem}
                 \item $\theta_{1}$ is the general trend adjustment, here selected as 1.
                 \item $\theta_{2}$ is the time-delay parameter
                 \item $\theta_{3}$ is the time acceleration parameter, here selected as 1.
                 \item $\theta_{4}$ is the vertical shift parameter
           \end{compactitem}
           
\begin{figure}[H]
\captionsetup[subfigure]{labelformat=empty}
\subfloat[]
  {\includegraphics[width=0.5\textwidth]{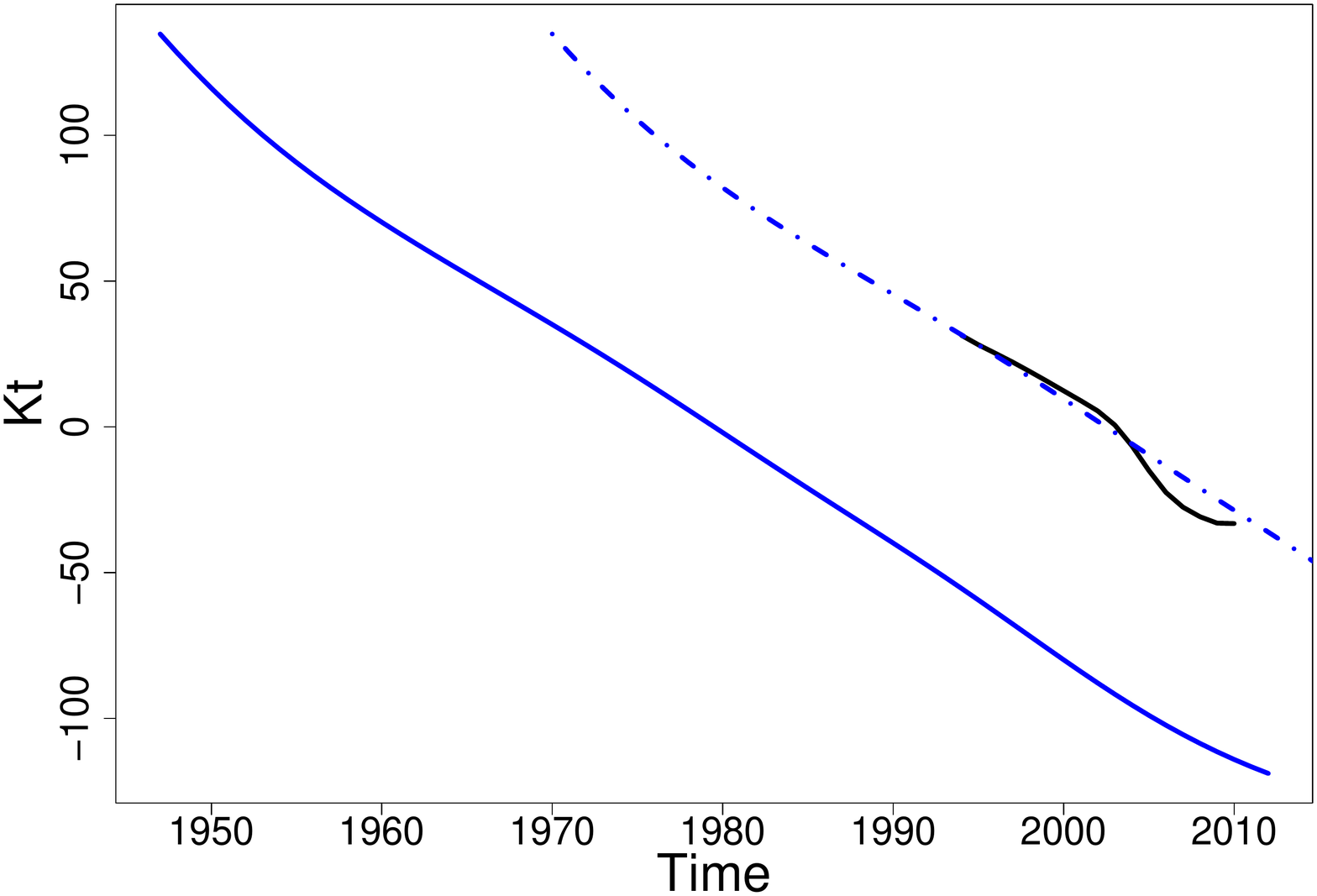}}\hfill
\subfloat[]
  {\includegraphics[width=0.5\textwidth]{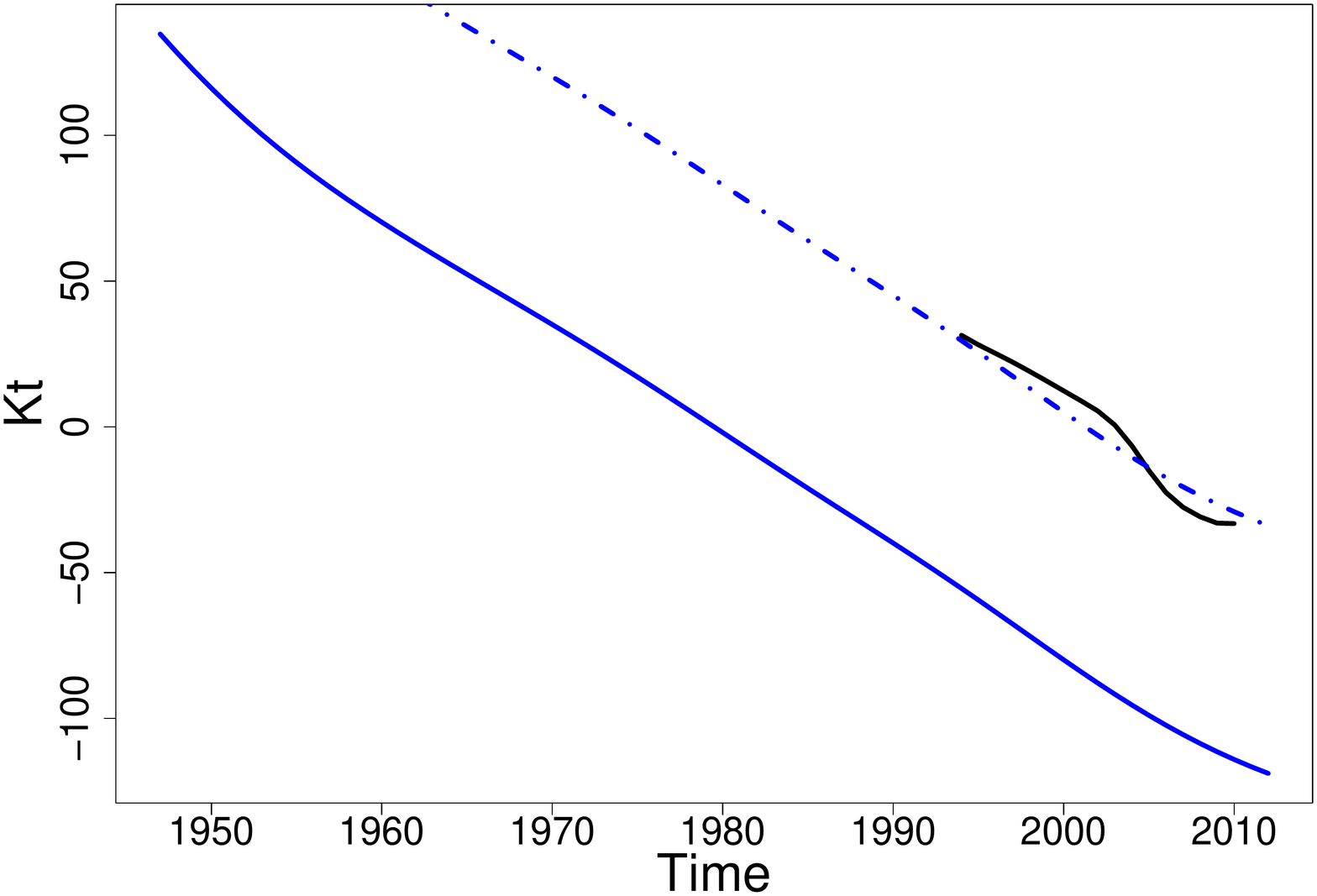}}
  \vspace{-0.5cm}
\caption{Time delay $\theta_{2}=-23$ (left) and vertical shift $\theta_{4}=85$ (right).\label{theta}}
\href{https://github.com/QuantLet/MuPoMo}{\quantnet MuPoMo}
\end{figure}

In Figure \ref{theta}, it demonstrates how $\theta$ influences shift of these two curves. In the left of Figure \ref{theta}, China's female $k_t$ can reach a similar behavioral area by shifting horizontally $\theta_{2} = -23$, while in the right it shows that another acceptable area could be obtained via approximately vertical shift of $\theta_{4} = 85$.

\subsection{Estimation}
In order to find the optimal solution for shape deviation parameters, we minimize the following loss function.
         \begin{eqnarray}
          \min_{\theta}\int_{t_c}\left\lbrace \hat{k}_c(u)-\theta_{1}\hat{k}_j\left( \frac{u-\theta_{2}}{\theta_{3}}\right) -\theta_{4}\right\rbrace ^{2}w(u)du, \label{min}
         \end{eqnarray} 
where $\hat{k}_c(t)$ and $\hat{k}_j(t)$ are the nonparametric estimates of the original time-varying indicators $k_c(t)$ and $k_j(t)$, and $t_c$ is the time interval of China mortality data. Here we have used local linear smoother for our implementation but any other smoothing methods could be used~\citep{Simonoff:1996}.

The comparison region needs to satisfy the following condition, in order to make sure the parameter estimation is compared only in the common region defined by $w(u)$.
 \begin{eqnarray*}
          w(u)= \prod_{t_j} 1_{[a,b]}\{(u-\theta_{2})/\theta_{3}\},
         \end{eqnarray*}
where $t_j$ is the time interval of Japan's mortality data, $a\geq inf(t_j)$ and $b\leq sup(t_j)$. To consider the importance of more recent data's impact on future trend, one could impose different weights on different support intervals. 

\paragraph{Algorithm}
To estimate the parameters by the nonlinear least squares estimation criterion given in~(\ref{min}), we first obtain the estimates of $k_c$ and $k_t$ by nonparametric local linear smoothing, denoted by $\hat k_c$ and $\hat k_t$ respectively. Then we set up the initial estimates $\theta^0 = (\theta_{1}^0,\theta_{2}^0,\theta_{3}^0, \theta_{4}^0)$ and solve the nonlinear least squares estimation problem by iteratively updating the estimates until convergence.    

\begin{figure}[H]
\captionsetup[subfigure]{labelformat=empty}
\subfloat[]
  {\includegraphics[width=0.5\textwidth]{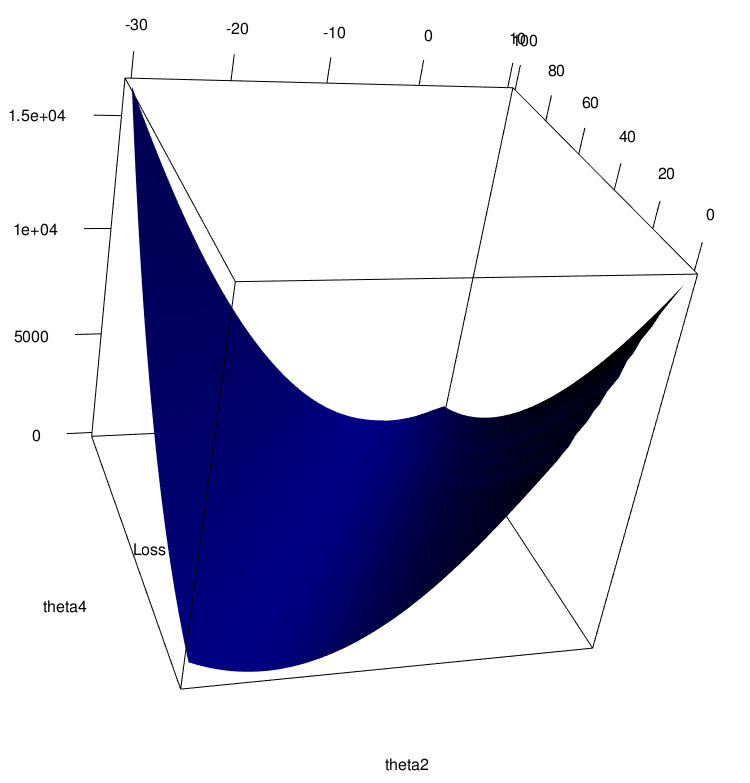}}\hfill
\subfloat[]
  {\includegraphics[width=0.45\textwidth]{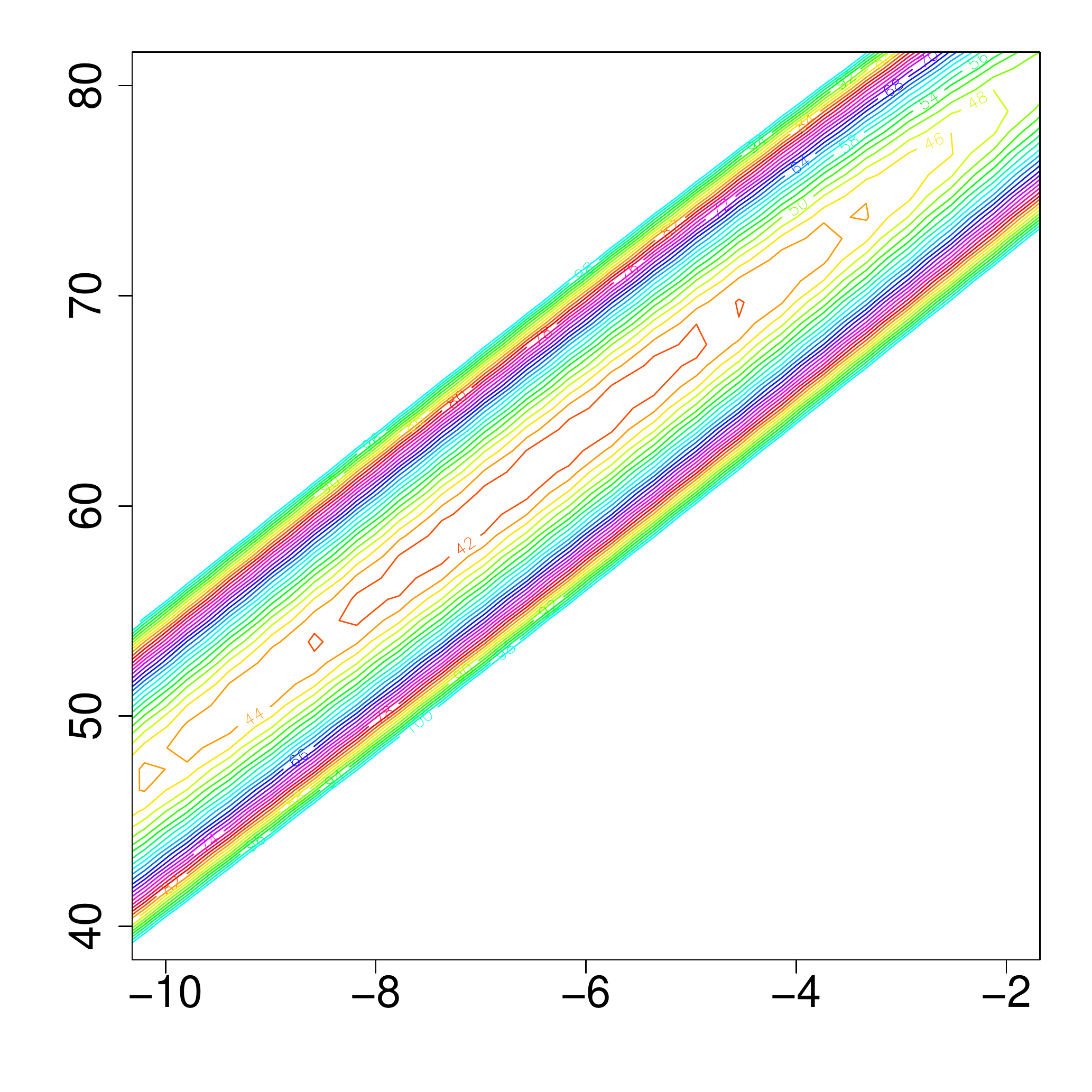}}
  \vspace{-0.5cm}
\caption{Loss surface of $\theta_{2}$ and $\theta_{4}$ (left) and Contour of $\theta_{2}$ and $\theta_{4}$ (right) with $\theta_1 = \theta_3 = 1$.\label{linear}}
\href{https://github.com/QuantLet/MuPoMo}{\quantnet MuPoMo}
\end{figure}

\paragraph{Initial choice of $\theta_{2}$ and $\theta_{4}$}

\begin{figure}[H]
	  \centering
	  \includegraphics[width=0.5\textwidth]{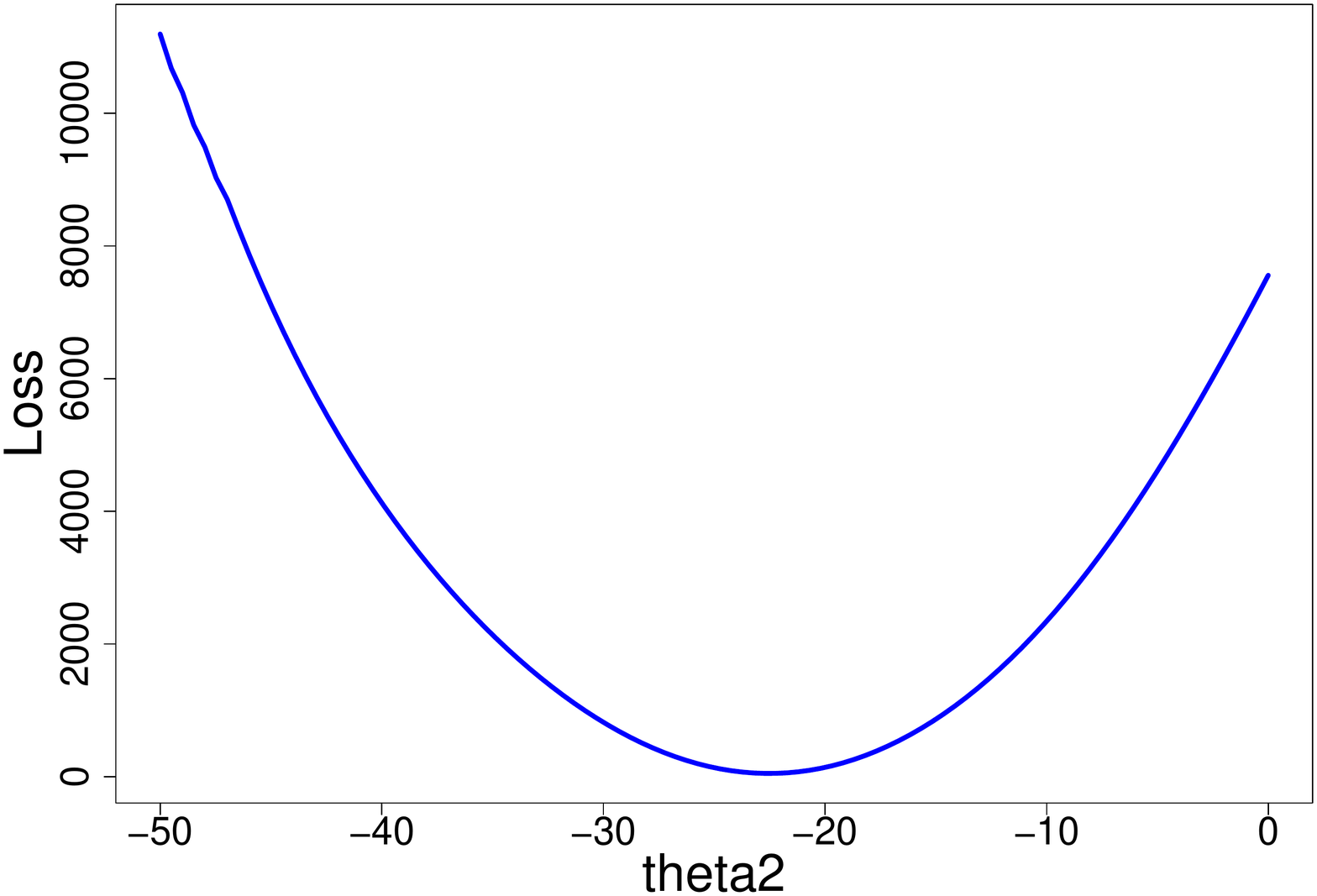}
	  \vspace{-0.5cm}
\caption{Loss function of $\theta_{2}$ with $(\theta_{1},\theta_{3},\theta_{4})^{\top}=(1,1,0)^{\top}$.\label{loss2}}
\href{https://github.com/QuantLet/MuPoMo}{\quantnet MuPoMo}
\end{figure}

From previous Figure \ref{ktcom} and in our initial analysis, we see that there is a potential ambiguity between $\theta_{2}$ and $\theta_{4}$. It is also clear that we can find the replacement relationship from Figure~\ref{theta}. To investigate further, we show the criterion function in Figure~\ref{linear} as a function of $(\theta_2,\theta_4)$. There is a valley area in the loss surface function of $\theta_{2}$ and $\theta_{4}$ in the left plot and also in the contour of $\theta_{2}$ and $\theta_{4}$ in the right one, which suggests that there exists an approximate linear combination of $\theta_{2}$ and $\theta_{4}$ in searching $\theta$ for an optimal solution. It will bring about difficulties in finding the optimal parameters $\theta$ in numerical optimization. This ambiguity in identifiability of the parameters is one of the ramifications of the partial observation in our setting.  

In order to find the optimal value, we should be very careful with selecting initial values of $\theta$ and in the consideration of identifiability issues we need to decide whether the analysis concentrates on time delay or vertical shift. In our analysis we stick with time delay influence $\theta_{2}$ since it is more valuable in prediction perspective, and thus the initial value of $\theta$ is determined as $(\theta_{1},\theta_{3})^{\top}=(1,1)^{\top}$ and set $\theta_4 = 0$. Hence the optimal initial $\theta_{2}$ is obtained around -23, see Figure \ref{loss2}.

\begin{figure}[H]
    \centering
	  \includegraphics[width=0.7\textwidth]{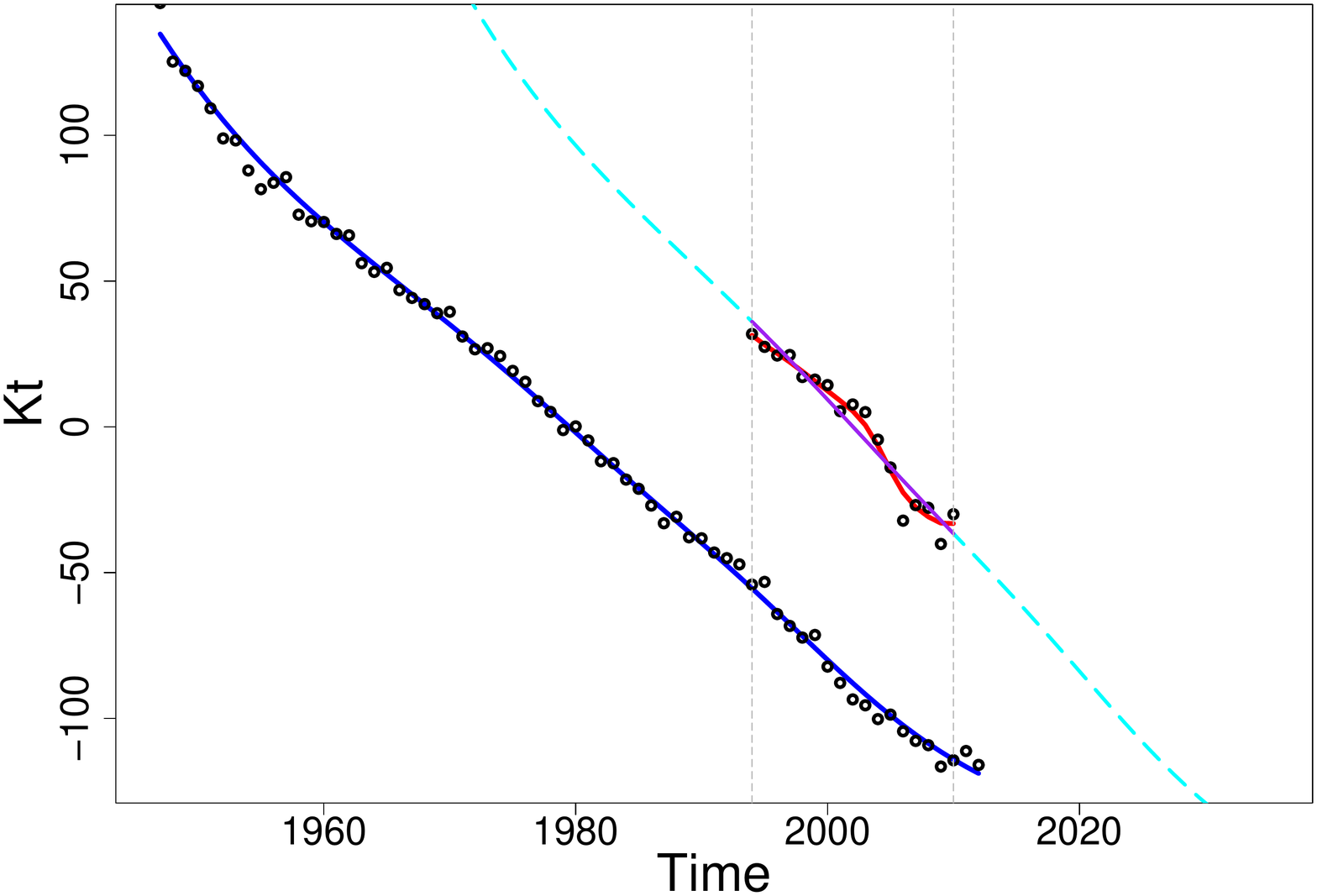}
	  \vspace{-0.5cm}
    \caption{Goodness of Fit and forecast of China's mortality trend from 2011 to 2030 via Japan's historical data: black dots represent the original $k_t$ from Japan and China, and \color{blue}{Japan smoothed trend}, \color{red}{China smoothed trend} \color{black} are displayed as well; the fitted trend is plotted as light blue dashed line, while the overlapping part is colored in purple; the light blue dashed line after year 2011 is the forecast part.\label{fore}}
\href{https://github.com/QuantLet/MuPoMo}{\quantnet MuPoMo}
\end{figure}

\paragraph{Goodness of Fit}
Based on the initial $\theta^{(0)}=(\theta_{1},\theta_{2},\theta_{3})^{\top}=(1,-23,1)^{\top}$ and algorithm, the optimal parameter $\theta$ is reached at $\hat{\theta}=(1.205,-22.621,1.000)^{\top}$. This is used in Figure~\ref{fore}, showing that after the optimal transformation of the curve of Japan (based on the optimal value of $\theta$) the $k_t$ of China fits quite well in the $k_t$ of Japan of years around from 1970 to 1990.

\subsection{Forecast}

Afterwards we can forecast $k_t$ for China via the data from Japan and the optimal estimated shape deviation parameter $\hat{\theta}$ to extend the forecasting horizon.
         \begin{eqnarray}
          k_c(t+h)=\hat{\theta}_{1}k_j\left\lbrace \frac{(t+h)-\hat{\theta}_{2}}{\hat{\theta}_{3}}\right\rbrace,
         \end{eqnarray}
where $\hat{\theta}=(1.205,-22.621,1.000)^{\top}$ and $t = 1994,1995,...,2010; h=1,2,...,20$.

Compared with the traditional forecasting method with time series analysis, our proposed method can extend the forecasting horizon from 5 years to 23 years, which is a big advantage from this semi-parametric comparison technique of regression curves, see Figure \ref{fore}. However, searching for numerical optimal solutions for seemingly linear regression curves is still a challenge in this context. This has motivated us to develop an extension to the multi-populations mortality models, because the identifiability issues can be better solved by borrowing information across multiple curves.

\section{Mortality forecasting: Multi-population models} \label{sec:multi-pop}

Based on the previous study of two countries under sparse data, investigation of multi-populations becomes promising and necessary since more information on nonlinear trend will be provided in case of multi-countries. At the same time it enables us to study global mortality trend in the past century and its future.

\begin{figure}[H]
\captionsetup[subfigure]{labelformat=empty}
\subfloat[]
  {\includegraphics[width=0.5\textwidth]{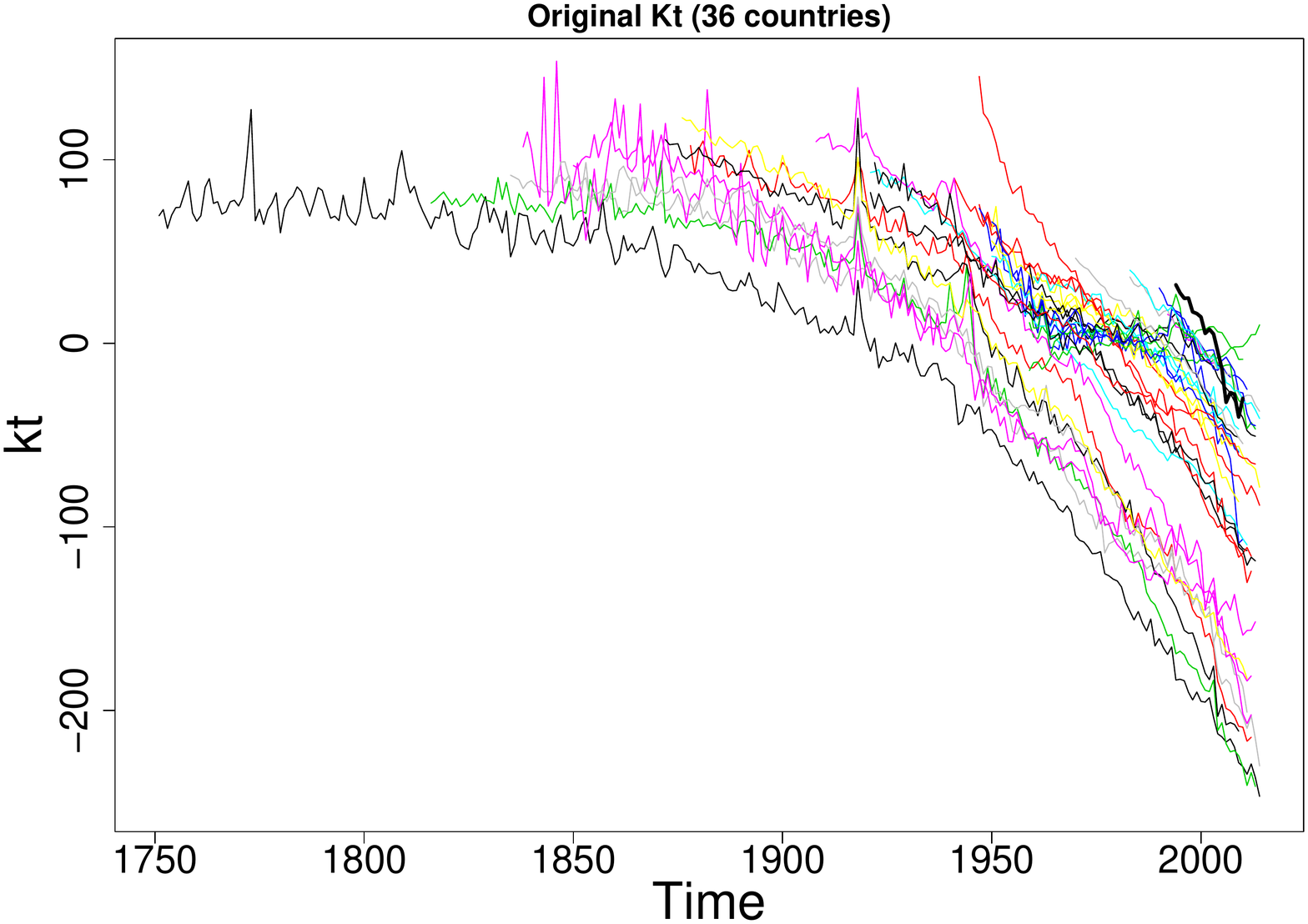}}\hfill
\subfloat[]
  {\includegraphics[width=0.5\textwidth]{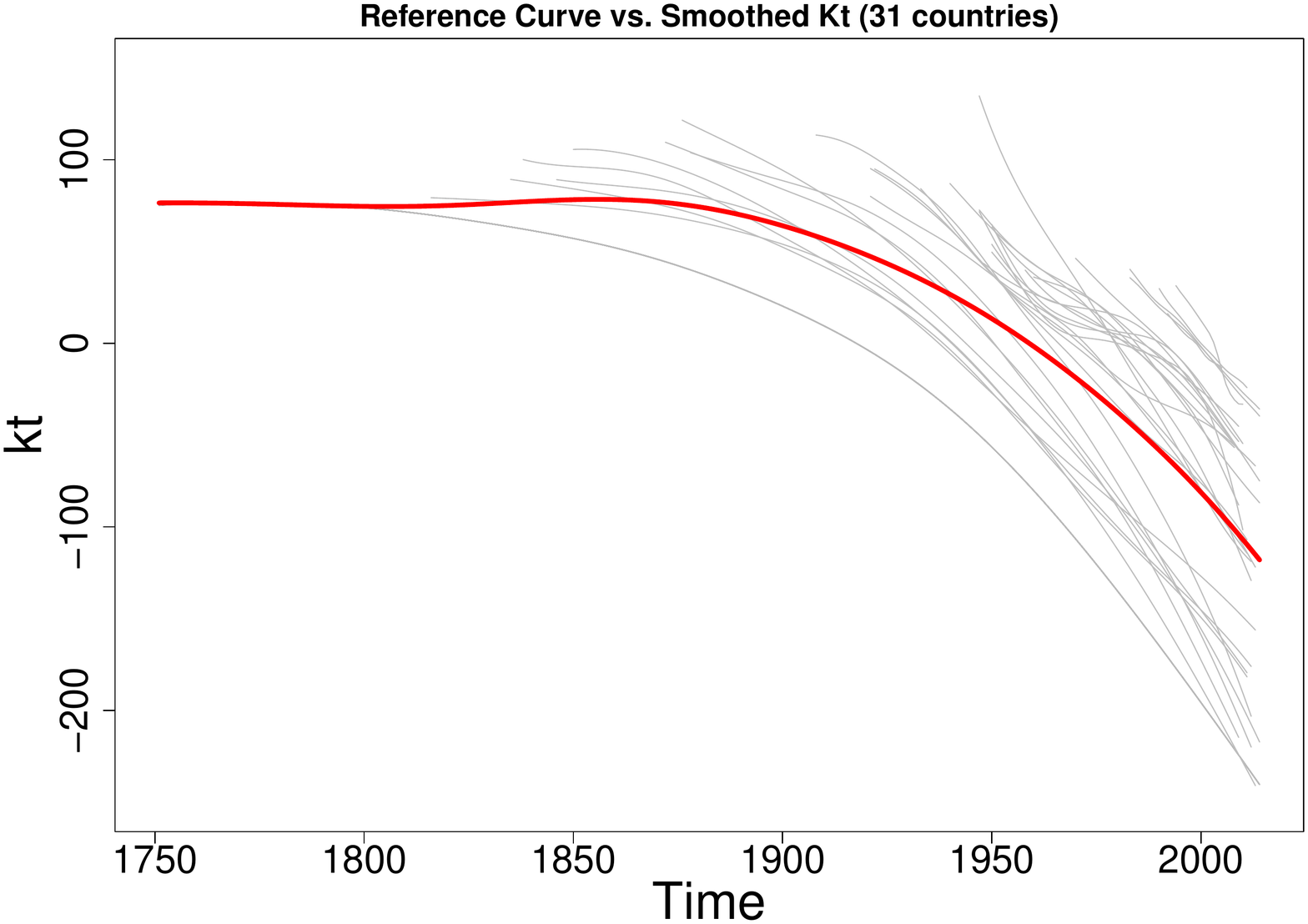}}
  \vspace{-0.5cm}
\caption{Similarities of mortality trend among countries: different colors represent different countries (left), and the red thick curve in the right plot stands for reference curve while grey ones are smoothed curves.\label{fig:multi-country}}
\href{https://github.com/QuantLet/MuPoMo}{\quantnet MuPoMo}
\end{figure}

Denote by $k_i$ the time-varying mortality indicator for country $i$, with $i\in \{1,...,N\}$.   
Figure~\ref{fig:multi-country} displays the estimates of $k_i$ for 36 countries in the database. The left panel shows the noisy curves that are originally estimated $k_t$ from the LC model without any further nonparametric smoothing, while on the right it shows the smoothed $k_t$ with an initial estimate of the reference curve overlayed. Later on, we will discuss why 31 countries are selected for analysis. By design, the available time measurements vary among countries. Nevertheless, we notice remarkable similarities in the trend across the countries, subject to individual variability.
 
In order to investigate this structural similarity and to borrow the information across the countries, we consider the shape invariant model introduced in Section~\ref{sec:commontrend} of methodology part.

\subsection{Model formulation}
Specifically we assume the additive noise model defined in~(\ref{e:additive}) for the derived $k_i$s from LC model to account for noise and suppose that the underlying curves share some common trend and can be represented in the form
         \begin{eqnarray}
          k_i(t)=\theta_{i1}k_0\left( \frac{t-\theta_{i2}}{\theta_{i3}}\right) +\theta_{i4}, \label{e:sim}
         \end{eqnarray}
where 
$k_0(t)$ is a reference curve, understood as common trend 
and $\theta_i=(\theta_{i1},\theta_{i2},\theta_{i3},\theta_{i4})^{\top}$ are shape deviation parameters. In order to be able to interpret the reference curve $k_0$ as a mean trend, we can use the normalizing constraints on the parameter $\theta_i$ as
         \begin{eqnarray}
          N^{-1}\sum_{i=1}^{N}\theta_{i1} = N^{-1}\sum_{i=1}^{N}\theta_{i3} =1,\\
          N^{-1}\sum_{i=1}^{N}\theta_{i2} = N^{-1}\sum_{i=1}^{N}\theta_{i4} =0 \,.
         \end{eqnarray}
         
Alternatively, we can use any country as a reference curve, for example, Sweden as the longest record holder could be a reasonable choice, in which case the reference curve is set to be $k_t$ of Sweden with $\theta_0=(1,0,1,0)$ and $\theta_i$ will measure the deviation with respect to the reference curve. In this work we will consider the mean curve as a reference curve and use the above normalization constraints. 

\subsection{Estimation}

\paragraph{Estimation of parameters}
Suppose that $k_0$ and $k_i$ are given. Then for each country $i$, the parameter $\theta_i$ can be determined by minimizing the least squares criterion as
\begin{equation}\label{e:nlse}
\int \Big\{k_i(t) - \theta_{i1} k_0\Big(\frac{t-\theta_{i2}}{\theta_{i3}}\Big) - \theta_{i4} \Big\}^2 w_i(t)\,dt
\end{equation}
where $w_i$ is chosen to ensure that the two functions are evaluated over the common domain as in the case of two countries. In practice, $k_0$ and $k_i$ are replaced by its nonparametric estimate. 

\paragraph{Estimation of Common Trend}

For given parameters $\theta_i, i=1,\ldots, n$, the functional relationship in~(\ref{e:sim}) implies that
         \begin{eqnarray}
          k_i(\theta_{i3}t+\theta_{i2})=\theta_{i1}k_0(t)+\theta_{i4} ,
         \end{eqnarray}
Thanks to the normalizing conditions on $\theta_{i1}$ and $\theta_{i4}$, this implies that
\begin{equation}\label{e:common-trend}
k_0(t) = N^{-1}\sum_{i=1}^N k_i(\theta_{i3}t+\theta_{i2}) \,.
\end{equation}
That is, if $k_i$ is appropriately transformed with respect to the individual parameters $\theta_i$, then $k_0$ is simply the average.  
In practice, we have noisy version of $k_i$ available at different number of time points. Then the functional mean can be estimated more efficiently with nonparametric smoothing, which essentially gives rises to a weighted average estimate.  

\paragraph{Estimation algorithm}

Combining the above two steps leads to the following iterative algorithm for estimation of the parameters. \begin{enumerate}
\item[(a)] Given $k_i$, obtain an initial estimate of $k_0$ based on all country-level mortality rates
\item[(b)] Given $k_0$, update $\theta_i$ by minimizing the nonlinear least squares criterion in (\ref{e:nlse}) for each $i=1,\ldots, N$.
\item[(c)] Normalize the parameters to satisfy the constraints. 
\item[(d)] Given $\theta_i, i=1,\ldots, N$, update $k_0$ by (\ref{e:common-trend}).
\item[(e)] Iterate (b)-(d) until convergence. 
\end{enumerate}
The general algorithm was proposed and studied by \citet{Kneip:Engel:1995}. We adapted accordingly to account for incomplete observations in our sample.

\paragraph{Initial values of $k_0$ and $k_i$}
To initialize $k_0$, we choose the trimmed mean of the sample estimates, based on the middle 50\% of the countries in terms of the length of the recording period. The estimation of $k_0$ and $k_i$ is done with local linear kernel smoothing method to account for measurement error. The smoothing parameter for $k_i$ was selected to maintain comparable smoothness across the samples using common degrees of freedom~\citep{Bowman:Azzalini:1997}.


\paragraph{Computational Issues with the Parameterization}
The shape invariant model implicitly assumes that there are identifiable features that are common across the sample. It is easy to check for densely observed curves (with non-monotone functions) by means of the derivative estimation, but for sparsely observed curves, there could be an ambiguity in identifying the parameters. In the case of the mortality curves, due to the limited measurements available, the ambiguity occurs in distinguishing the differential effect of vertical shift ($\theta_4)$ and horizontal shift $(\theta_2)$ in time. In this case, we choose to attribute the effect as horizontal shift and set $\theta_4=0$, as this is more amenable to interpretation and meanwhile promising to extend forecast horizon. Comparison of these two parameterization cases will be illustrated in Section~\ref{sec:global-trend}. 

\paragraph{Bootstrap for Prediction Interval}

The forecasting for Chinese mortality~(\ref{e:sim}) is now updated based on the common reference curve $k_0$.
In order to construct a prediction interval of China's mortality trend, we also need a reliable estimate of the measure of variability.
Unlike the standard setting studied by \citet{Kneip:Engel:1995} for relatively densely observed data on common intervals, it is difficult to derive an asymptotic distribution of the estimators for sparse and incomplete data as ours. Here, we opt for a bootstrap method to approximate the uncertainties in estimation and prediction.  

The standard bootstrap techniques relying on identically independent distributed (i.i.d.) observations are not appropriate here. 
Recently re-sample methods for dependent data have considered several options: bootstrap with i.i.d. innovations, bootstrap with block segments and model-based bootstrap. 
Due to limited sample of China's mortality time series, bootstrap with block segments do not create the ideal re-sampled time series. Alternatively we bootstrap the mortality data based on i.i.d. innovations obtained from fitting time series model, and afterwards we carry out estimation on the re-sampled data and generate prediction interval at different levels.

Suppose that for the time series $k_1,..., k_n$ and some fixed $p \in N$, there exists a parametric estimator of the conditional expectation $E(k_t \vert k_{t-1},..., k_{t-p})$ denoted by $\widehat{m}_n(k_{t-1},..., k_{t-p})$. This estimator leads to residuals
\begin{equation}\label{boot}
\widehat{e_t}:= k_t - \widehat{m}_n(k_{t-1},..., k_{t-p}), t = p+1,...,n.
\end{equation}
Resampling from these residuals leads to a bootstrap sample of time series
\begin{equation}\label{boot_ts}
k_t^{*} = \widehat{m}_n(k_{t-1}^{*},..., k_{t-p}^{*}) + e_t^{*}, t = p+1,...,n.
\end{equation}
The idea of parametric fit to the conditional expectation can be executed by ARIMA models. \citet{Kreiss:Lahiri:2012} discuss different situations with parametric and nonparametric modeling of predictor $k_t$ and respective asymptotic consistence properties.

\subsection{Global Trend}

\subsubsection{Outlying countries}

In light of similarities across countries, seeking global mortality trend becomes very reasonable and natural. Some research also find out mortality trend has connection with economic development level, GDP for instance, see \citet{Hanewald:2011}. In the remaining section, we are going through the empirical analysis of common trend in different groups via slightly different models.
\begin{figure}[H]
    \centering
	  \includegraphics[width=0.7\textwidth]{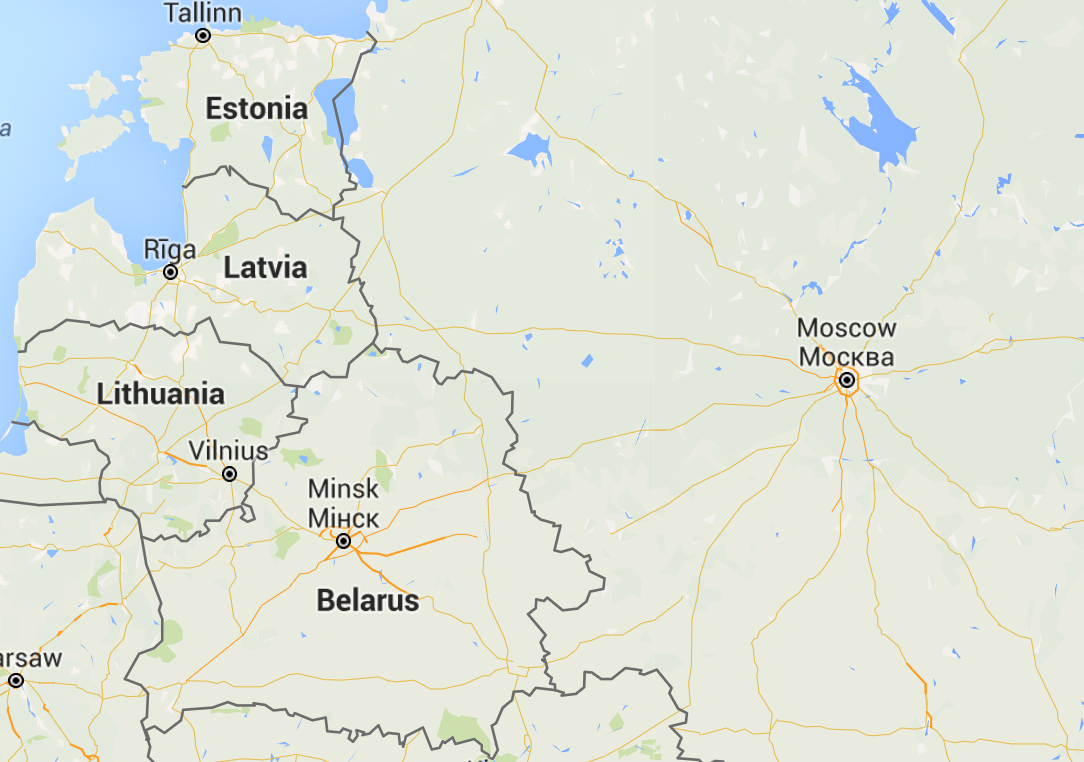}
    \caption{Five outlying countries: geographically neighbors in east Europe. Source: Google Map.\label{map}}
\end{figure}

Evidence tells us that the mortality rate is decreasing as time evolves, due to medical improvement, economic development and social stability. However, we notice that there are some remarkable outliers like Russia, Estonia, Latvia, Lithuania and Belarus in the database. 

\begin{figure}[H]
\captionsetup[subfigure]{labelformat=empty}
\subfloat[]
  {\includegraphics[width=0.5\textwidth]{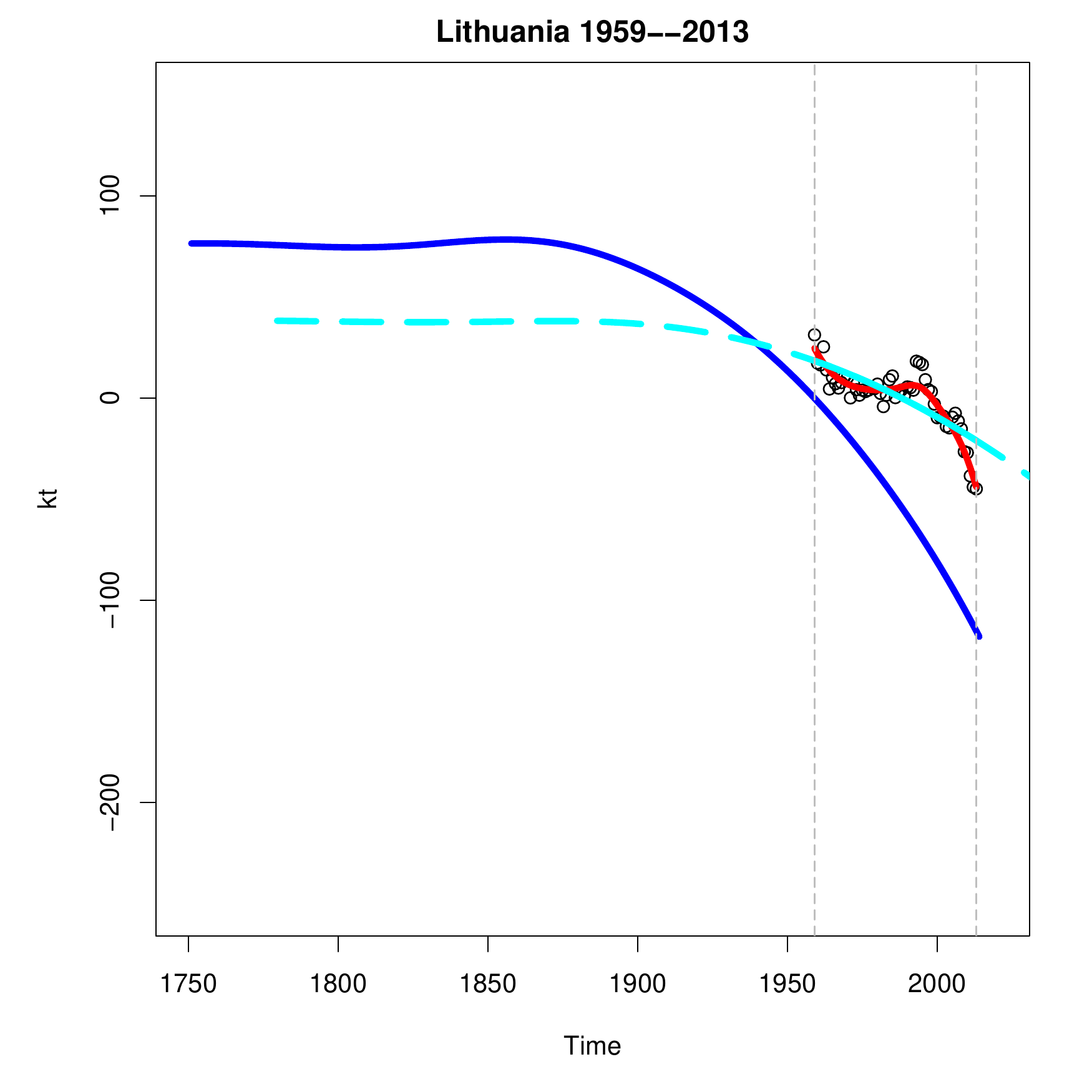}}\hfill
\subfloat[]
  {\includegraphics[width=0.5\textwidth]{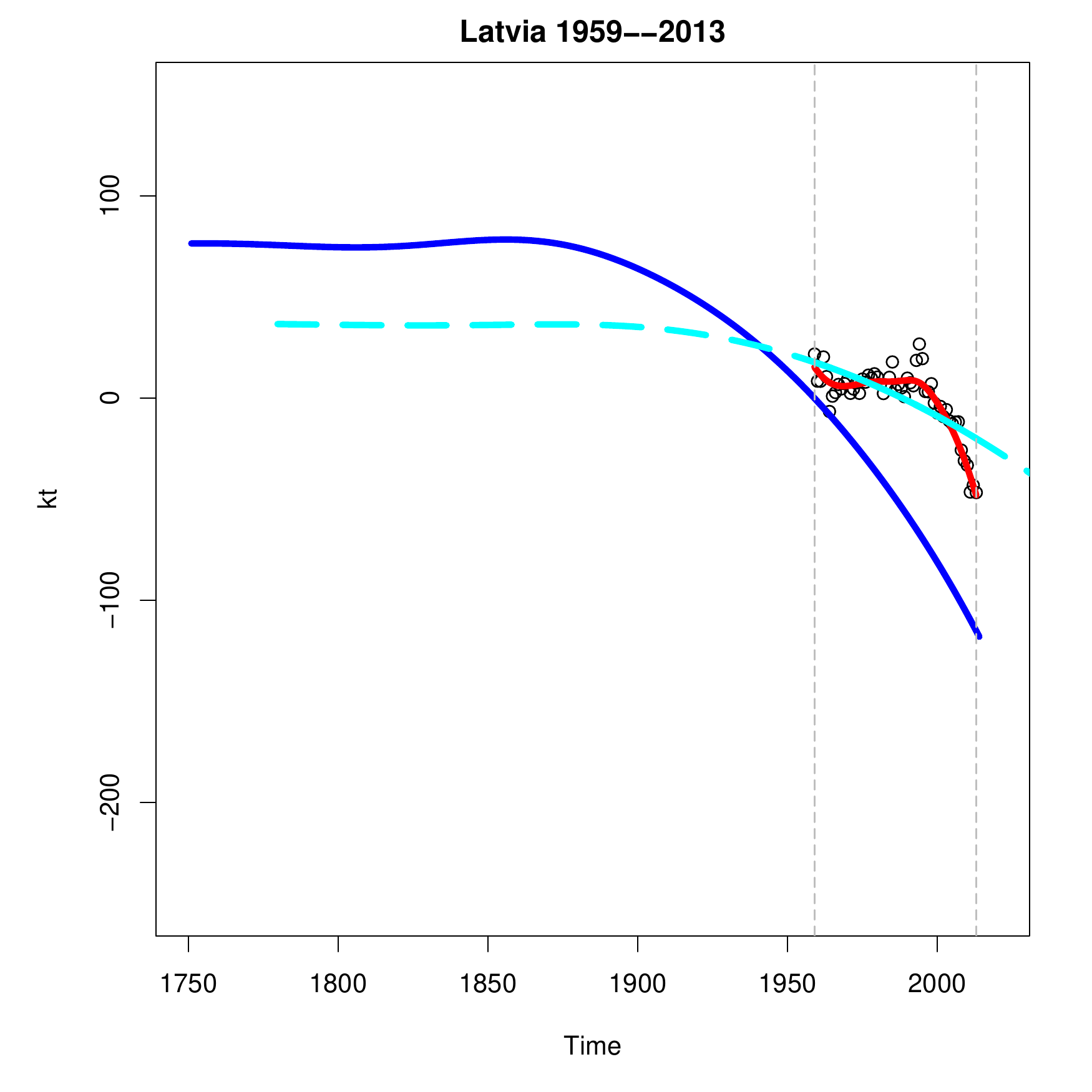}}
\caption{Different mortality movements of Lithuania and Latvia: blue curves represent global mortality trend, light blue curves are representing estimated individual country-level mortality trends based on global trend, red curves are smoothed original individual country-level mortality trends and black dots are original individual country-level mortality trends.\label{outly}}
\href{https://github.com/QuantLet/MuPoMo}{\quantnet MuPoMo}
\end{figure}

As shown in Figure \ref{map}, all these five countries locate themselves in east Europe and are used to be members of the Soviet Union. They share similar geographic characteristics and meanwhile experienced parallel economic and social progress, and within our expectation they reflect comparable mortality moving path as well, as displayed in Figure \ref{outly} and Appendix 1. 
Note that solid (blue) curves represent global mortality trend, dashed (yan) curves are representing estimated individual country-level mortality trends based on global trend, the short solid (red) curves are smoothed original individual country-level mortality trends and (black) circles are original individual country-level mortality trends.

Surprisingly, they exhibit a quite opposite tendency in contrast with other 31 countries. The mortality rates go through a short period of decrease, then stay stable or slightly increase for several years and afterwards go back to declining path again. One possible reason on this different phenomenon perhaps is connected with political event of dissolution of the soviet union.

To ideally demonstrate global mortality movements in majority of countries, we remove these five countries for remaining study to reduce influences from minor outlying ones.

\subsubsection{Mortality Trend among the Majority} \label{sec:global-trend}

As discussed previously, for sparsely observed curves there could be an ambiguity in identifying all of the four parameters. 
Therefore, we choose to compare the original 4-parameters model with a simplified 3-parameter model of setting $\theta_4=0$.
\begin{figure}[H]
\captionsetup[subfigure]{labelformat=empty}
\subfloat[]
  {\includegraphics[width=0.5\textwidth]{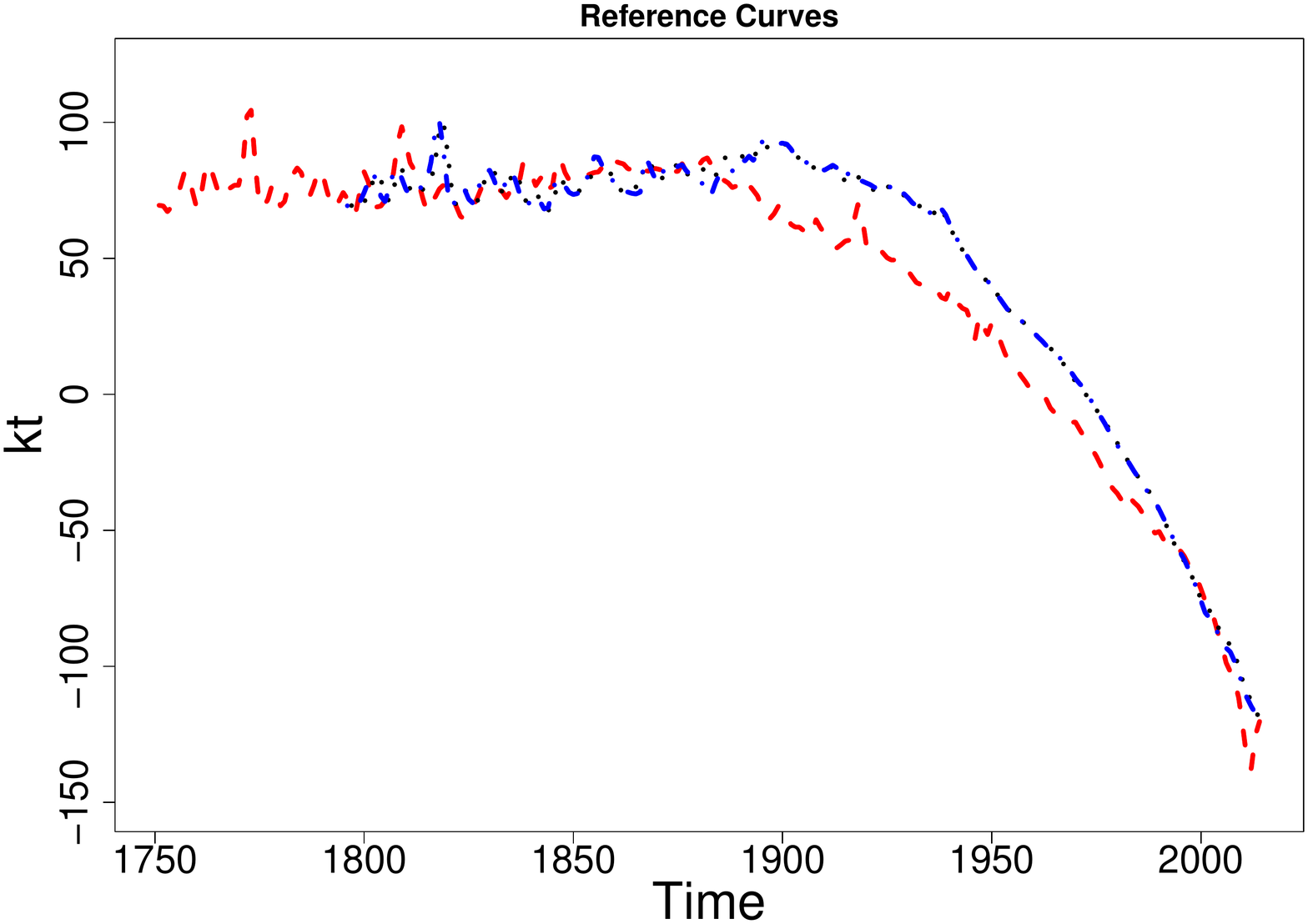}}\hfill
\subfloat[]
  {\includegraphics[width=0.5\textwidth]{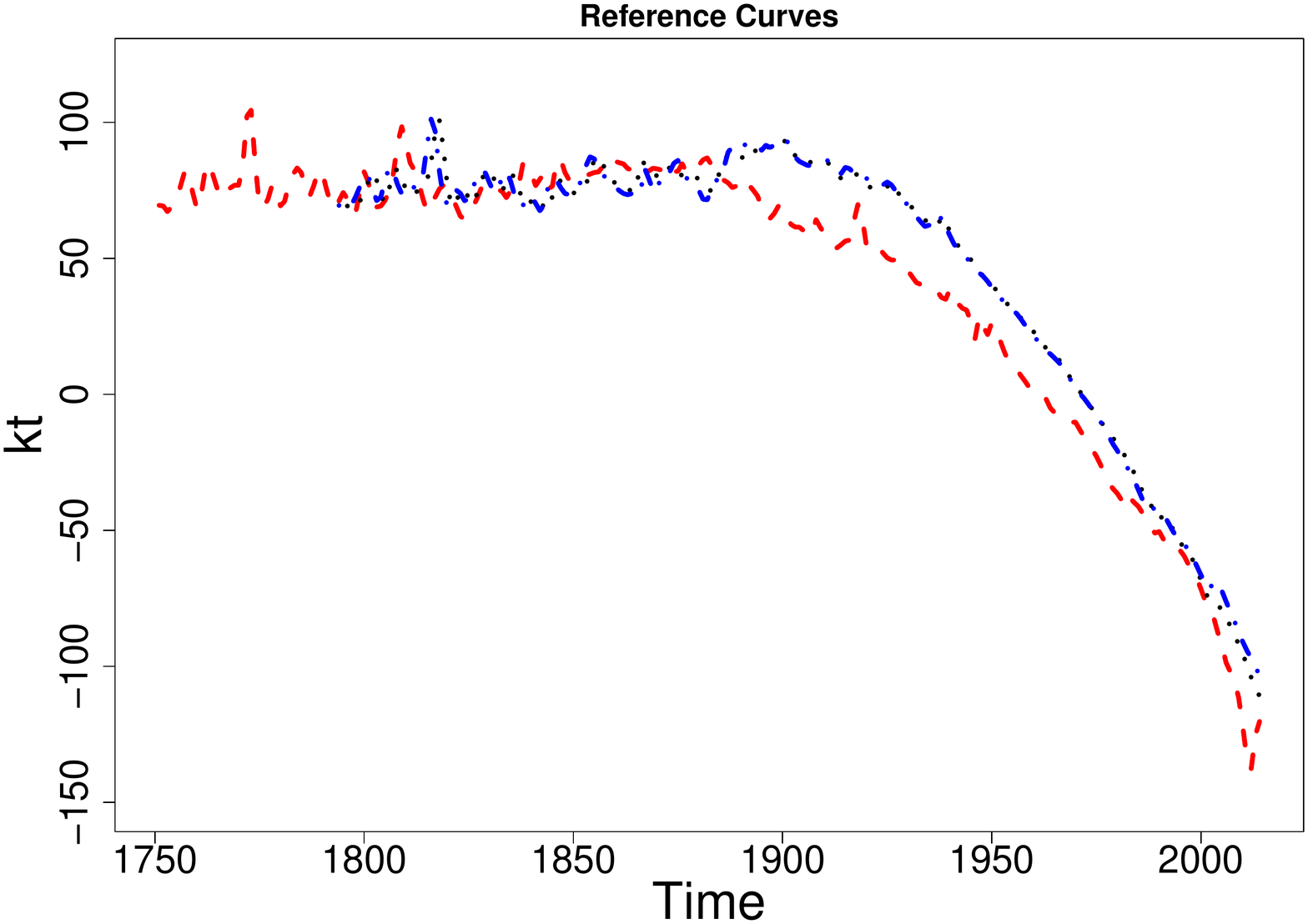}}
\caption{Common mortality trends estimated by 3-parameters model (left) vs. 4-parameters model (right).\label{3vs4}}
\href{https://github.com/QuantLet/MuPoMo}{\quantnet MuPoMo}
\end{figure}
Under two different parameterization, we estimate the parameters and reference curve respectively. In Figure \ref{3vs4}, the reference curves and common trend for these two cases are plotted: the left one is calculated from the 3-parameters model and the right one is generated from the 4-parameters one. The red curve is the initial reference curve, blue ones are one-step ahead updated reference curves. 
From these two plots, no clear and obvious difference can be seen. But from the viewpoint of analytic thinking, we choose to attribute the effect as horizontal shift and set $\theta_4=0$, as this is more amenable to interpretation and meanwhile promising to extend forecast horizon. 

\begin{figure}[H]
    \centering
	  \includegraphics[width=0.7\textwidth]{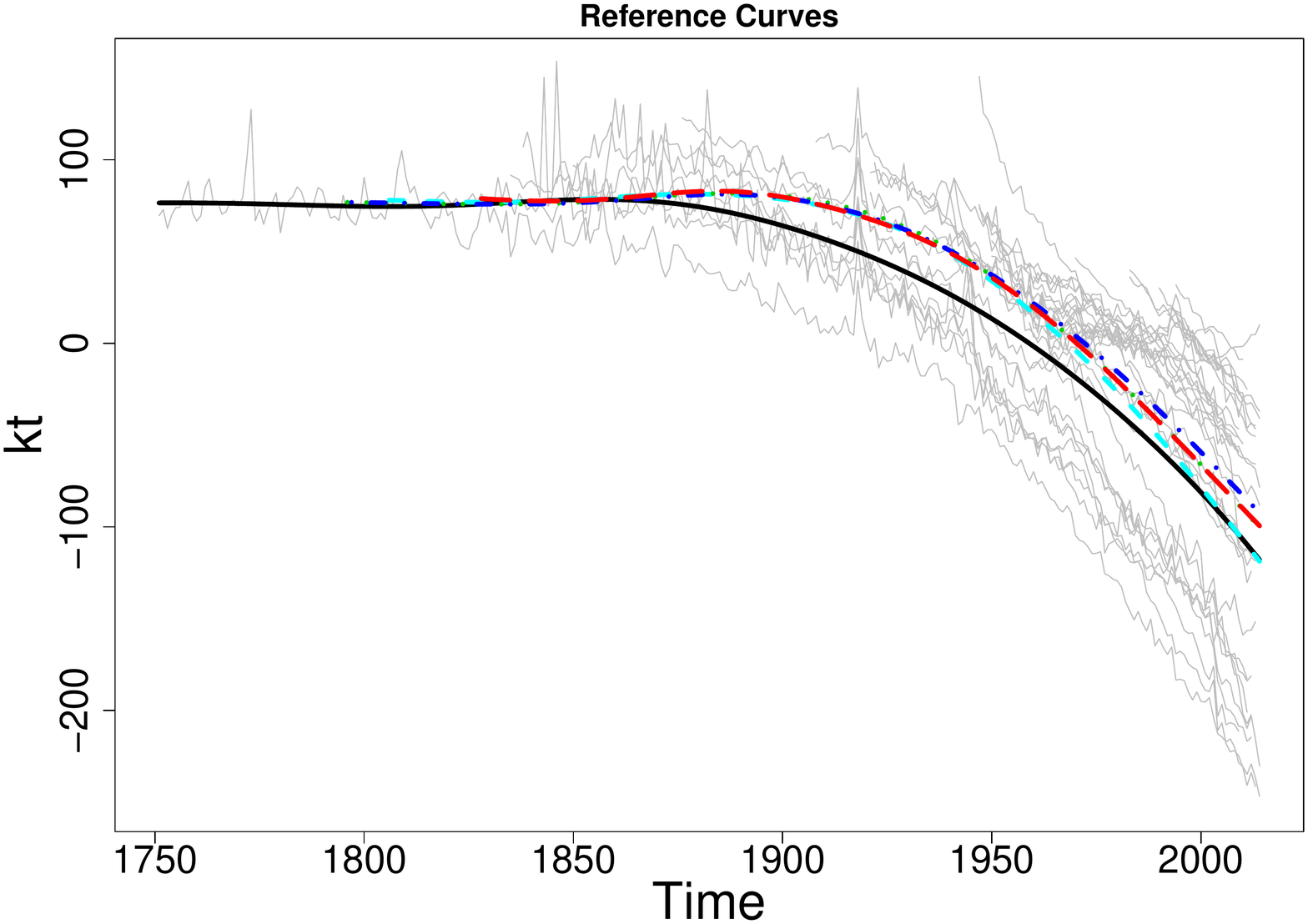}
    \caption{Common mortality trend compared with individual nation-level mortality trends: red curve is the initial reference curve, blue and black ones are updated reference curves convergent to common trend, while the grey lines represent individual country. \label{ref3vsnations}}
    \href{https://github.com/QuantLet/MuPoMo}{\quantnet MuPoMo}
\end{figure}

Figure \ref{ref3vsnations} explains the common mortality trend generated from 3-parameters model compared with individual nation-level mortality trend. In this graph, the black solid curve is the initial reference curve, cyan, green, blue and red ones represent the updated ones at different iteration stage while the grey ones are the non-smoothed mortality trend from each country. It is obvious that after one step optimization, reference curves are already showing a quite similar pattern. The common mortality trend is adjusted to an upper level, mostly because more developing countries (Czech Republic, Hungary and China, for example) started collecting demographic data at a later time period in contrast with more developed countries (such as Sweden, Norway and France), and also developing countries have higher mortality rates generally. The figures on illustrating individual case are provided in Appendix 2.

\subsection{China and Global Mortality Trend}

Since a common mortality trend is available, it could be applied to help improve estimation and forecasting of individual case. Especially when sample size from individual country is relatively smaller than anticipated, semi-parametric comparison of common mortality trend with each individual nation-level one will be a promising way with respect to forecasting. At the same time, it helps to reduce ambiguity in identifying the parameters in case of seemingly linear co-movement between regression curves, like the case of comparing China and Japan.

\begin{figure}[H]
    \centering
	  \includegraphics[width=0.7\textwidth]{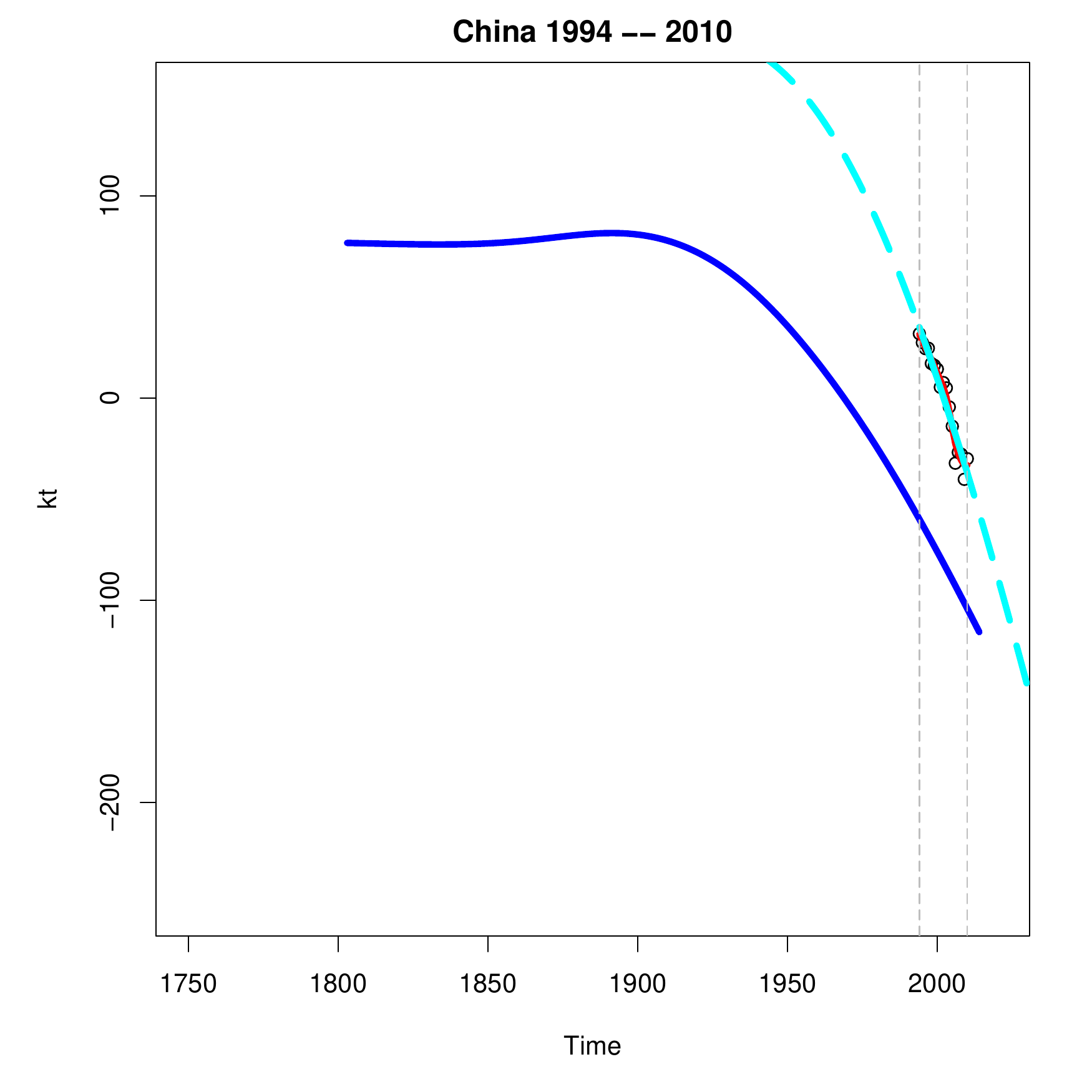}
    \caption{Common mortality trend and estimated China mortality trend based on common trend.\label{ref4vschina}}
    \href{https://github.com/QuantLet/MuPoMo}{\quantnet MuPoMo}
\end{figure}

Figure \ref{ref4vschina} displays newly estimated China mortality trend via semi-parametric comparison with common trend. The (blue) solid line is the common trend or updated reference curve and the (cyan) dashed line is the estimated China mortality trend based on the common trend. In comparison, the raw China mortality curve estimate is marked by black circles, with the individual smoothing estimate overlayed in short (red) line. Thanks to parameters deviation on time axis $\frac{t-\theta_2}{\theta_3}$, we could extend forecasting horizon of China approximately 40 years through the information from common trend, see Figure \ref{ref4vschina}. Referring to estimation and forecasting of the other 30 countries, they are arranged in Appendix 2.

With model-based bootstrap approach, we simulate 500 re-sampled China's mortality time series from 1994 to 2010 based on ARIMA model. From each simulation, we estimate the optimal shape deviation parameters $\theta$ and accordingly calculate the estimated China mortality with longer time horizon based on the common trend. 

\begin{figure}[H]
\captionsetup[subfigure]{labelformat=empty}
\subfloat[]
  {\includegraphics[width=0.33\textwidth]{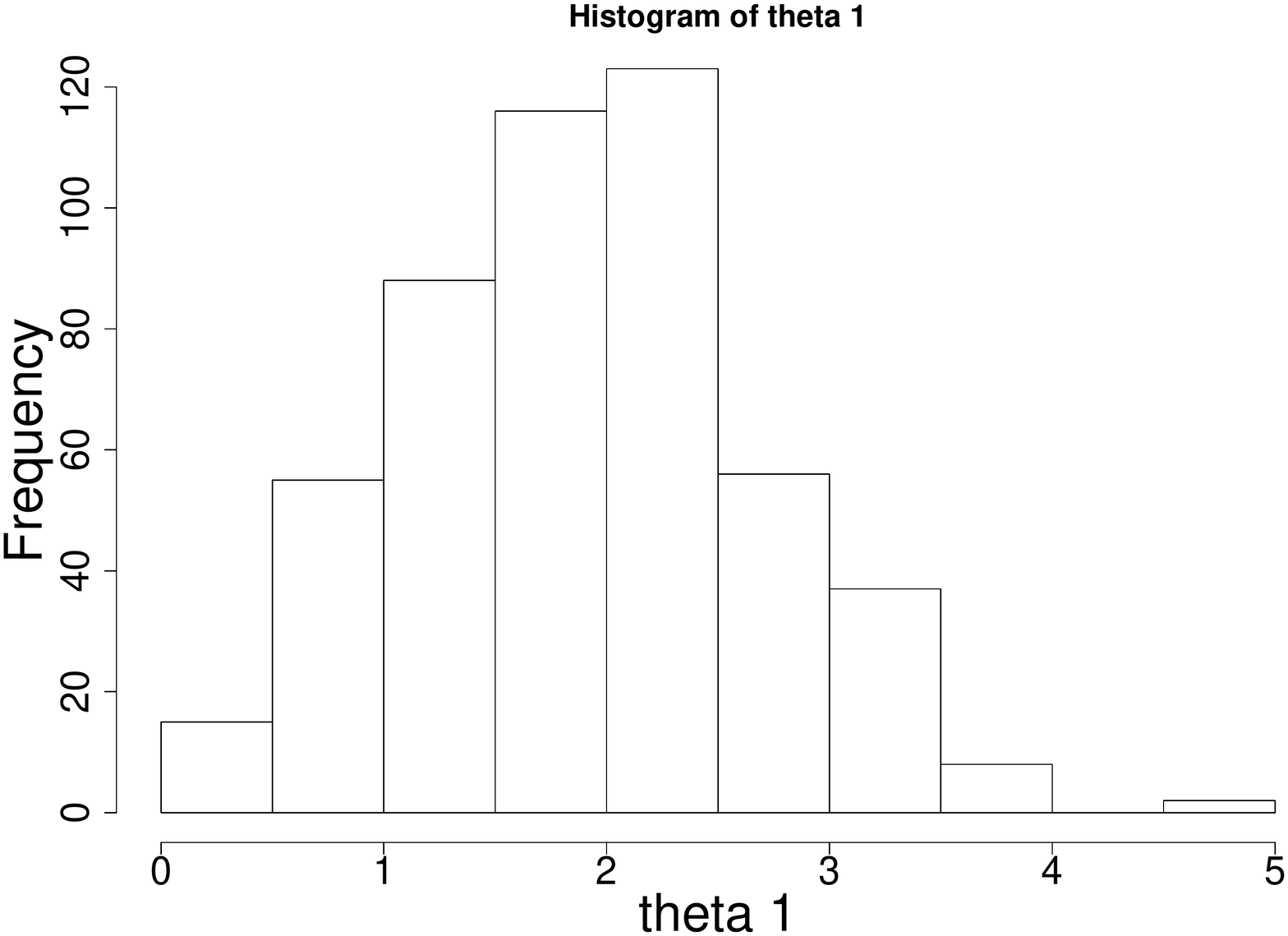}}\hfill
\subfloat[]
{\includegraphics[width=0.33\textwidth]{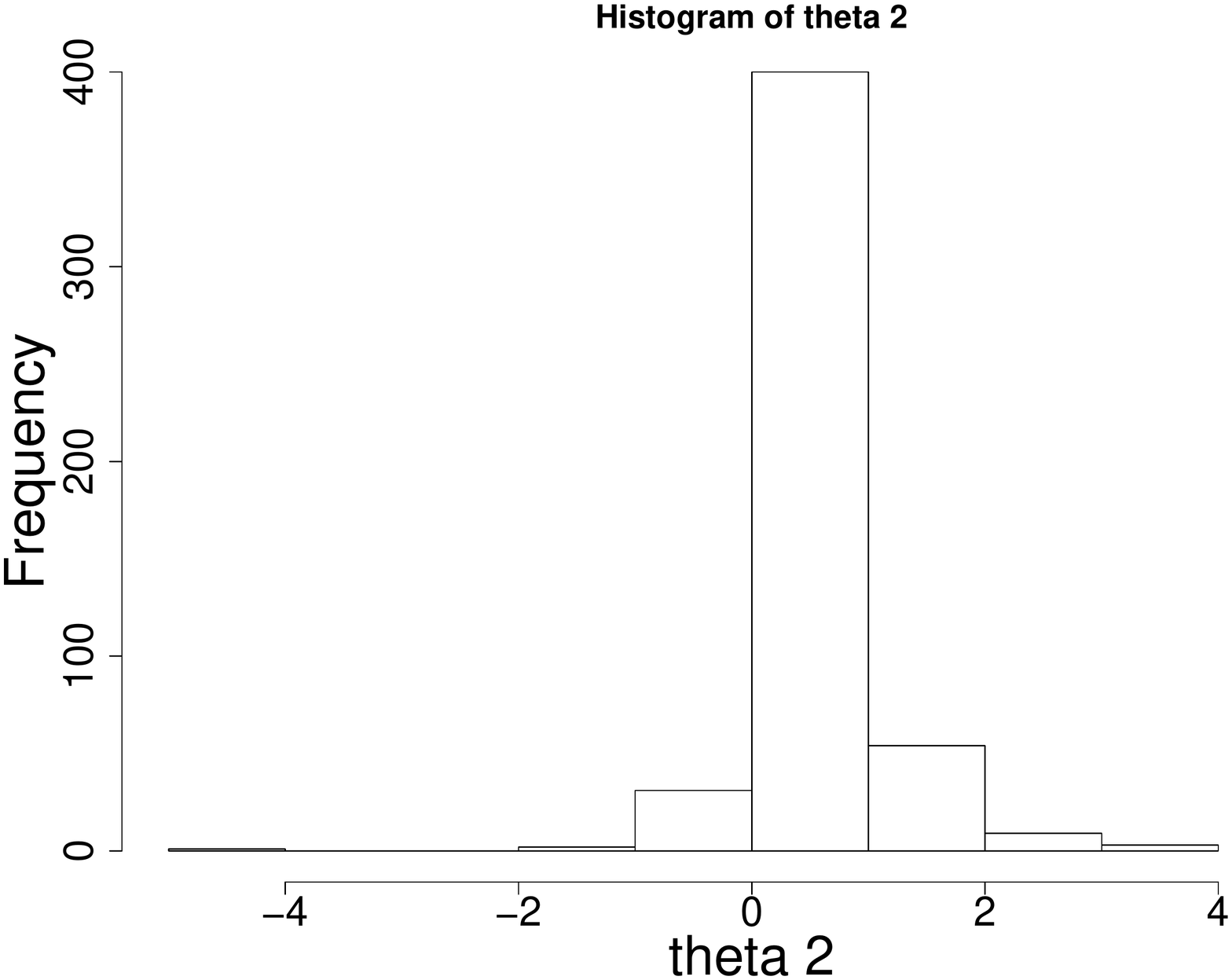}}\hfill
\subfloat[]
  {\includegraphics[width=0.33\textwidth]{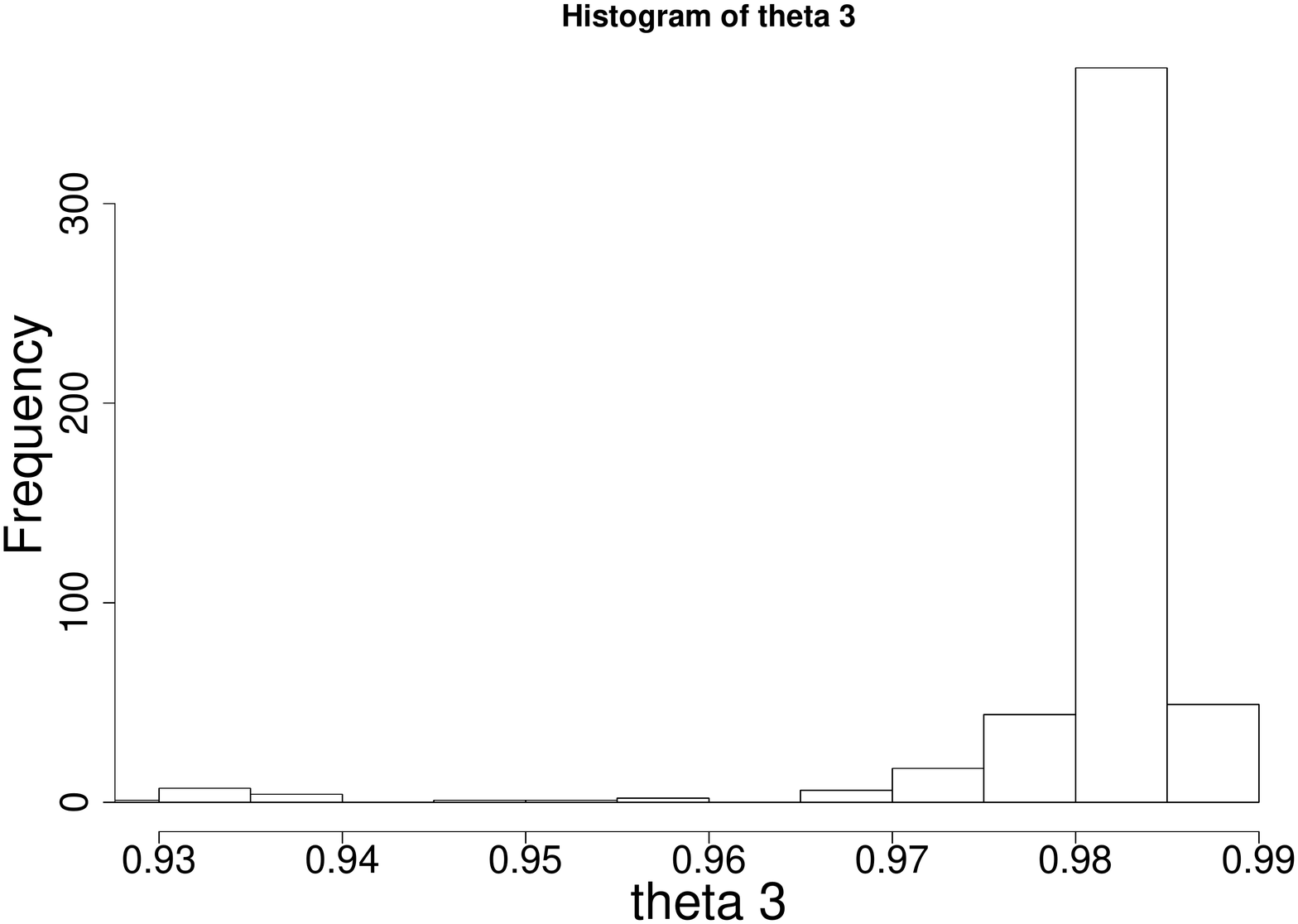}}
\caption{Histograms of $\hat{\theta}_1$,$\hat{\theta}_2$ and $\hat{\theta}_3$.\label{histogram}}
\href{https://github.com/QuantLet/MuPoMo}{\quantnet MuPoMo}
\end{figure}

In Figure \ref{histogram}, we display the variation of $\hat{\theta}_1$,$\hat{\theta}_2$ and $\hat{\theta}_3$ across the countries. From histogram of $\hat{\theta}_1$, $50\%$ of $\hat{\theta}_1$ lies between 1.5 and 2.5, which indicates the overall accelerating declining mortality trend of China compared with global trend. More than $95\%$ of $\hat{\theta}_2$, the parameter describing time delay, falls into the interval of $(0,10)$, which further confirms that there exists a time delay of China's mortality trend around 10 years later than global situation. Majority of $\hat{\theta}_3$ ranges from 0.98 to 1, which reveals a little time acceleration in China's mortality trend.

\begin{figure}[H]
\captionsetup[subfigure]{labelformat=empty}
\subfloat[]
  {\includegraphics[width=0.5\textwidth]{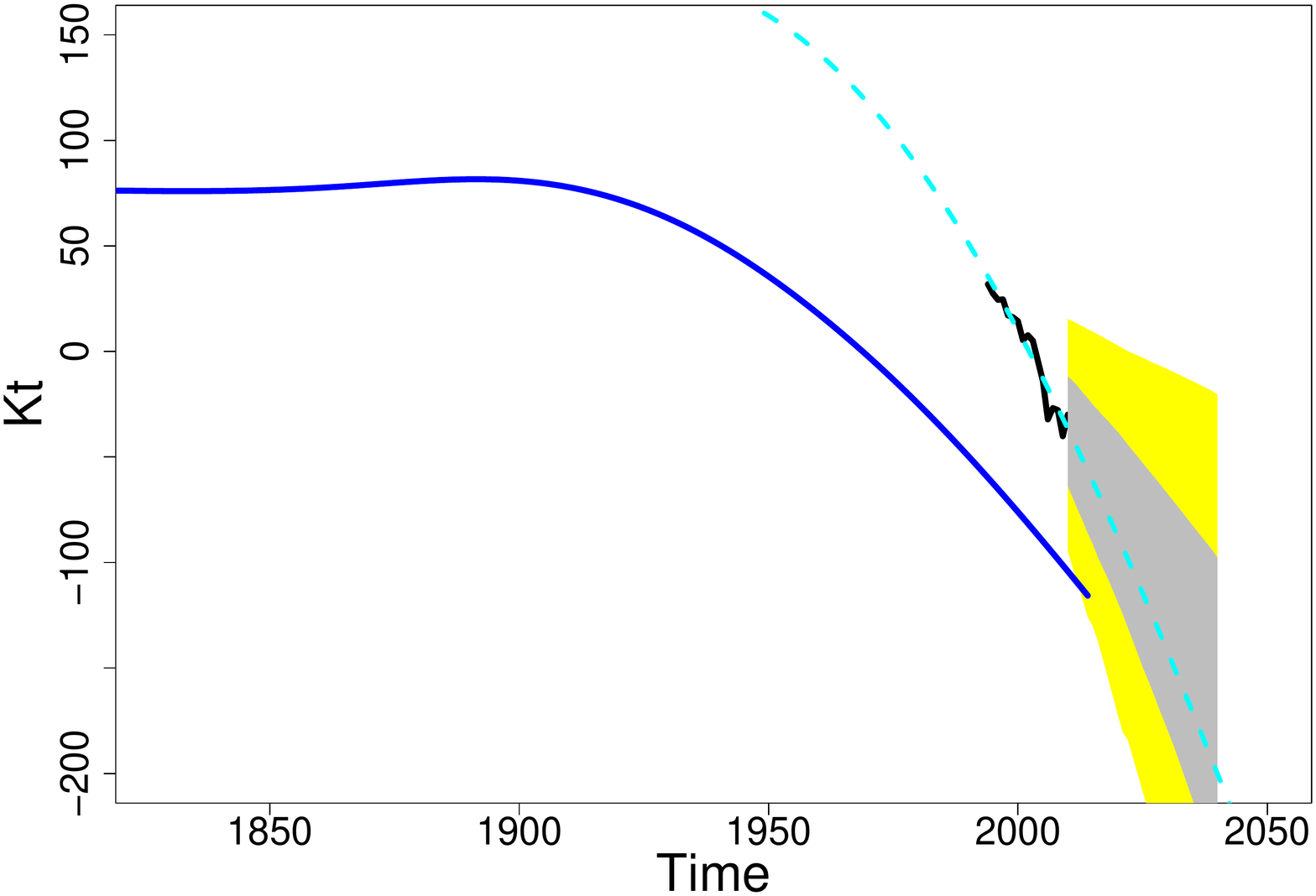}}\hfill
\subfloat[]
  {\includegraphics[width=0.5\textwidth]{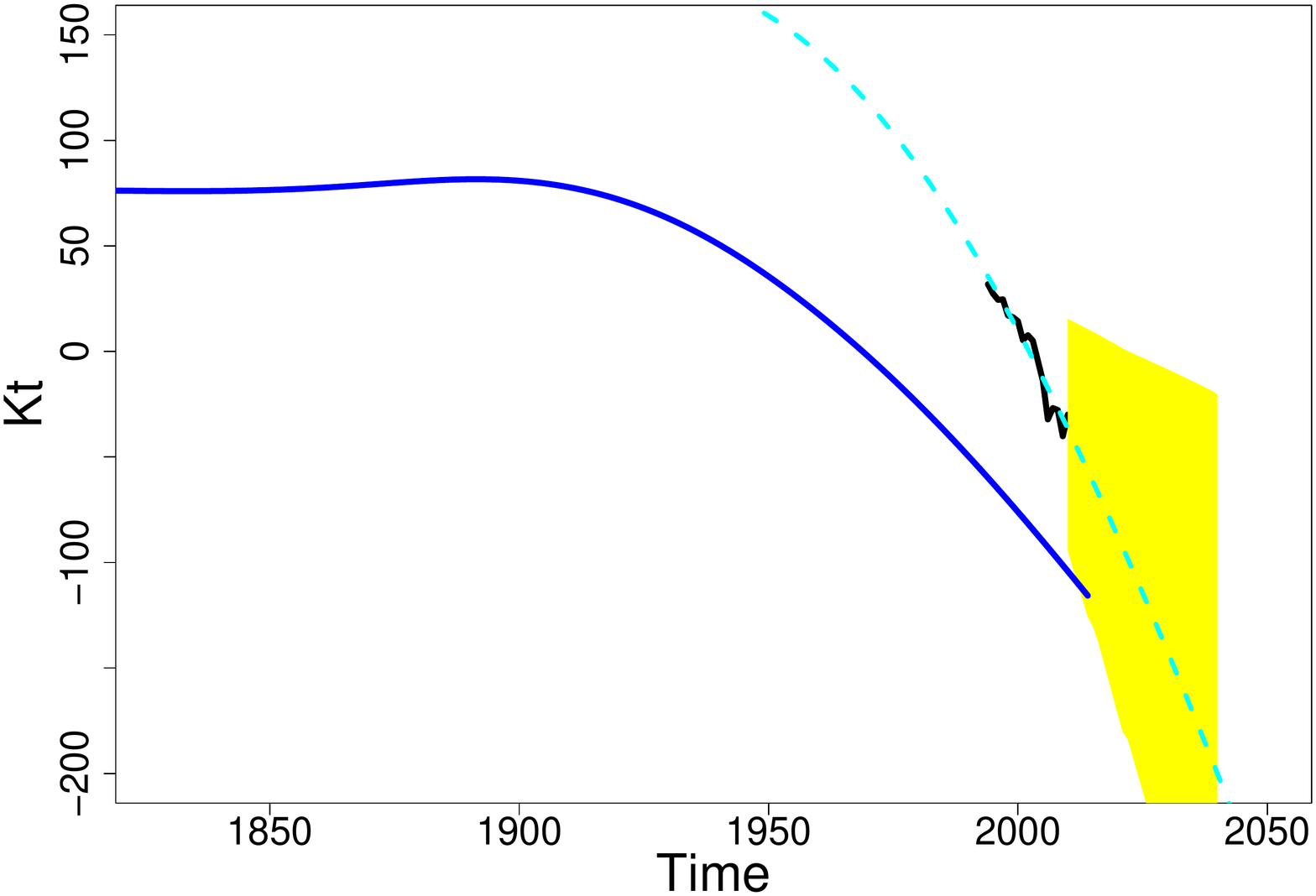}}
\caption{Confidence intervals at different levels: 80\% (left) vs. 90\% confidence interval (right).\label{CI}}
\href{https://github.com/QuantLet/MuPoMo}{\quantnet MuPoMo}
\end{figure}

In Figure \ref{CI}, confidence intervals at different levels are displayed. On the left part, confidence intervals at $80\%$ and $90\%$ are plotted in grey zone and blue zone respectively, while yellow zone highlights the central area of possible forecast path. On the right one, only $90\%$ confidence interval is presented. Black line stands for global mortality trend, red one is original China's mortality trend and light blue curve shows the estimated China's mortality trend based common trend and original China's data.

\section{Discussion}\label{sec:discussion}

The global mortality, as expected, is undergoing a shift toward exhibiting declining tendency, and with a dramatic decreasing movement in the last several decades in contrast with hundreds years ago. It also depicts that most of countries are converging with a similar mortality pattern of decreasing over time. The improvement possibly results from economic development and medical improvement, and it is also not difficult to imagine that there will be gradually declining mortality rate in the near future due to technology progress.

In addition to the global mortality trend, each country still behaves differently from others to some extent. From this perspective, it might be possible for life insurance companies to design insurance products among different countries to hedge global longevity risk. That is, if wider range of countries is covered by a particular insurance company, it is possible to redistribute longevity risk among them.

Another advantage from this research is to establish a better forecasting regime to foresee mortality change in longer time horizon, particularly for countries with limited historical mortality data, such as China and Chile.


\bibliography{mortality-ref}

\newpage
\section{Appendices}

\subsection{Appendix 1}

\begin{figure}[H]
\captionsetup[subfigure]{labelformat=empty}
\subfloat[]
  {\includegraphics[width=0.5\textwidth]{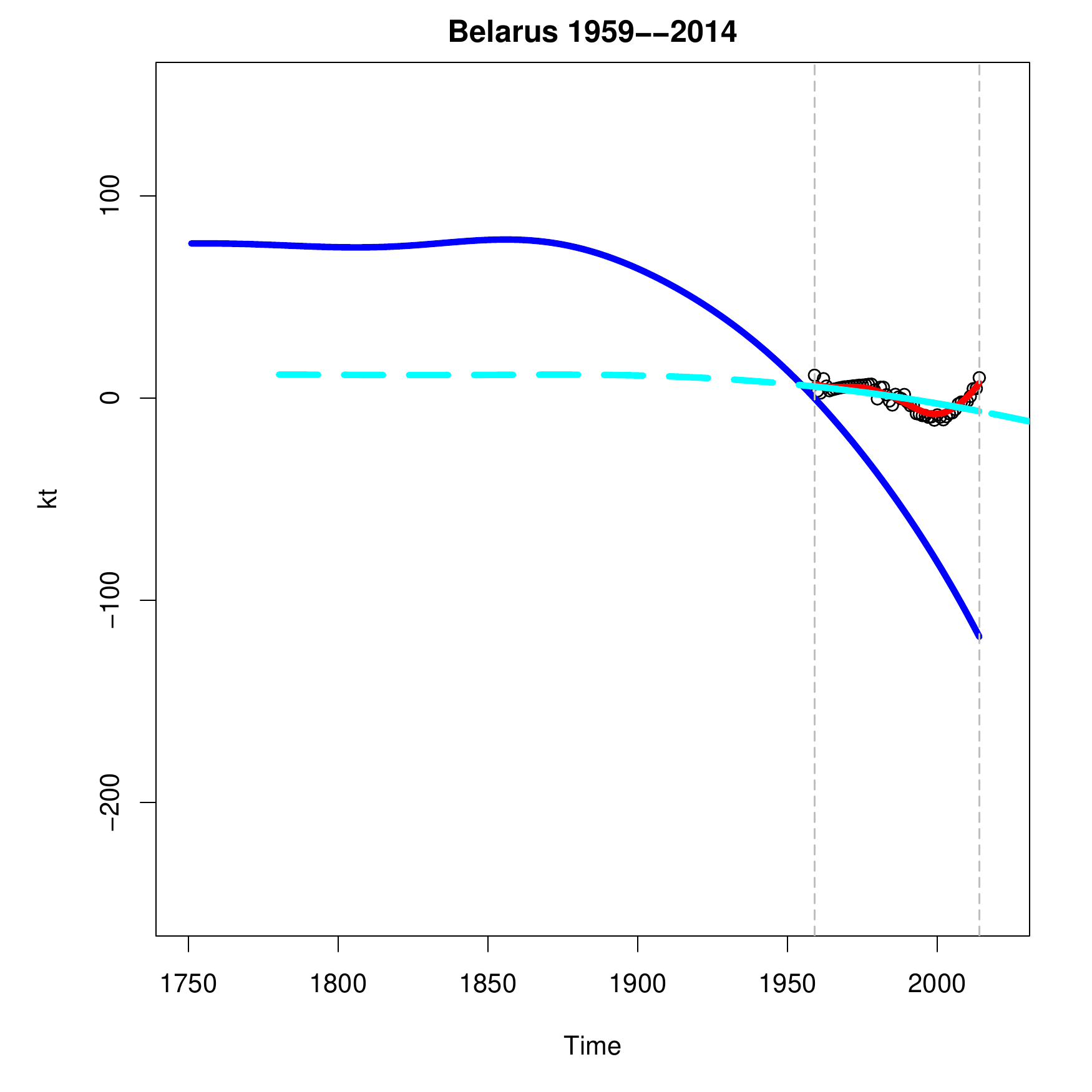}}\hfill
\subfloat[]
  {\includegraphics[width=0.5\textwidth]{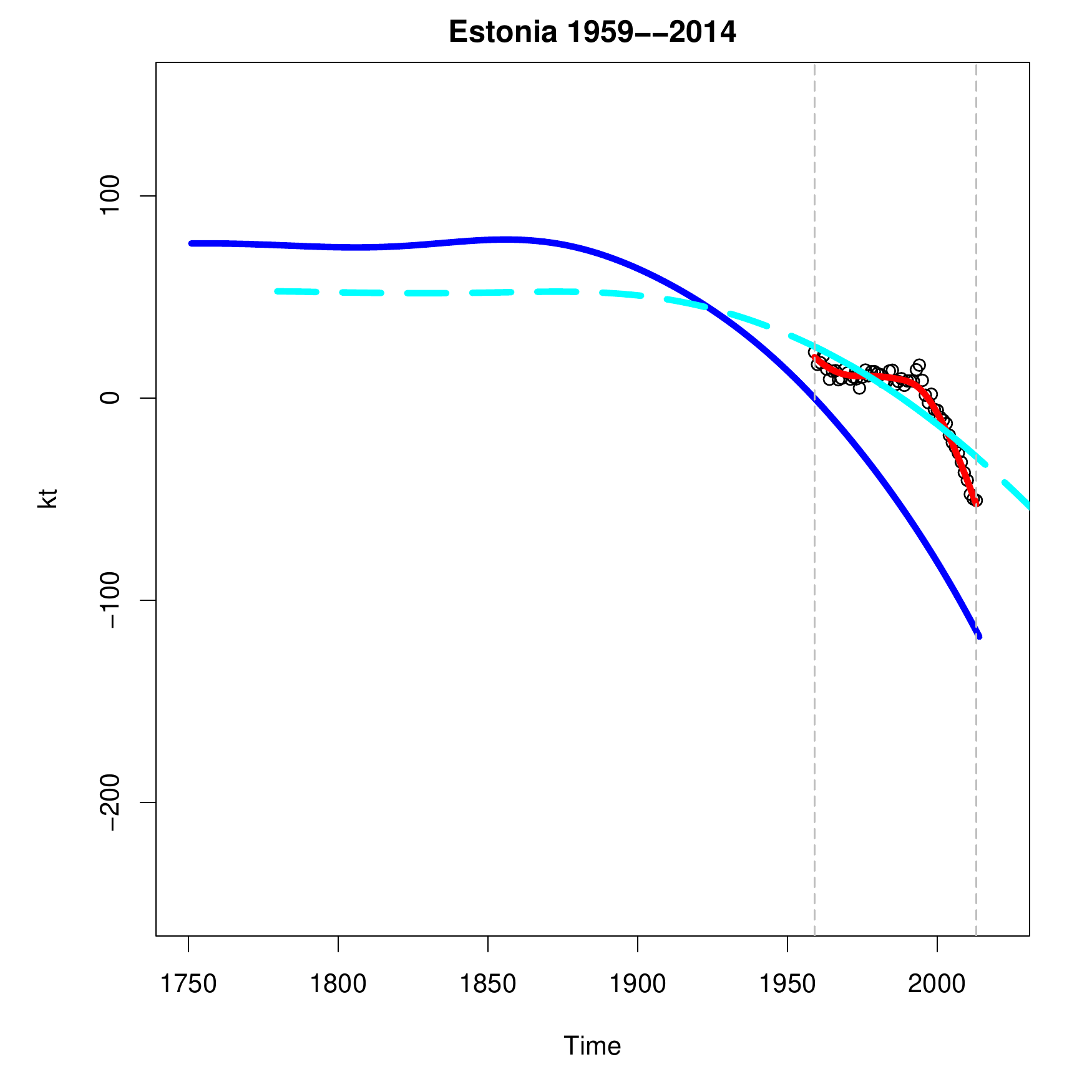}}\\
  \subfloat[]
  {\includegraphics[width=0.5\textwidth]{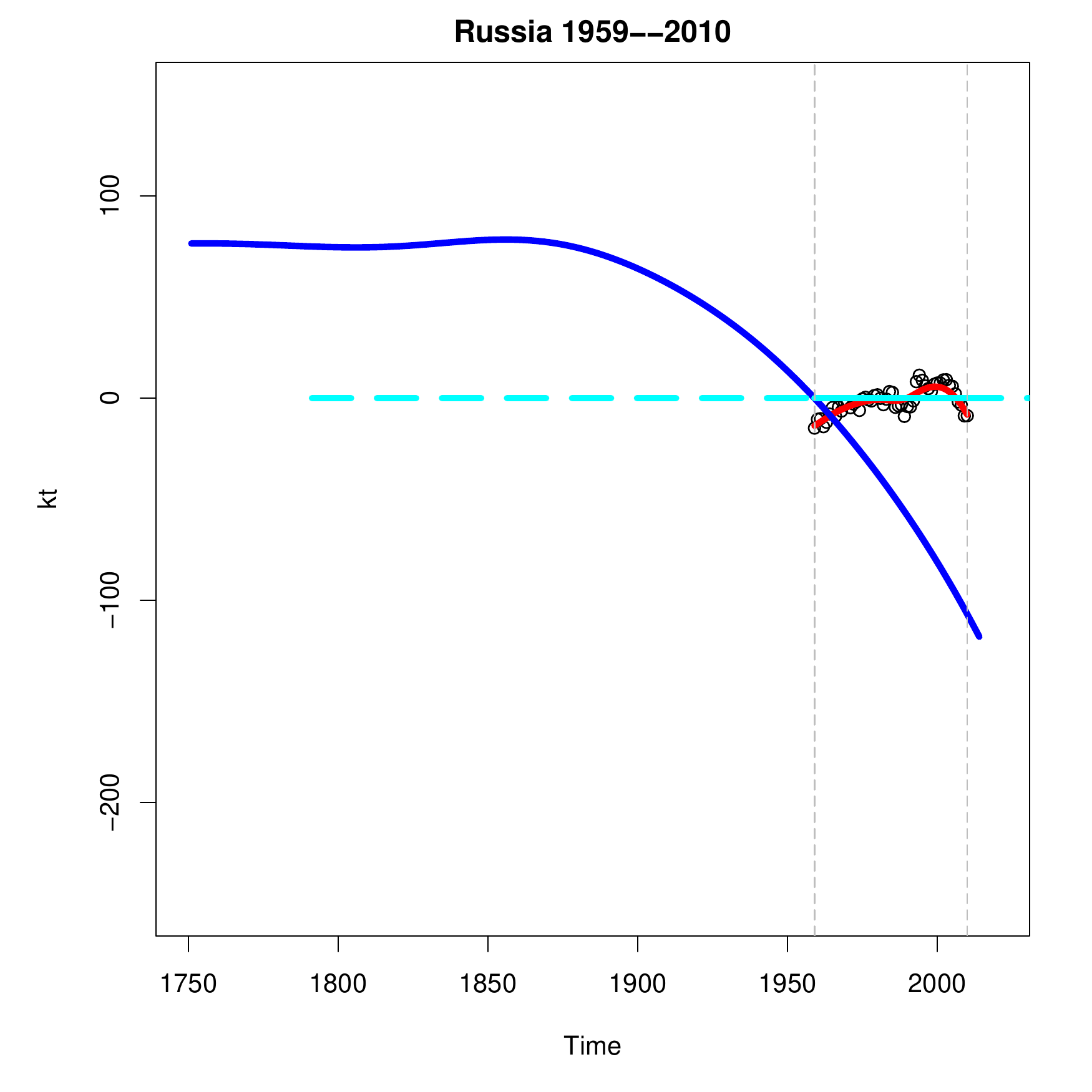}}\hfill
  \vfill
  \href{https://github.com/QuantLet/MuPoMo}{\quantnet MuPoMo}
\end{figure}

\vspace*{-1cm}
\subsection{Appendix 2}
\begin{figure}[H]
\captionsetup[subfigure]{labelformat=empty}
\subfloat[]
  {\includegraphics[width=0.5\textwidth]{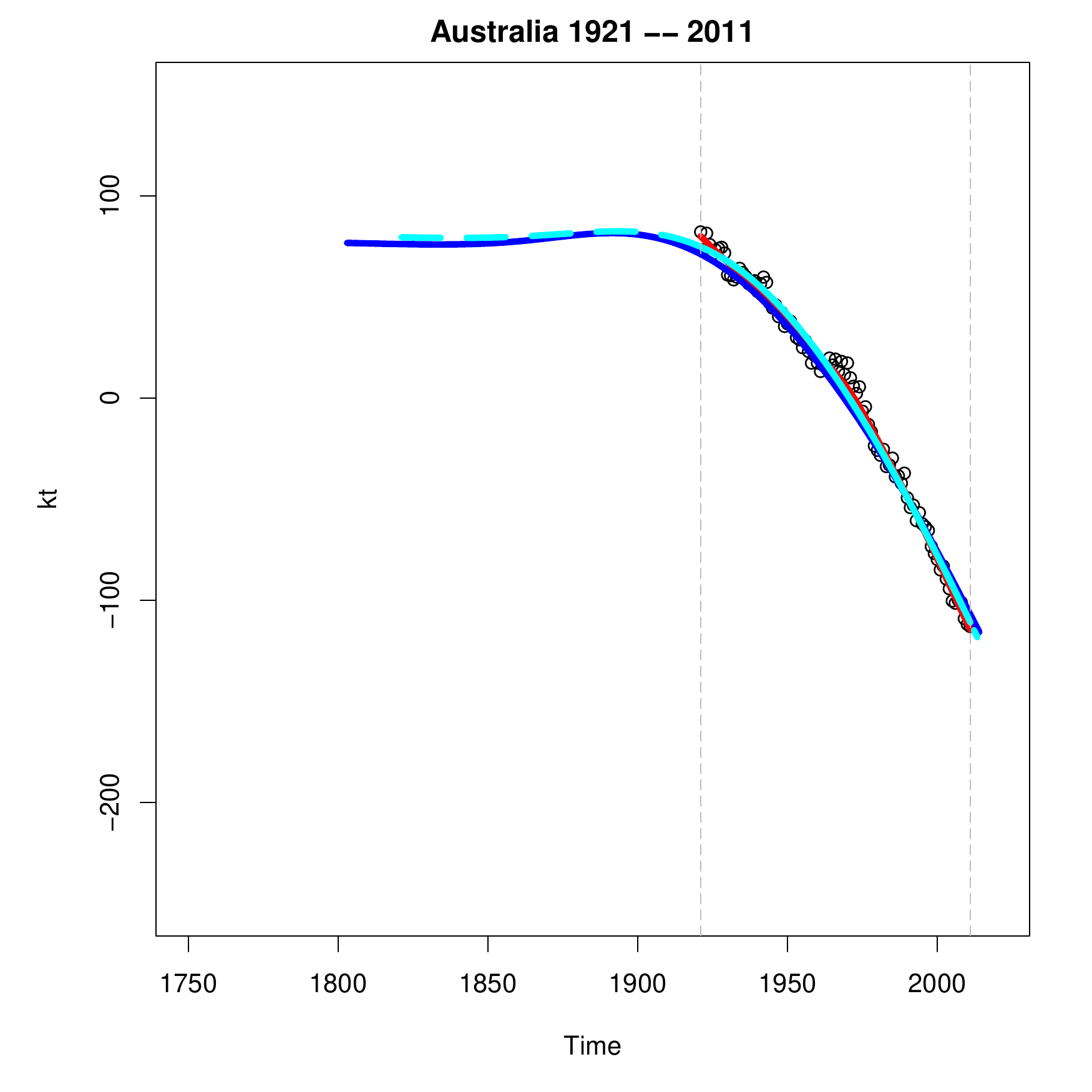}}\hfill
\subfloat[]
  {\includegraphics[width=0.5\textwidth]{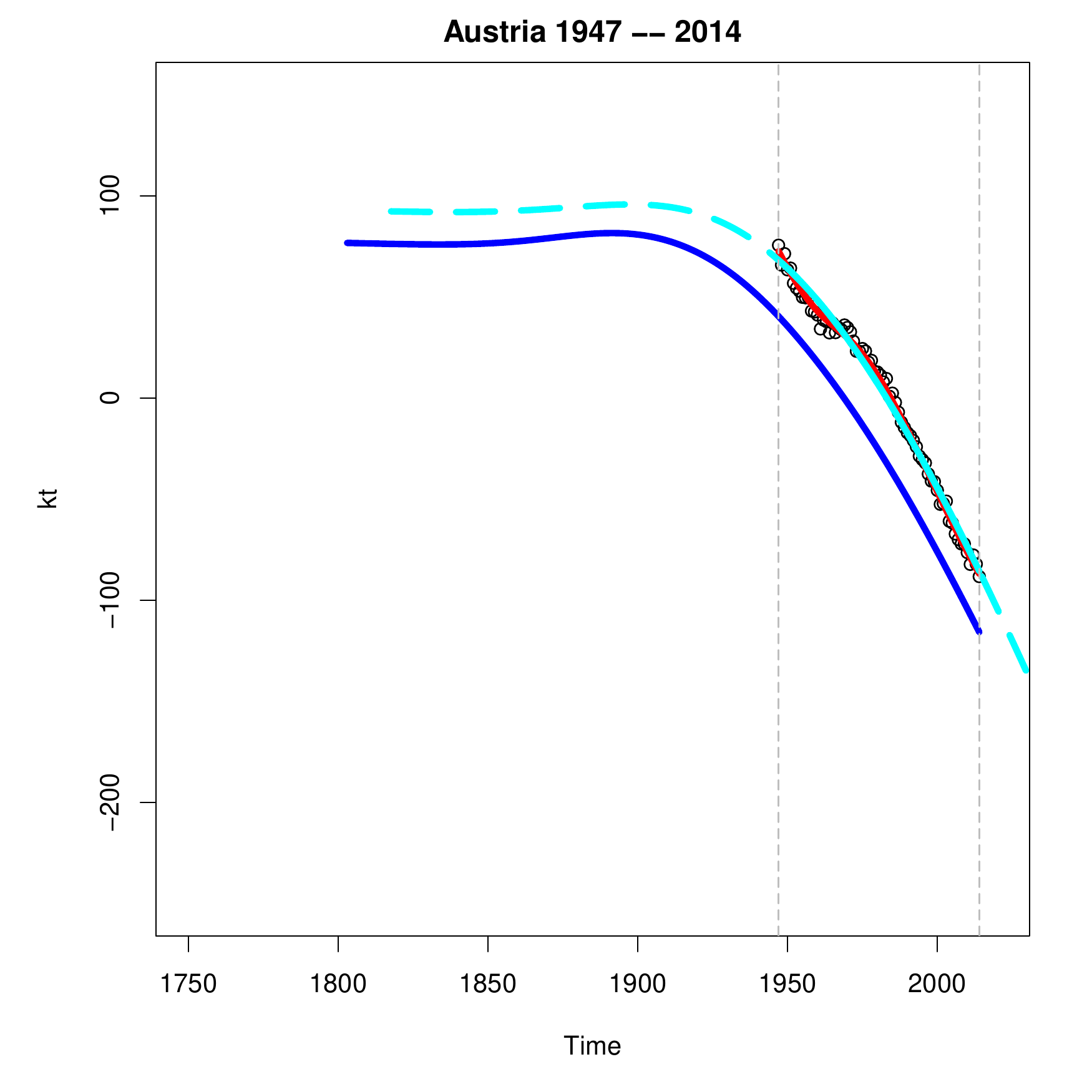}}\\
\vspace{-1.2 cm}
\subfloat[]
  {\includegraphics[width=0.5\textwidth]{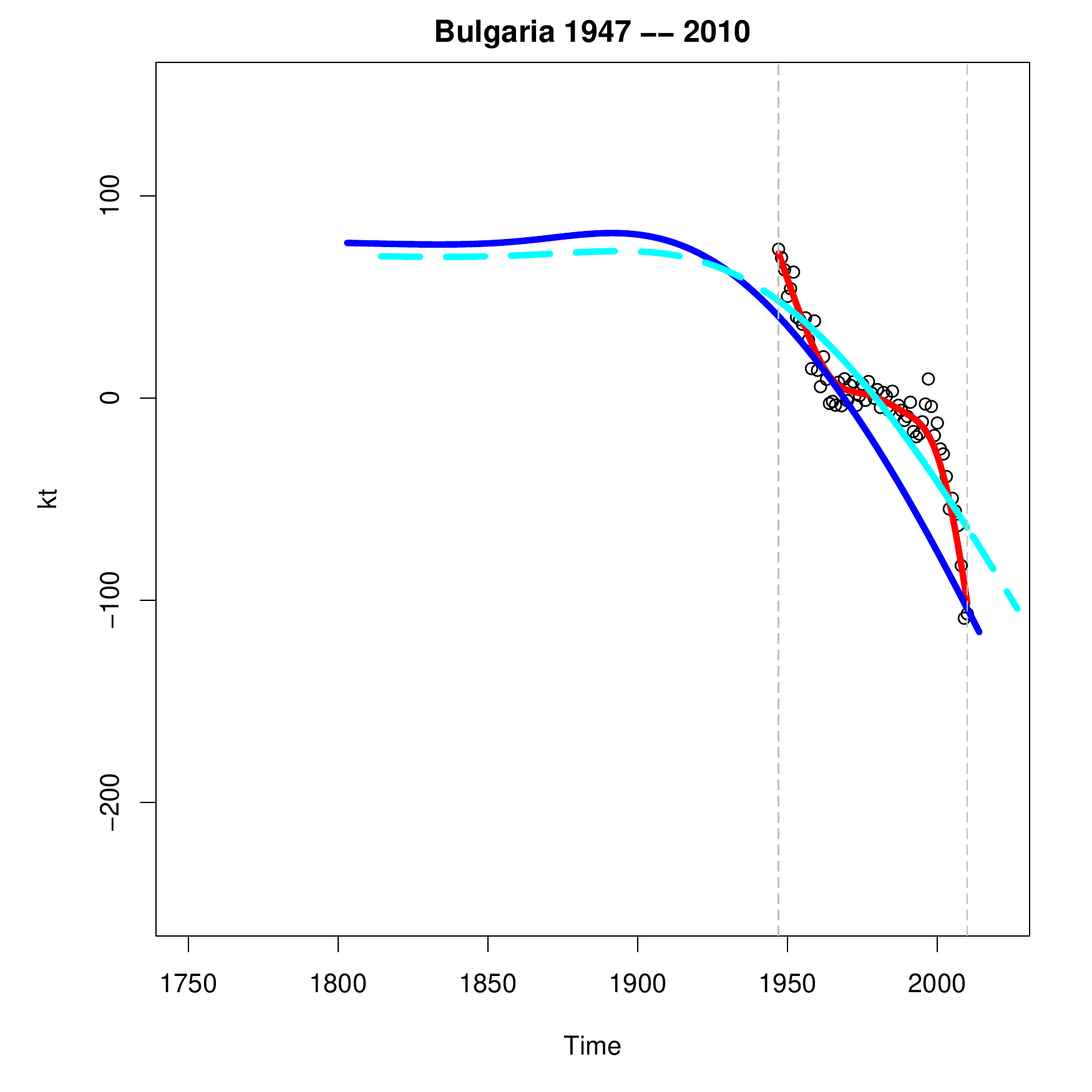}}\hfill
\subfloat[]
  {\includegraphics[width=0.5\textwidth]{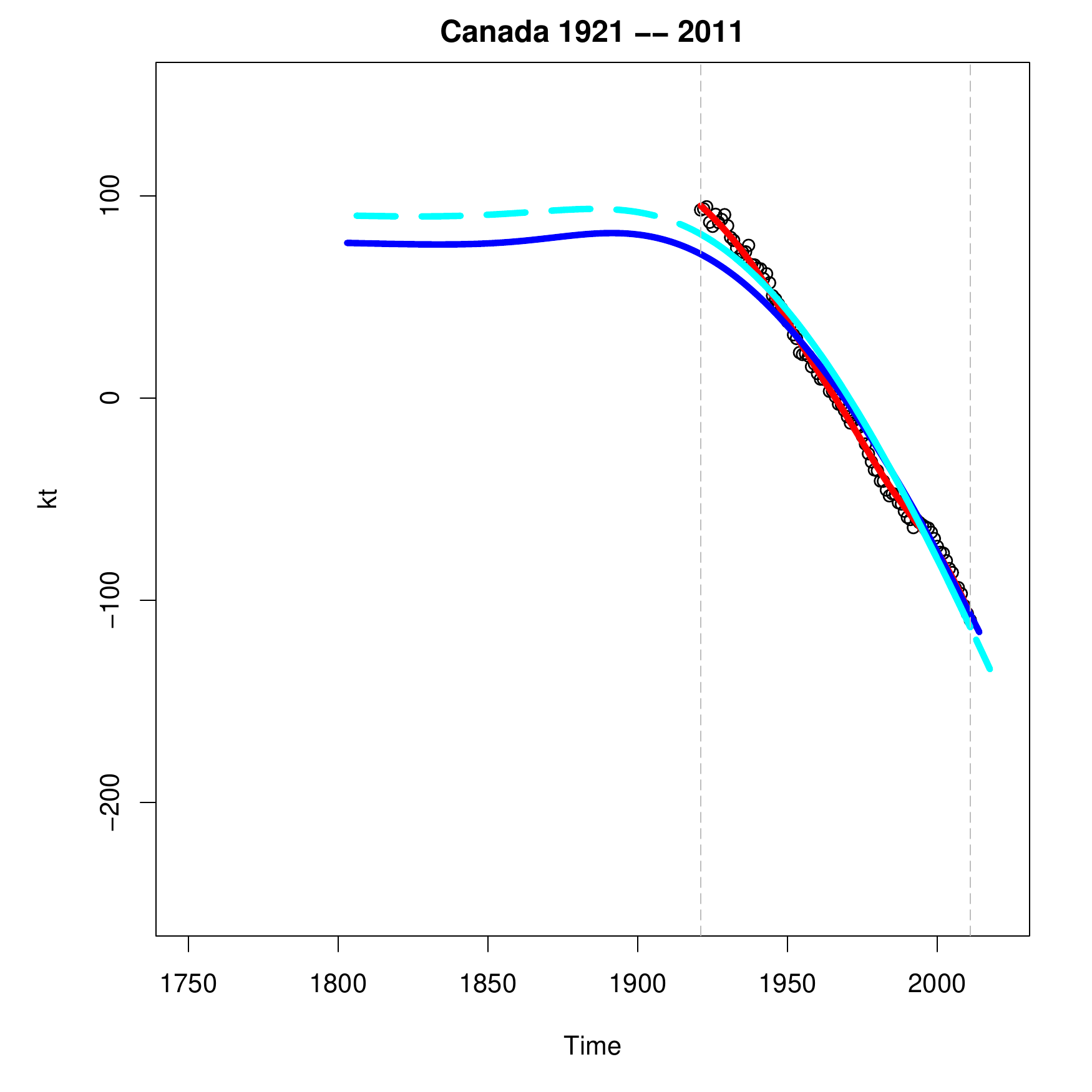}}\\
\vspace{-1.2 cm}
  \subfloat[]
  {\includegraphics[width=0.5\textwidth]{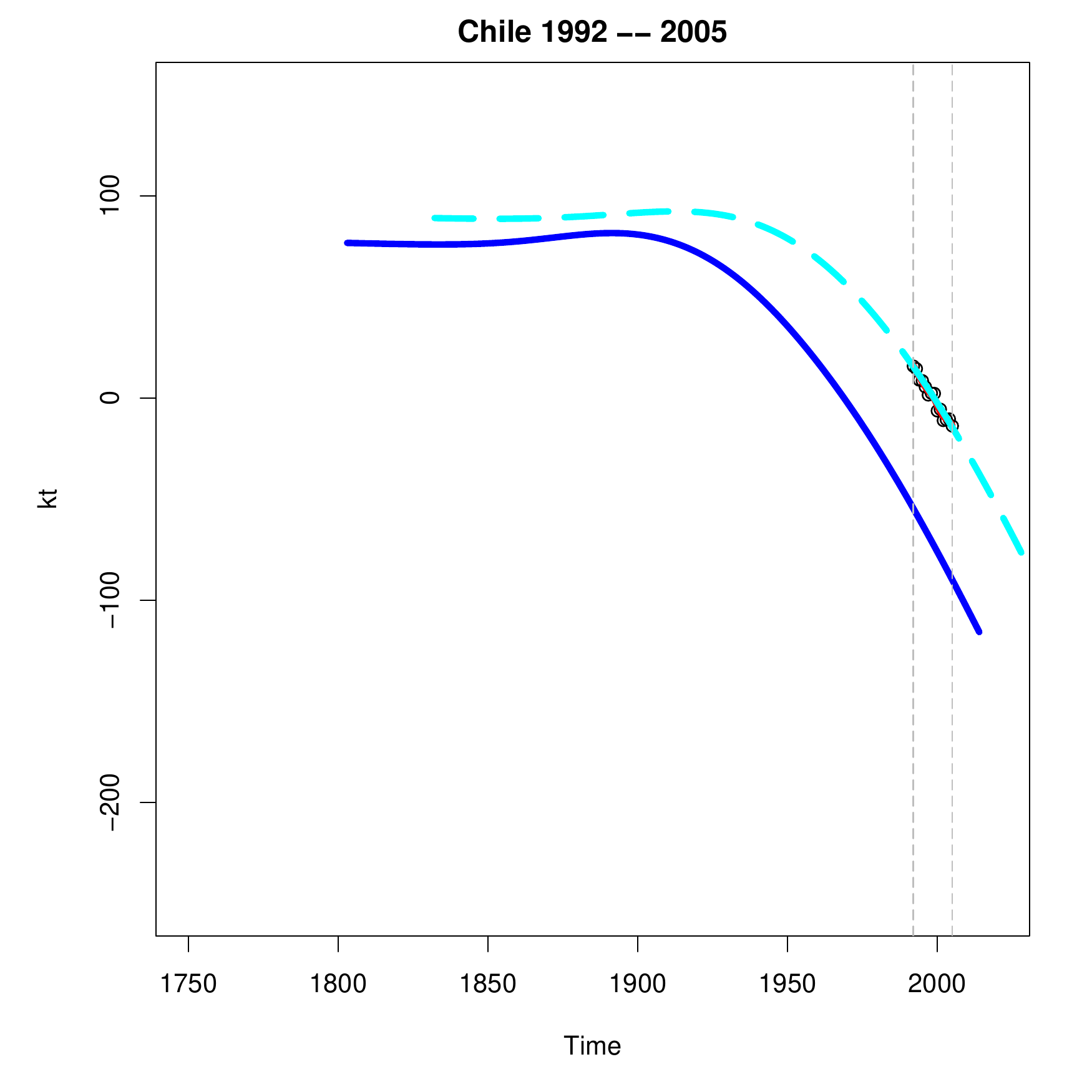}}\hfill
\subfloat[]
  {\includegraphics[width=0.5\textwidth]{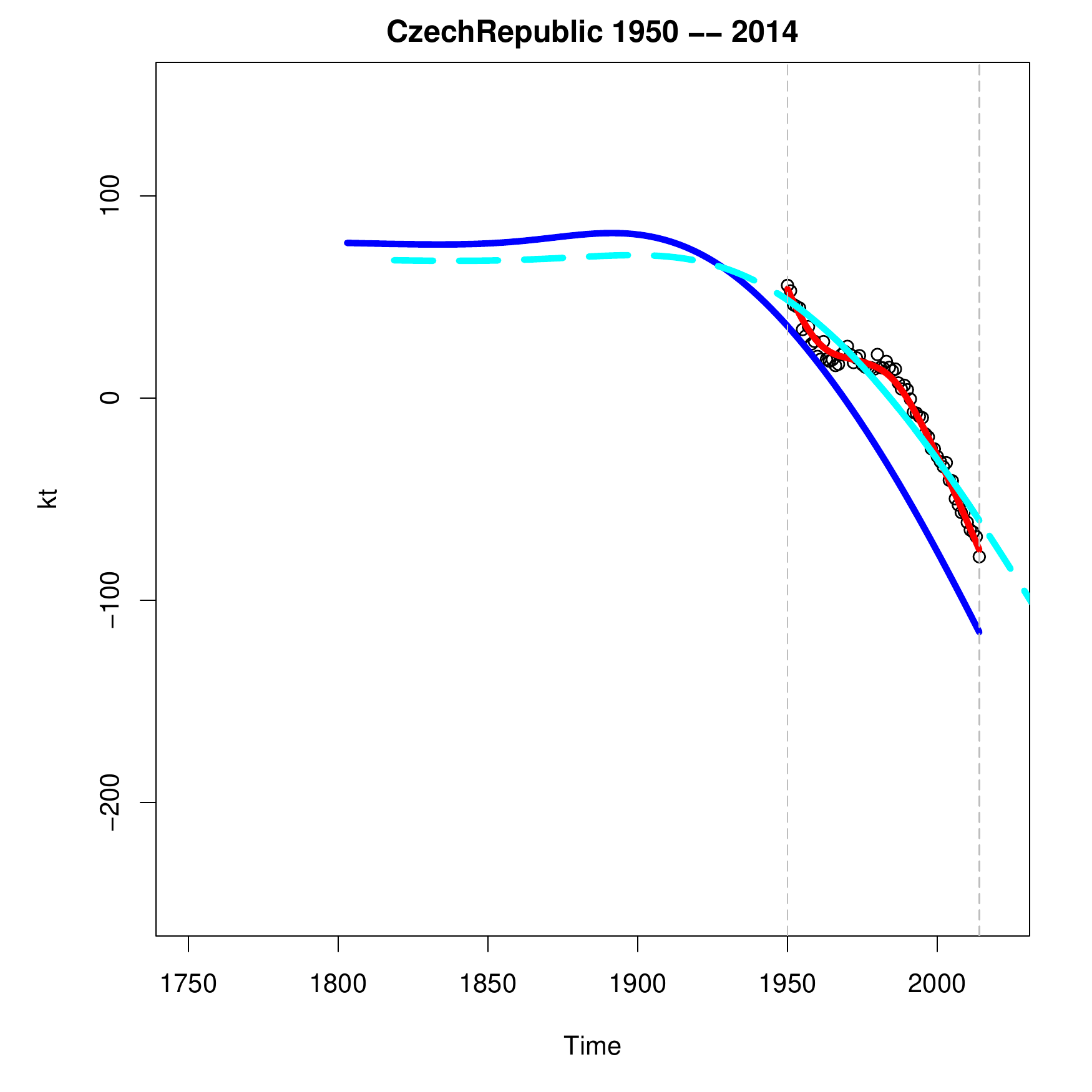}}\\
\vspace{-1.2 cm}
\href{https://github.com/QuantLet/MuPoMo}{\quantnet MuPoMo}
\end{figure}

\begin{figure}[H]
\captionsetup[subfigure]{labelformat=empty}
\subfloat[]
  {\includegraphics[width=0.5\textwidth]{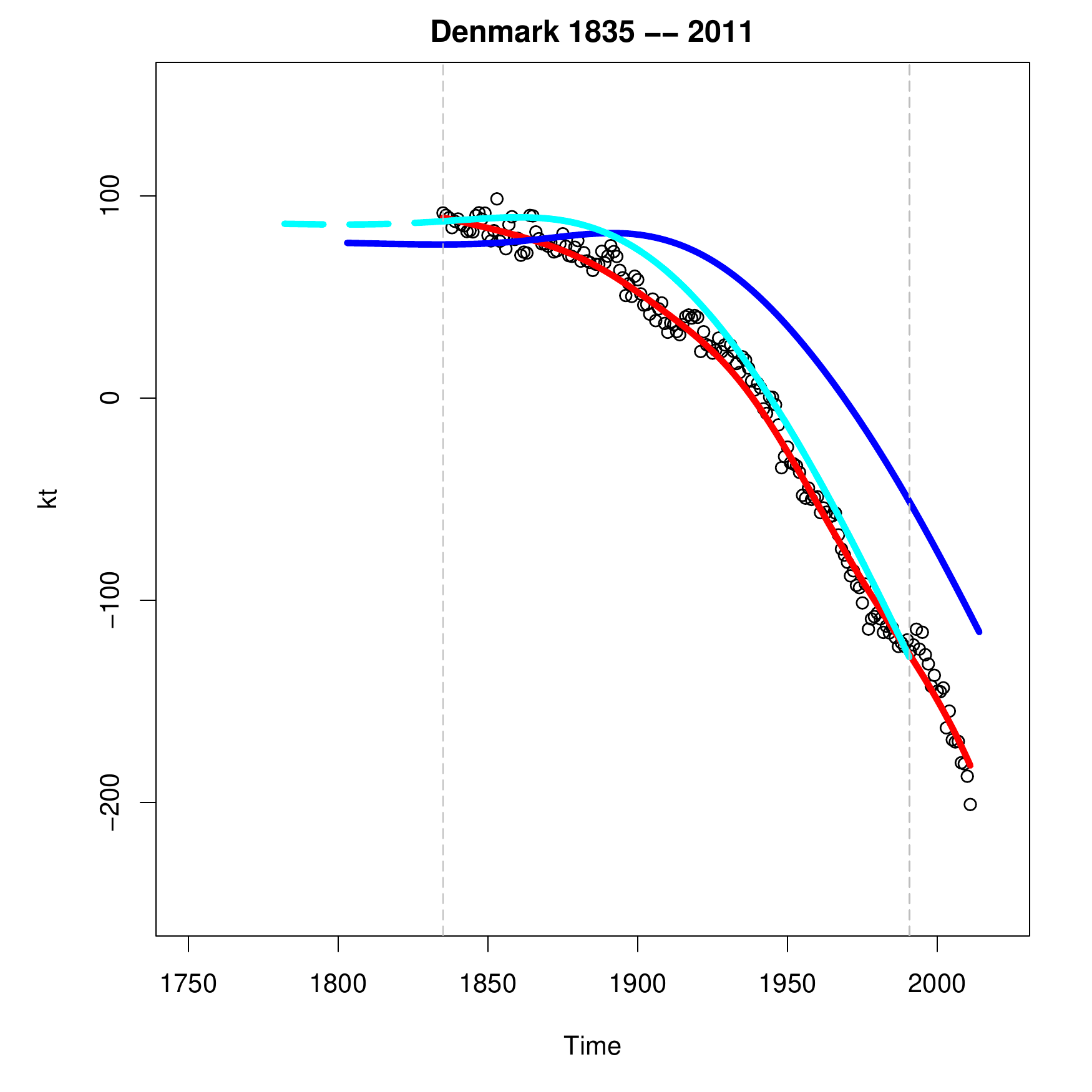}}\hfill
\subfloat[]
  {\includegraphics[width=0.5\textwidth]{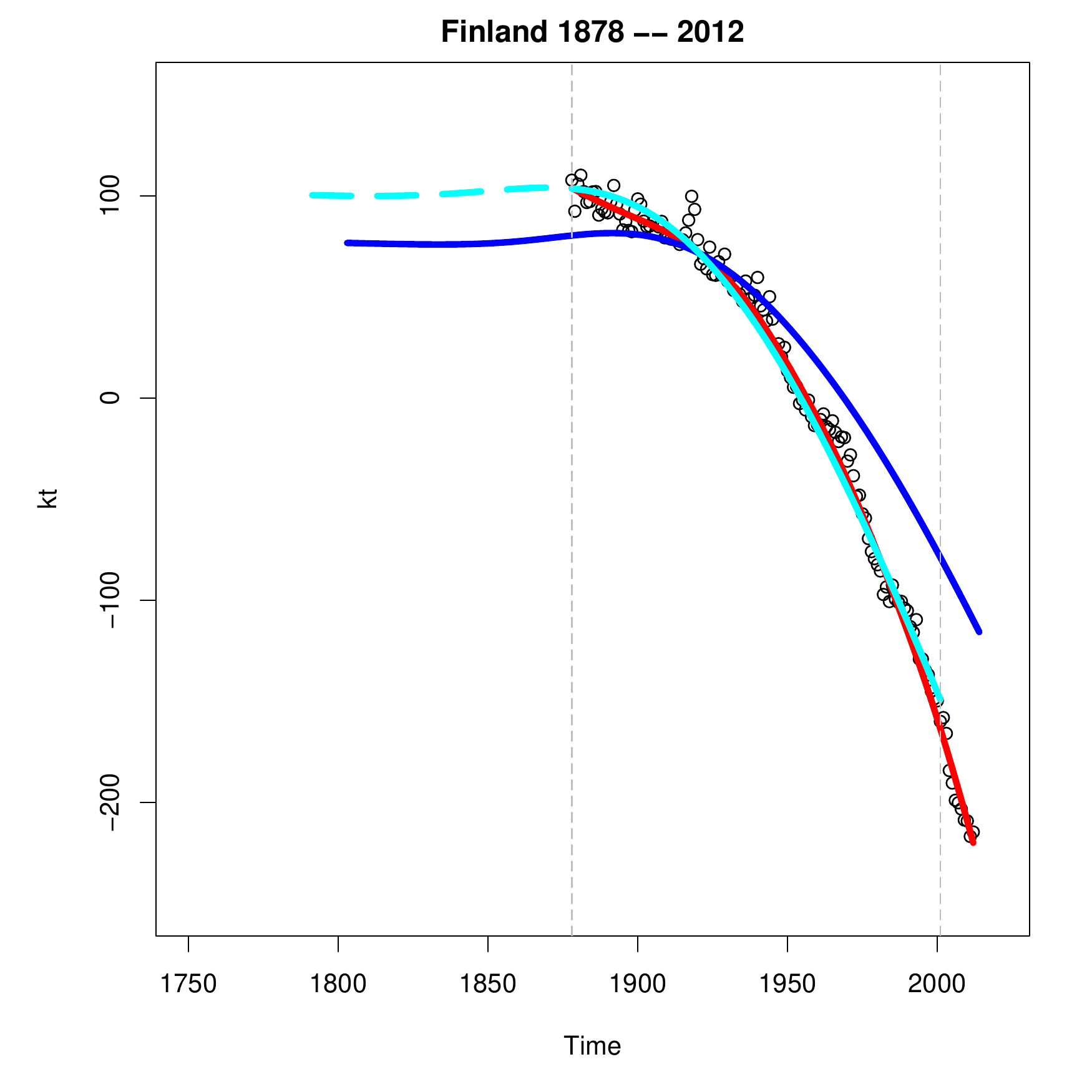}}\\
\vspace{-1.2 cm}
\subfloat[]
  {\includegraphics[width=0.5\textwidth]{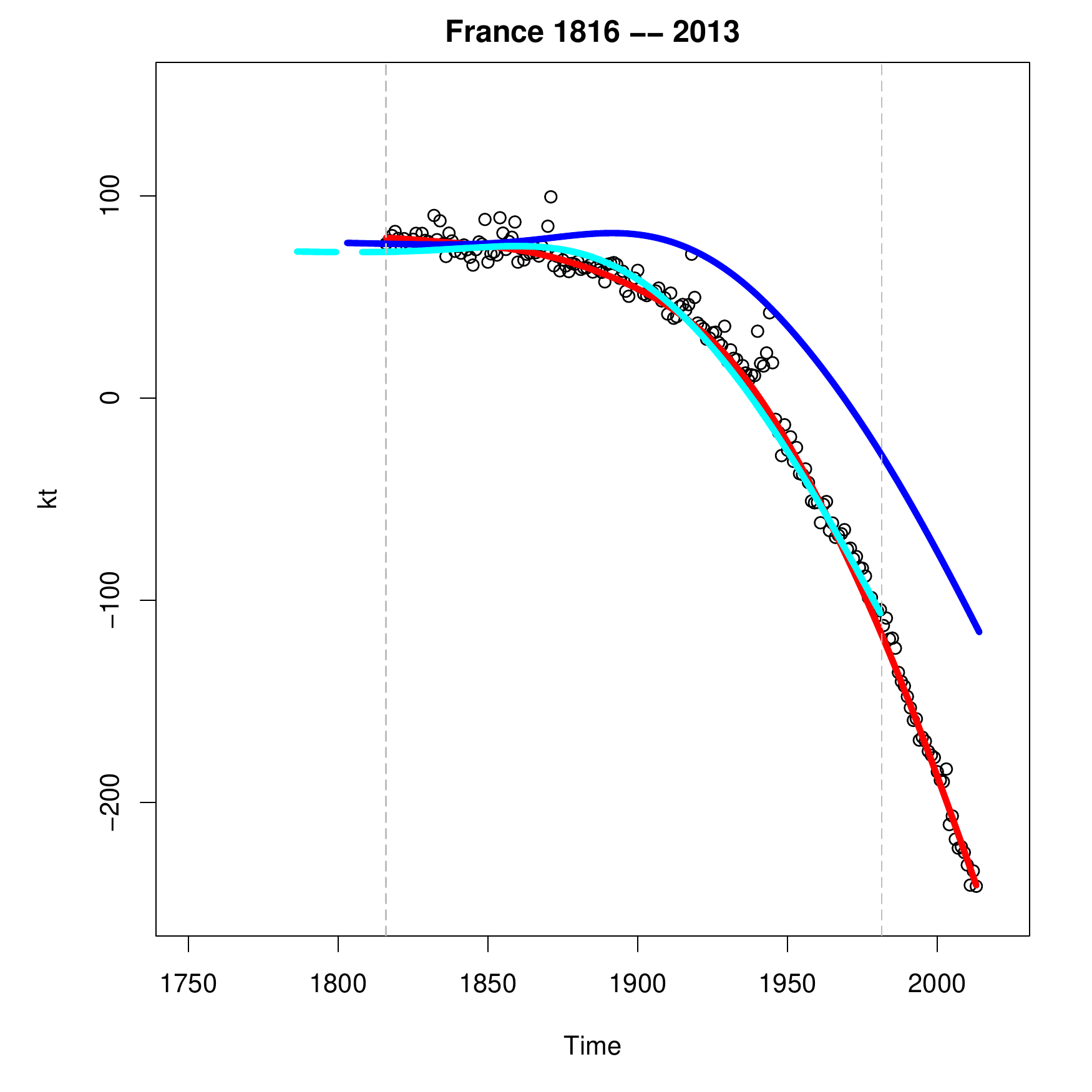}}\hfill
\subfloat[]
  {\includegraphics[width=0.5\textwidth]{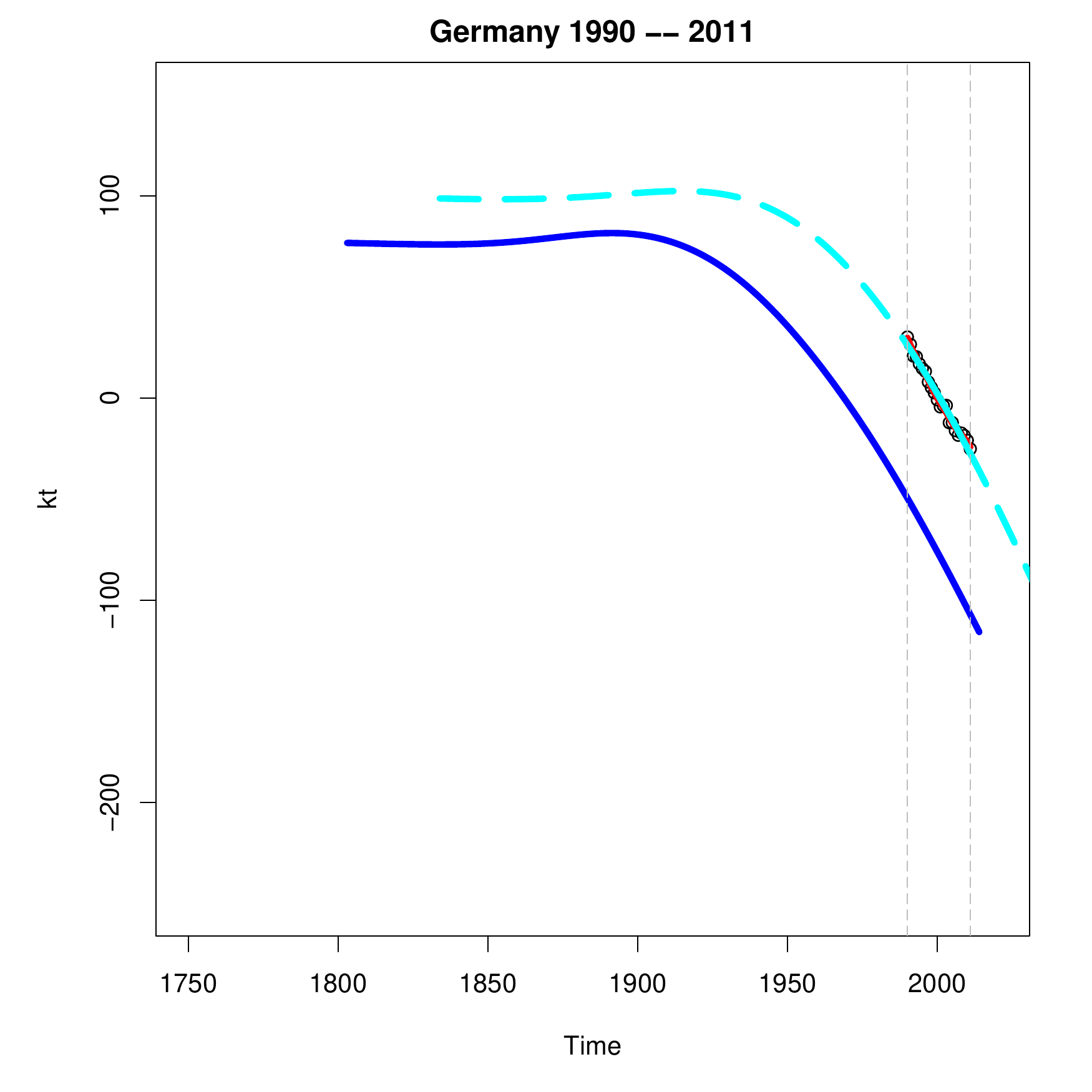}}\\
\vspace{-1.2 cm}
  \subfloat[]
  {\includegraphics[width=0.5\textwidth]{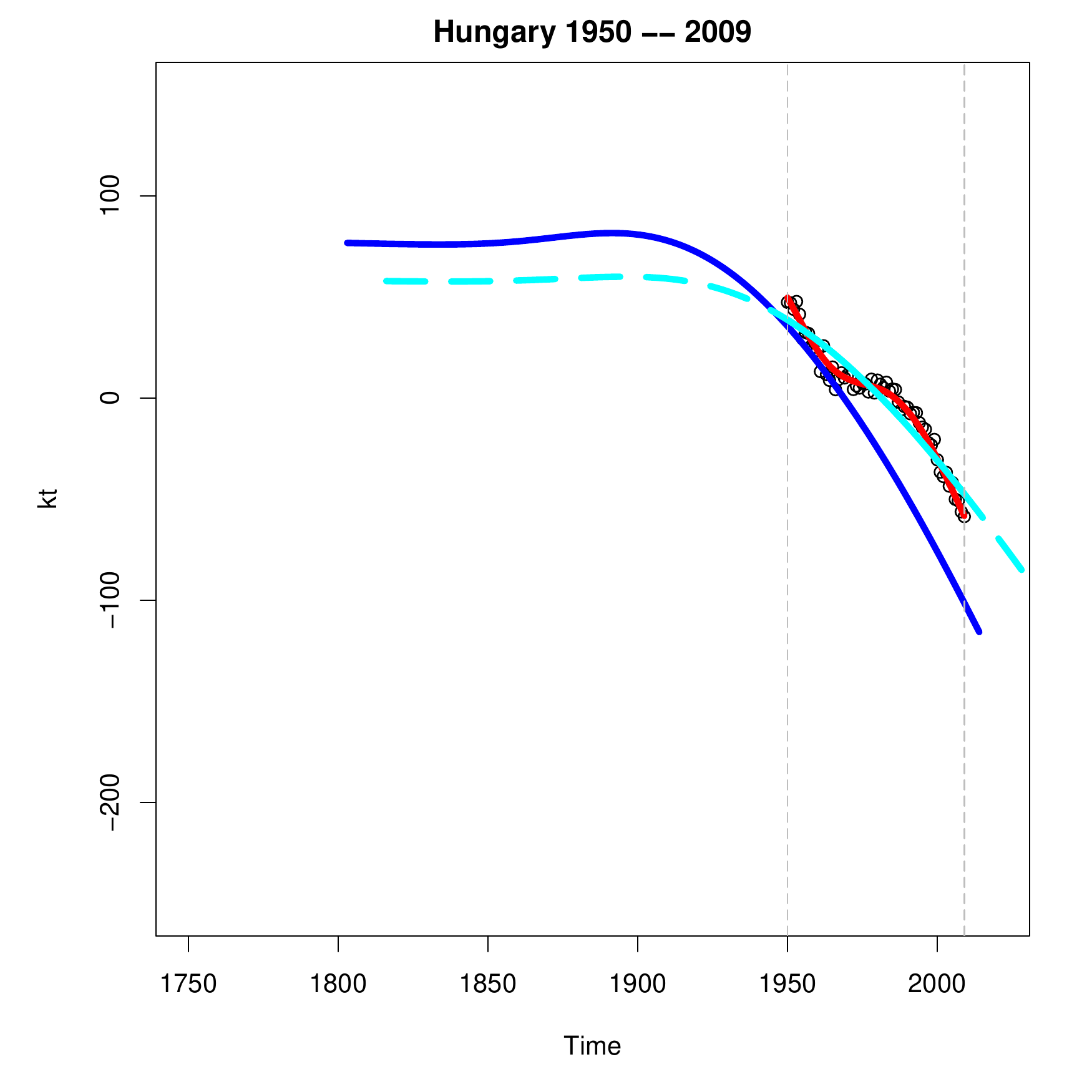}}\hfill
\subfloat[]
  {\includegraphics[width=0.5\textwidth]{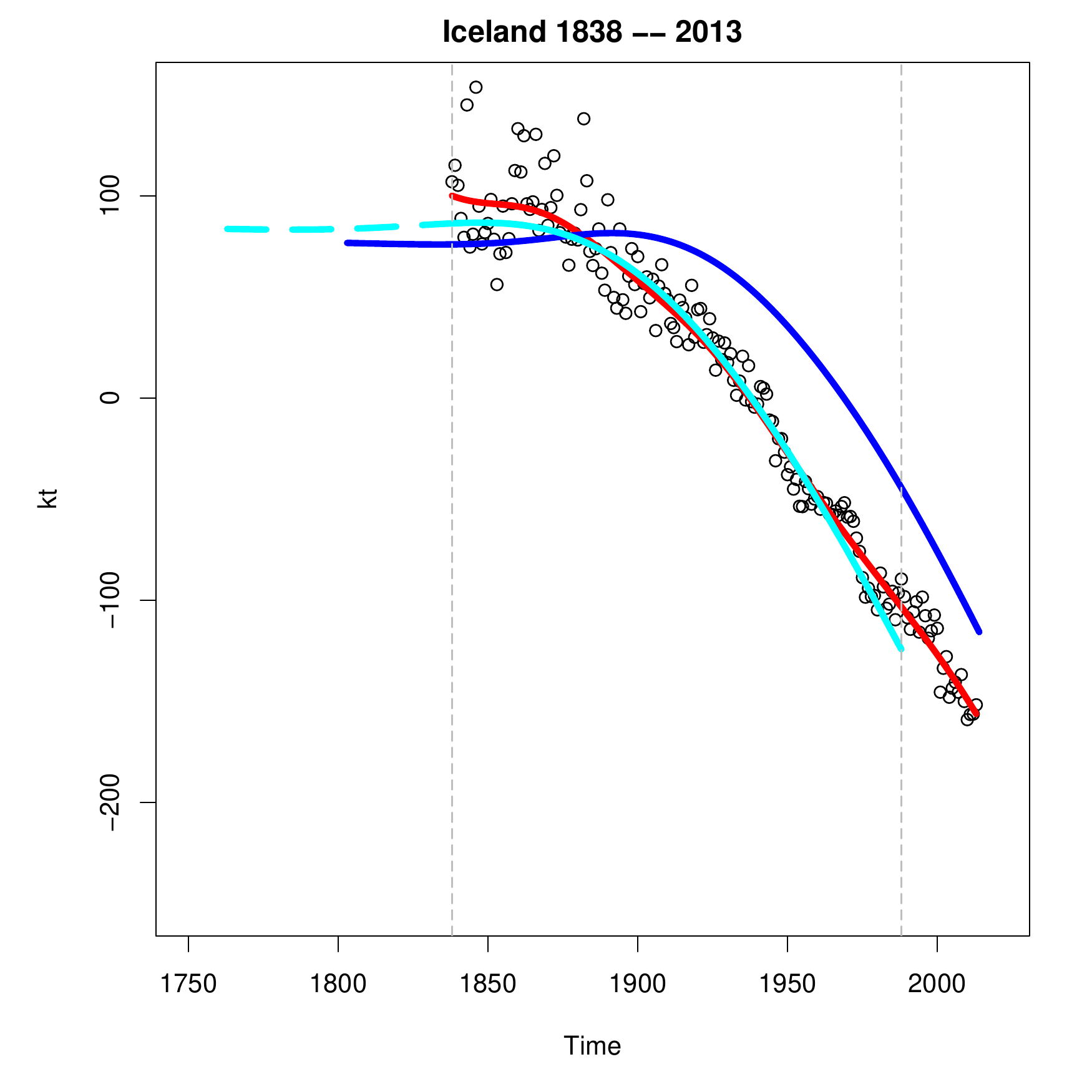}}\\
\vspace{-1.2 cm}
\href{https://github.com/QuantLet/MuPoMo}{\quantnet MuPoMo}
\end{figure}

\begin{figure}[H]
\captionsetup[subfigure]{labelformat=empty}
\subfloat[]
  {\includegraphics[width=0.5\textwidth]{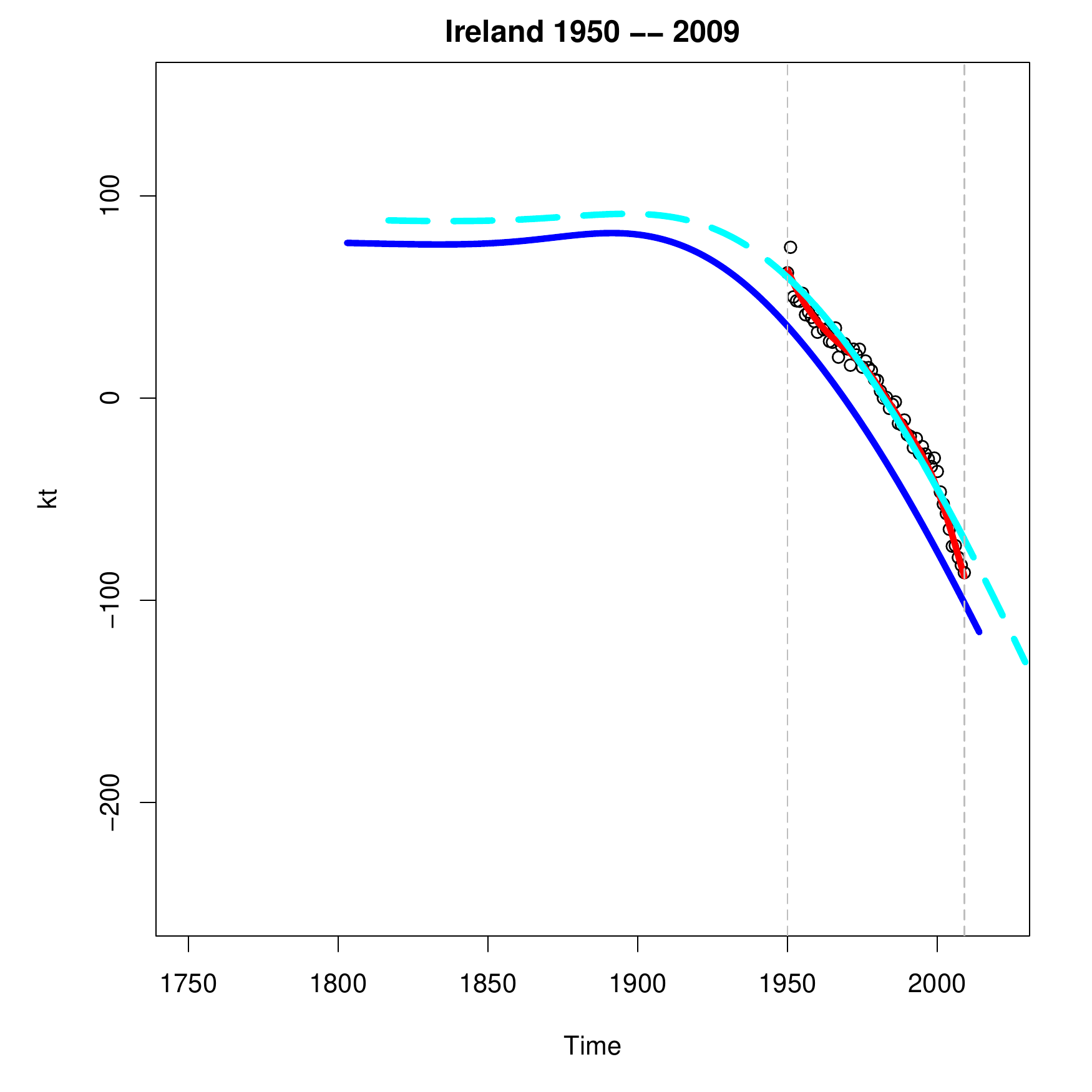}}\hfill
\subfloat[]
  {\includegraphics[width=0.5\textwidth]{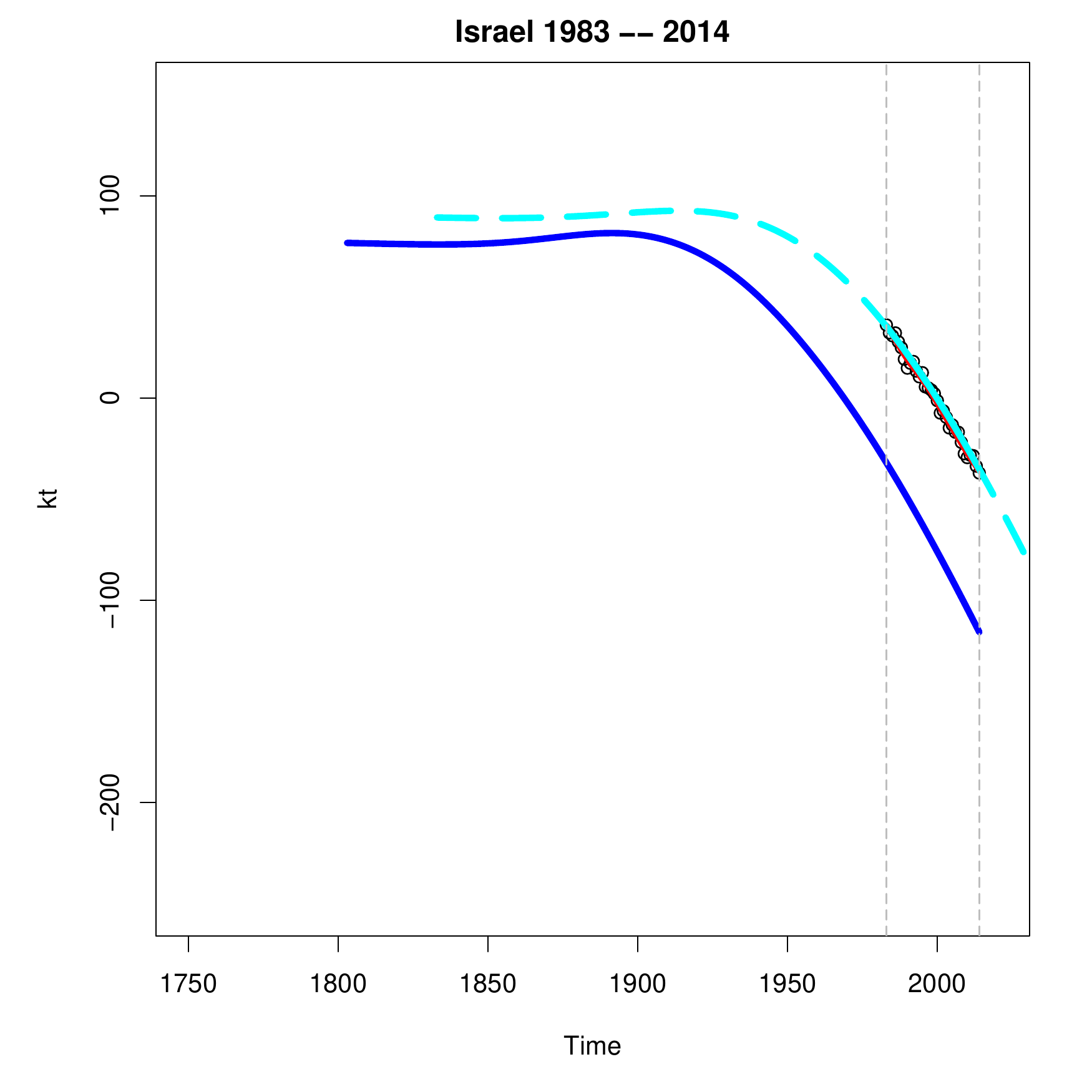}}\\
\vspace{-1.2 cm}
\subfloat[]
  {\includegraphics[width=0.5\textwidth]{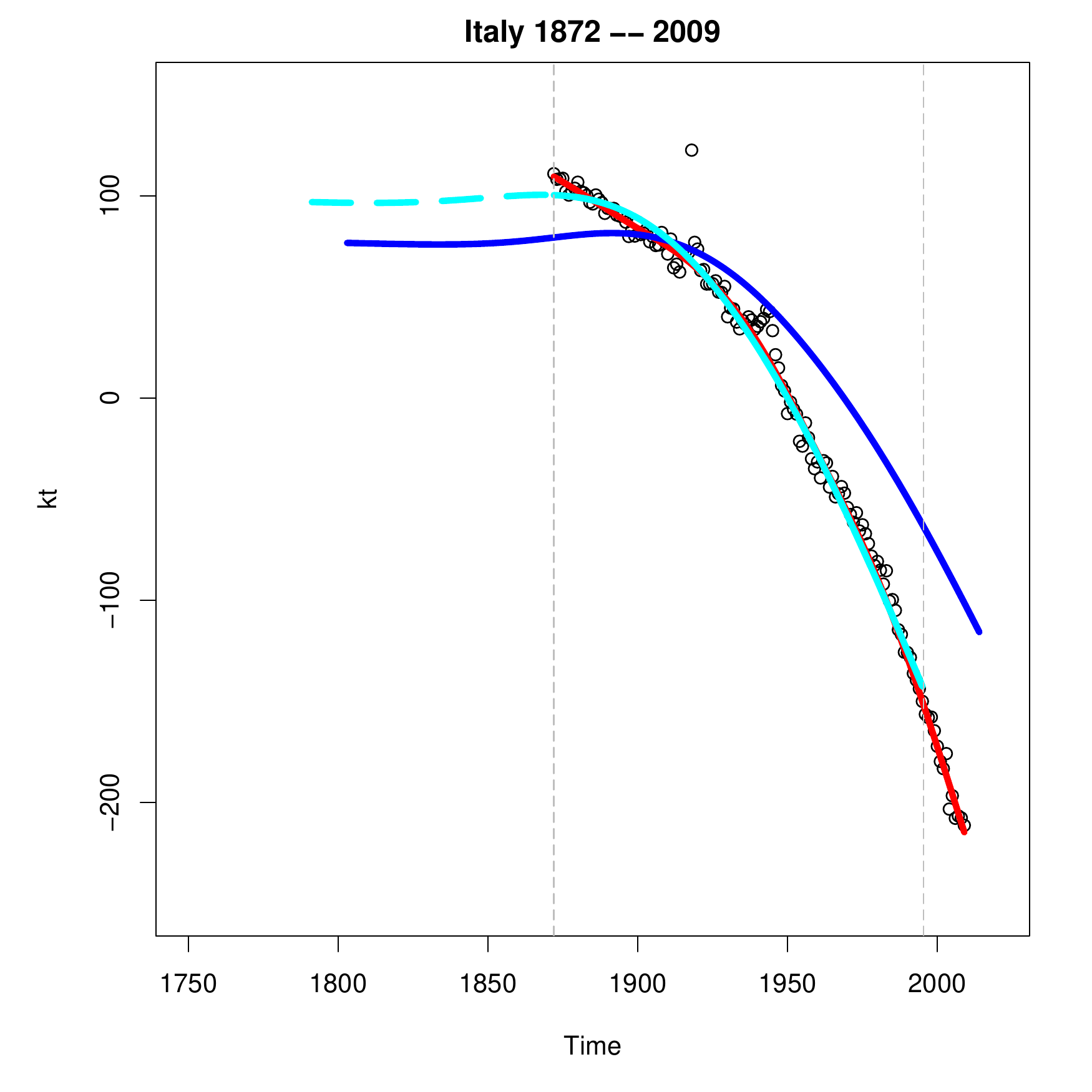}}\hfill
\subfloat[]
  {\includegraphics[width=0.5\textwidth]{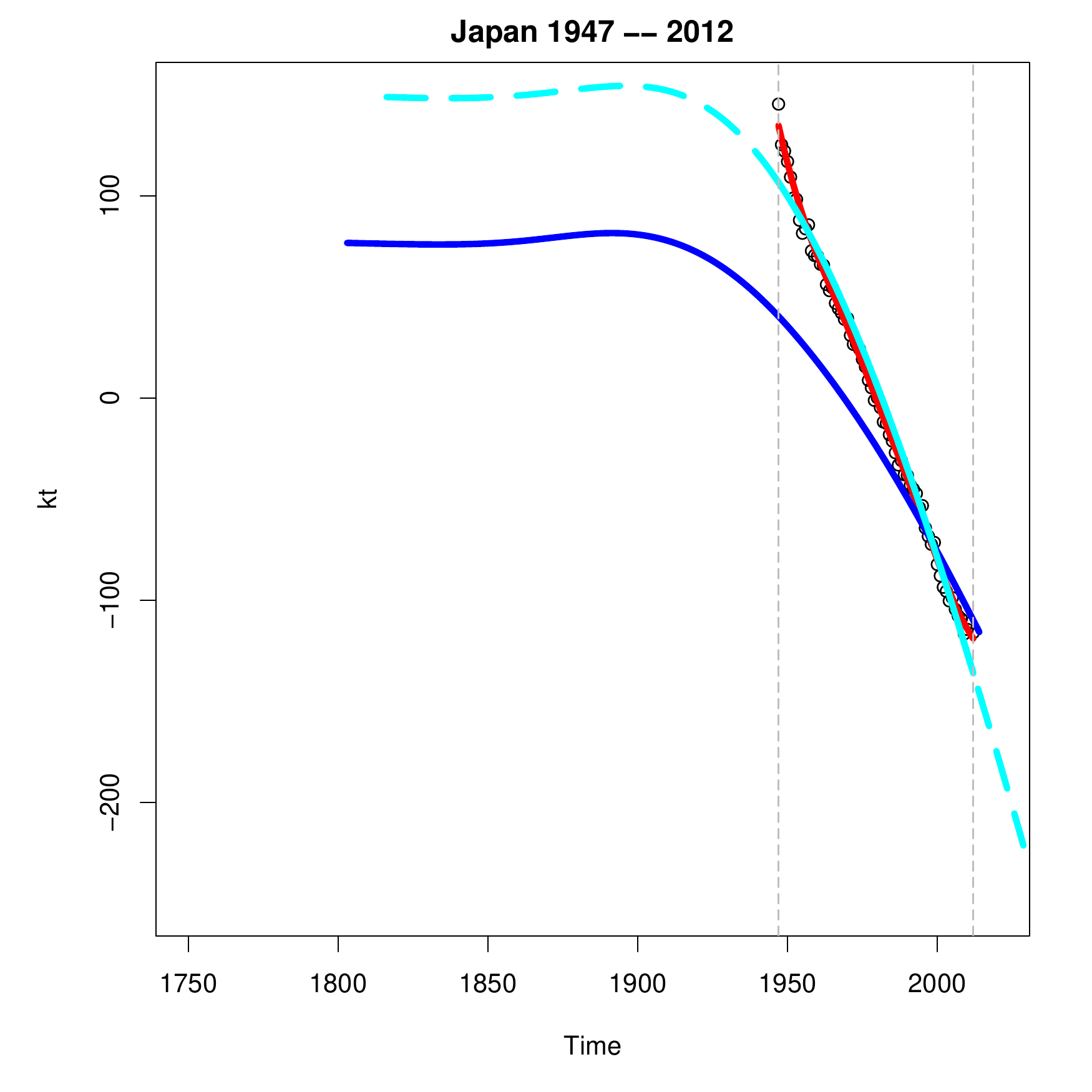}}\\
\vspace{-1.2 cm}
  \subfloat[]
  {\includegraphics[width=0.5\textwidth]{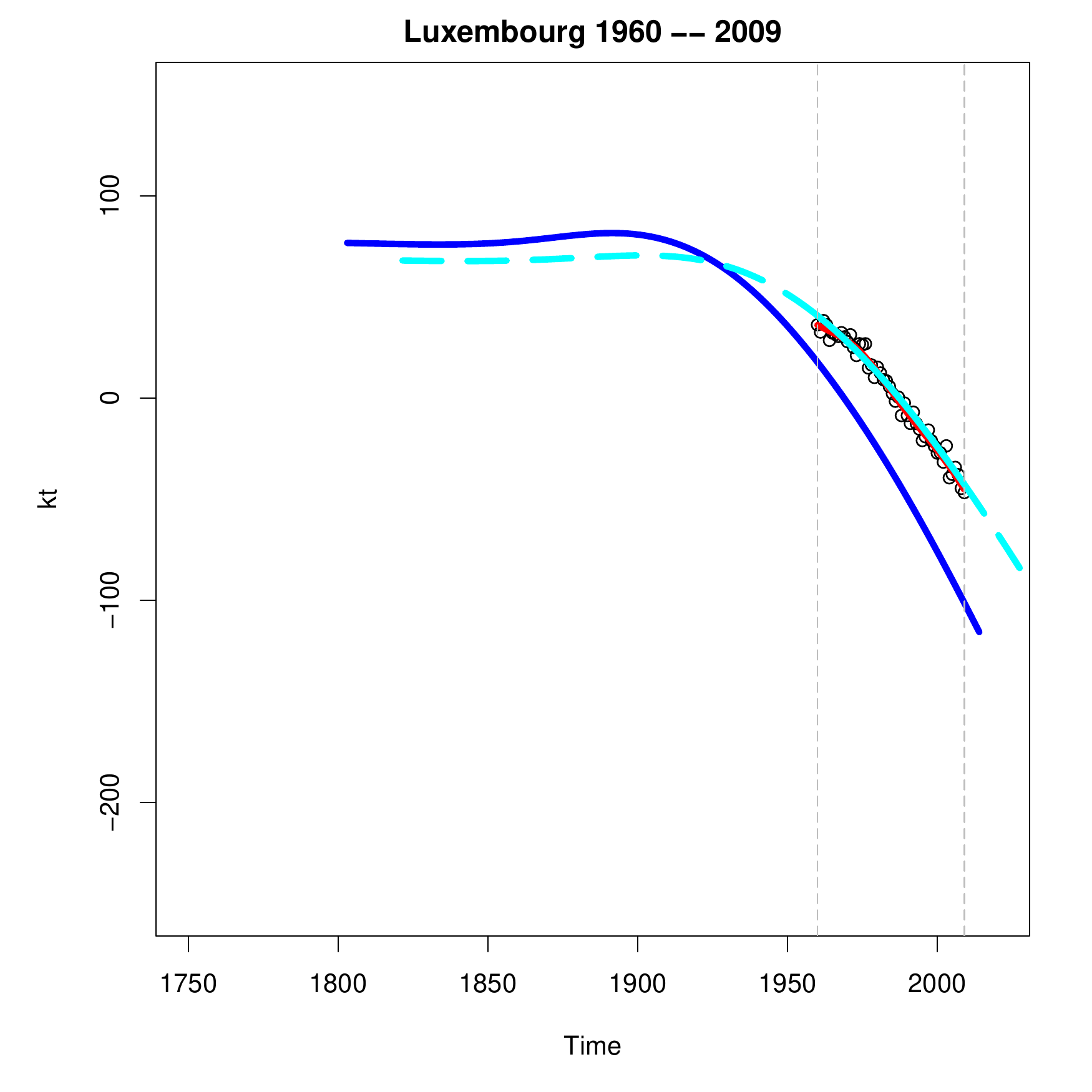}}\hfill
\subfloat[]
  {\includegraphics[width=0.5\textwidth]{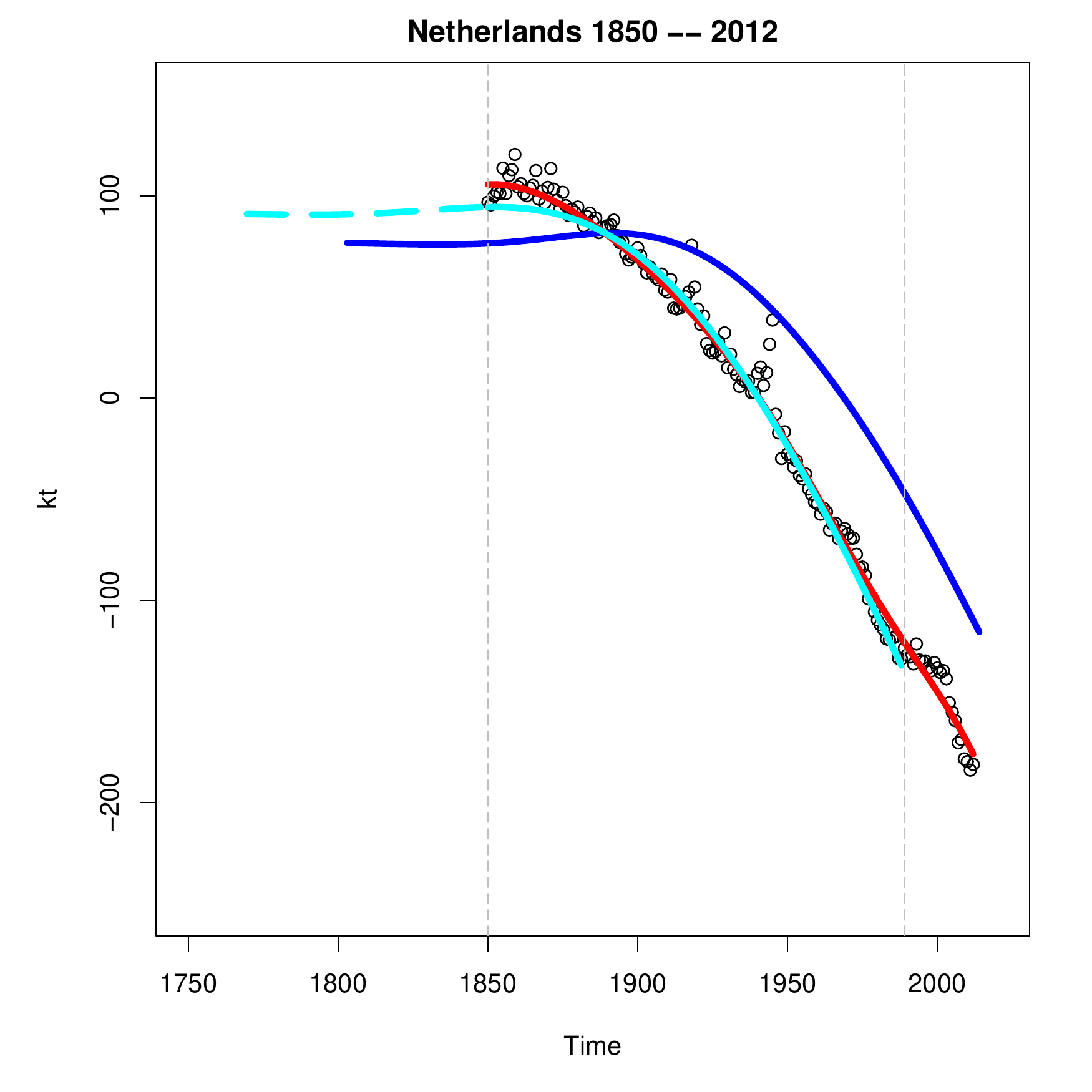}}\\
\vspace{-1.2 cm}
\href{https://github.com/QuantLet/MuPoMo}{\quantnet MuPoMo}
\end{figure}

\begin{figure}[H]
\captionsetup[subfigure]{labelformat=empty}
\subfloat[]
  {\includegraphics[width=0.5\textwidth]{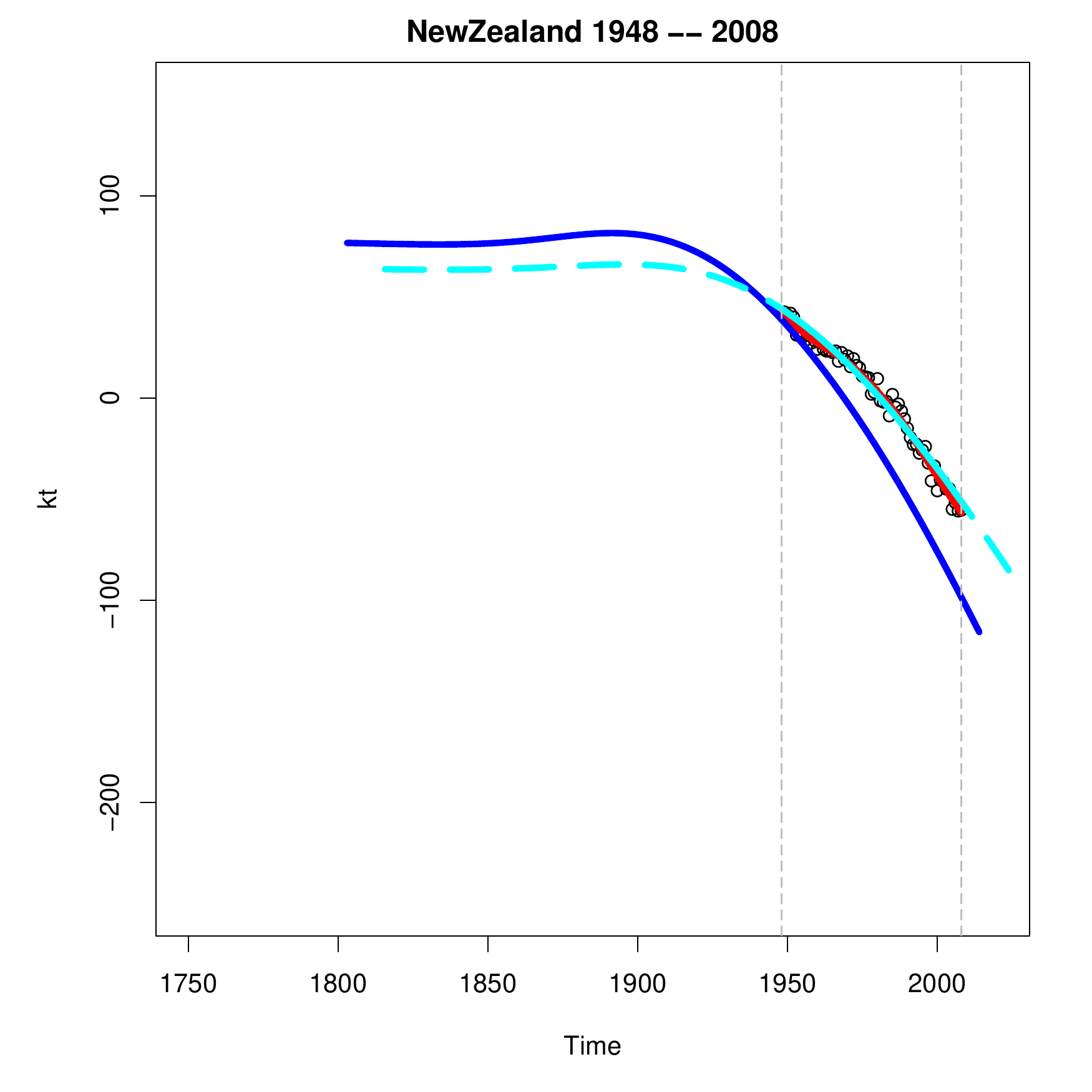}}\hfill
\subfloat[]
  {\includegraphics[width=0.5\textwidth]{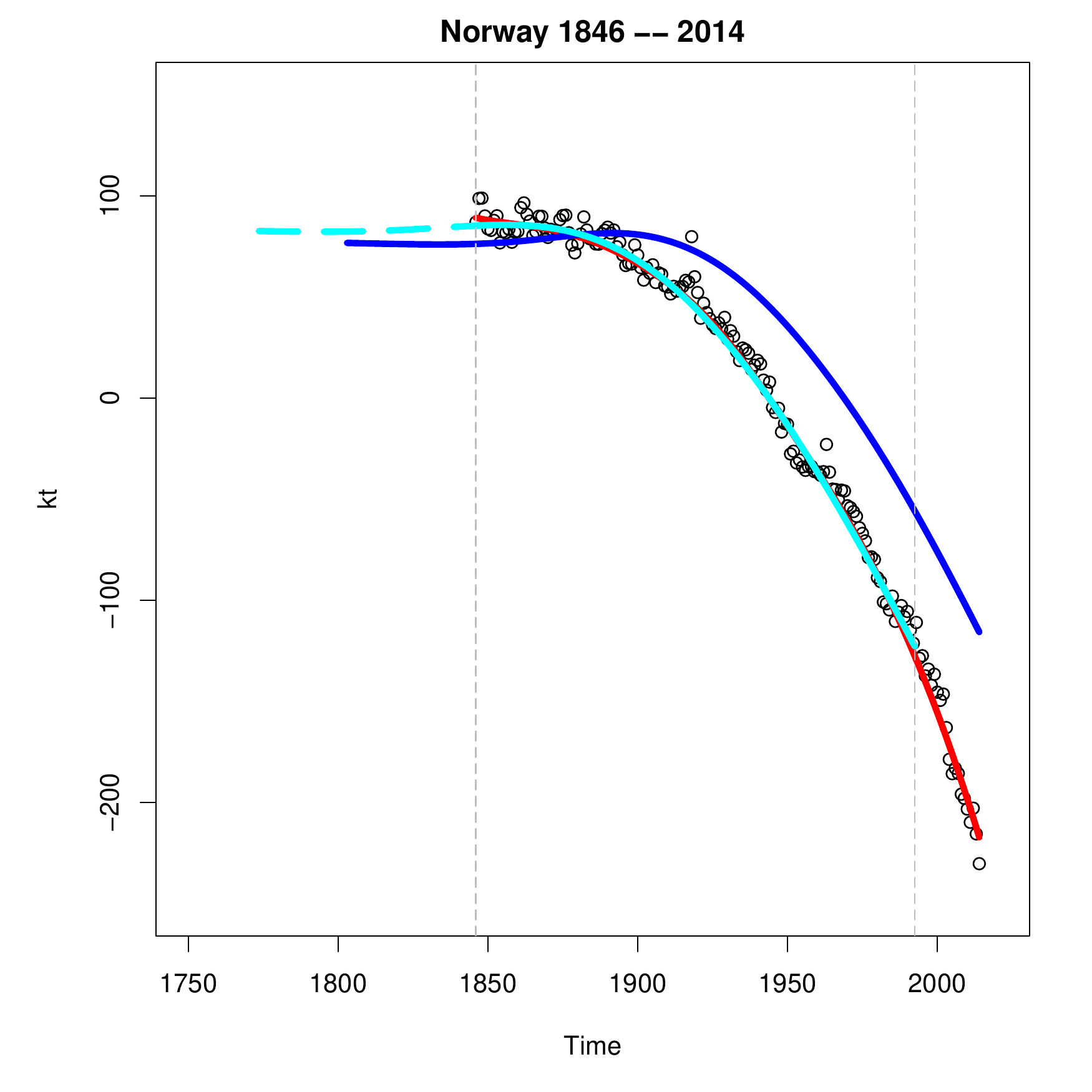}}\\
\vspace{-1.2 cm}
\subfloat[]
  {\includegraphics[width=0.5\textwidth]{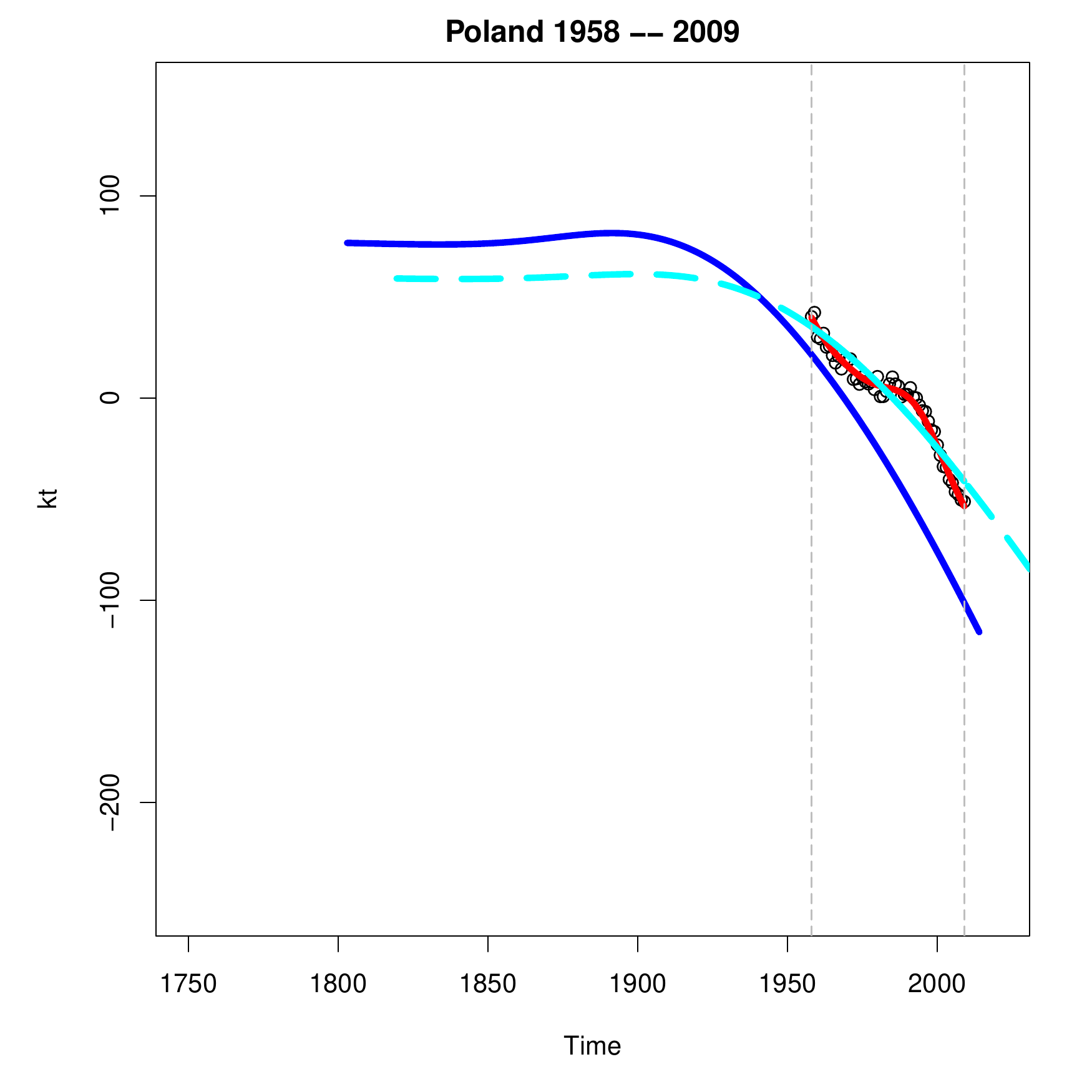}}\hfill
\subfloat[]
  {\includegraphics[width=0.5\textwidth]{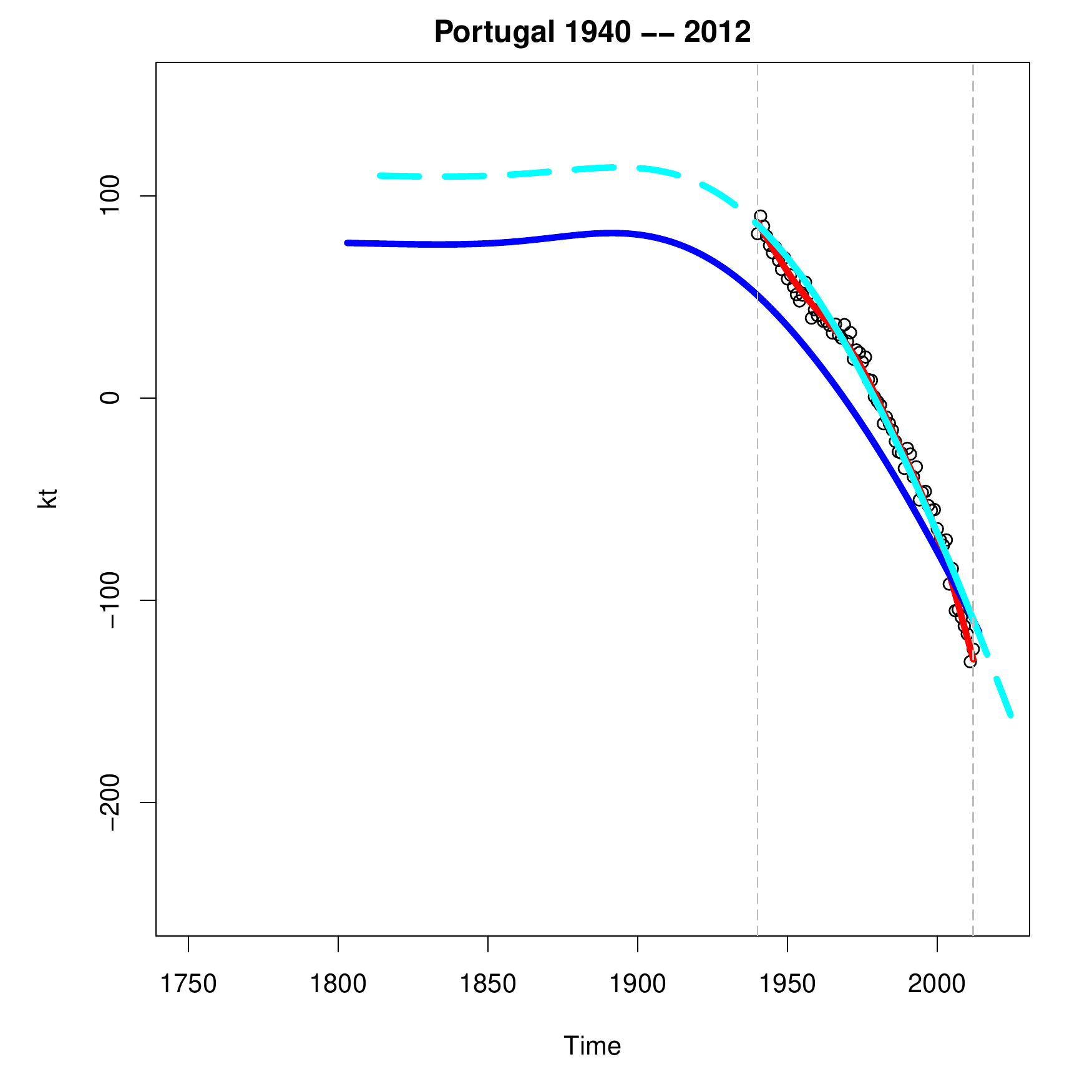}}\\
\vspace{-1.2 cm}
  \subfloat[]
  {\includegraphics[width=0.5\textwidth]{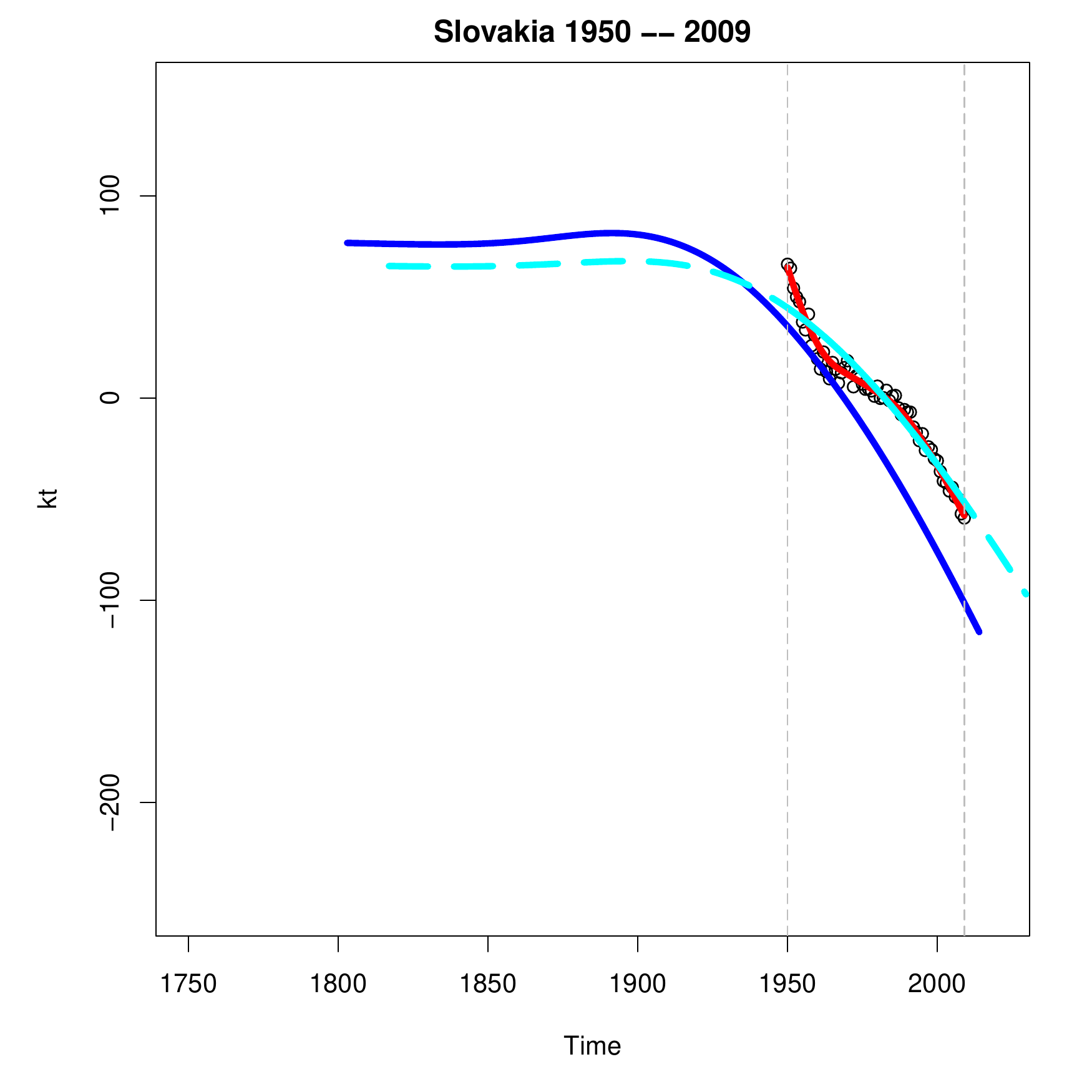}}\hfill
\subfloat[]
  {\includegraphics[width=0.5\textwidth]{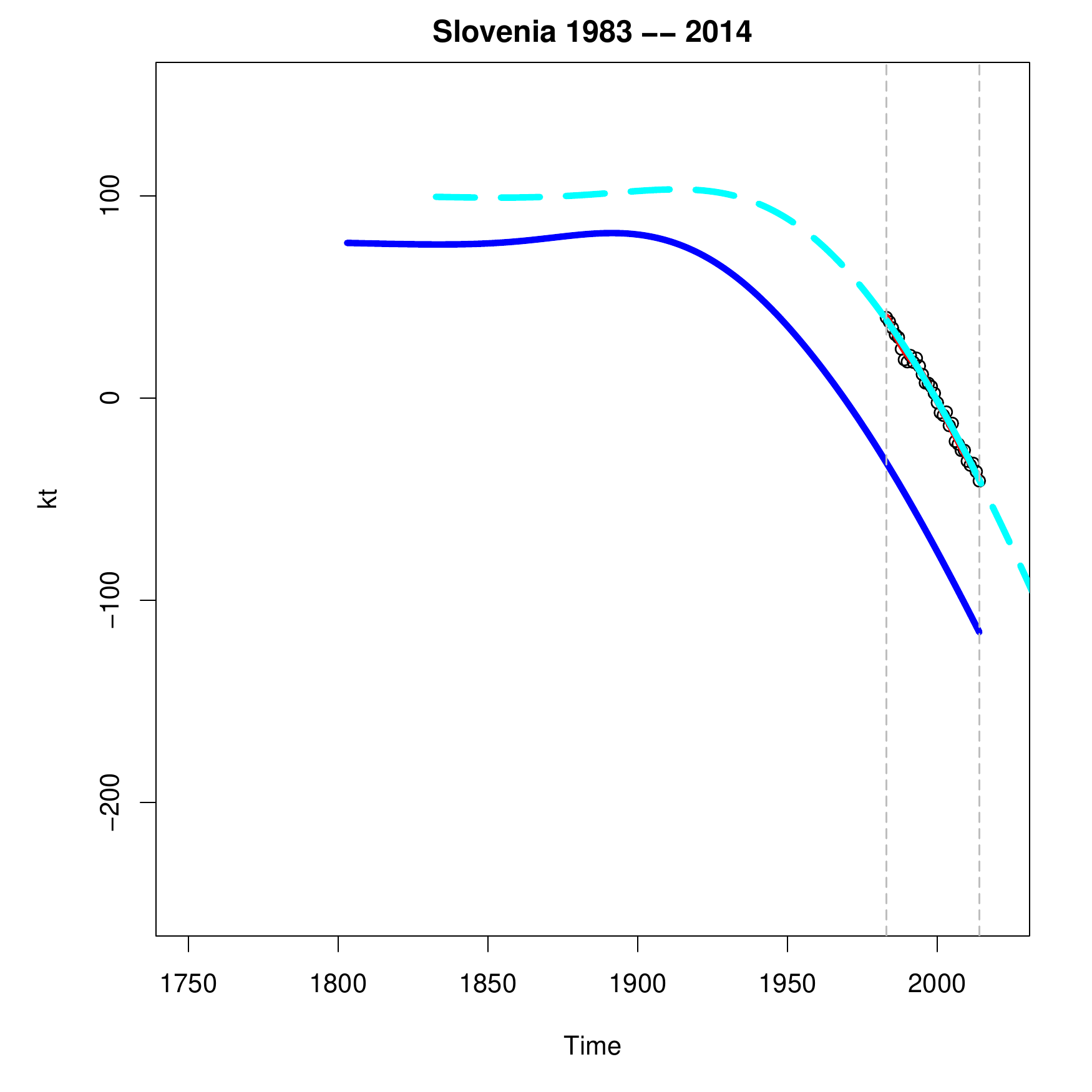}}\\
\vspace{-1.2 cm}
\href{https://github.com/QuantLet/MuPoMo}{\quantnet MuPoMo}
\end{figure}

\begin{figure}[H]
\captionsetup[subfigure]{labelformat=empty}
\subfloat[]
  {\includegraphics[width=0.5\textwidth]{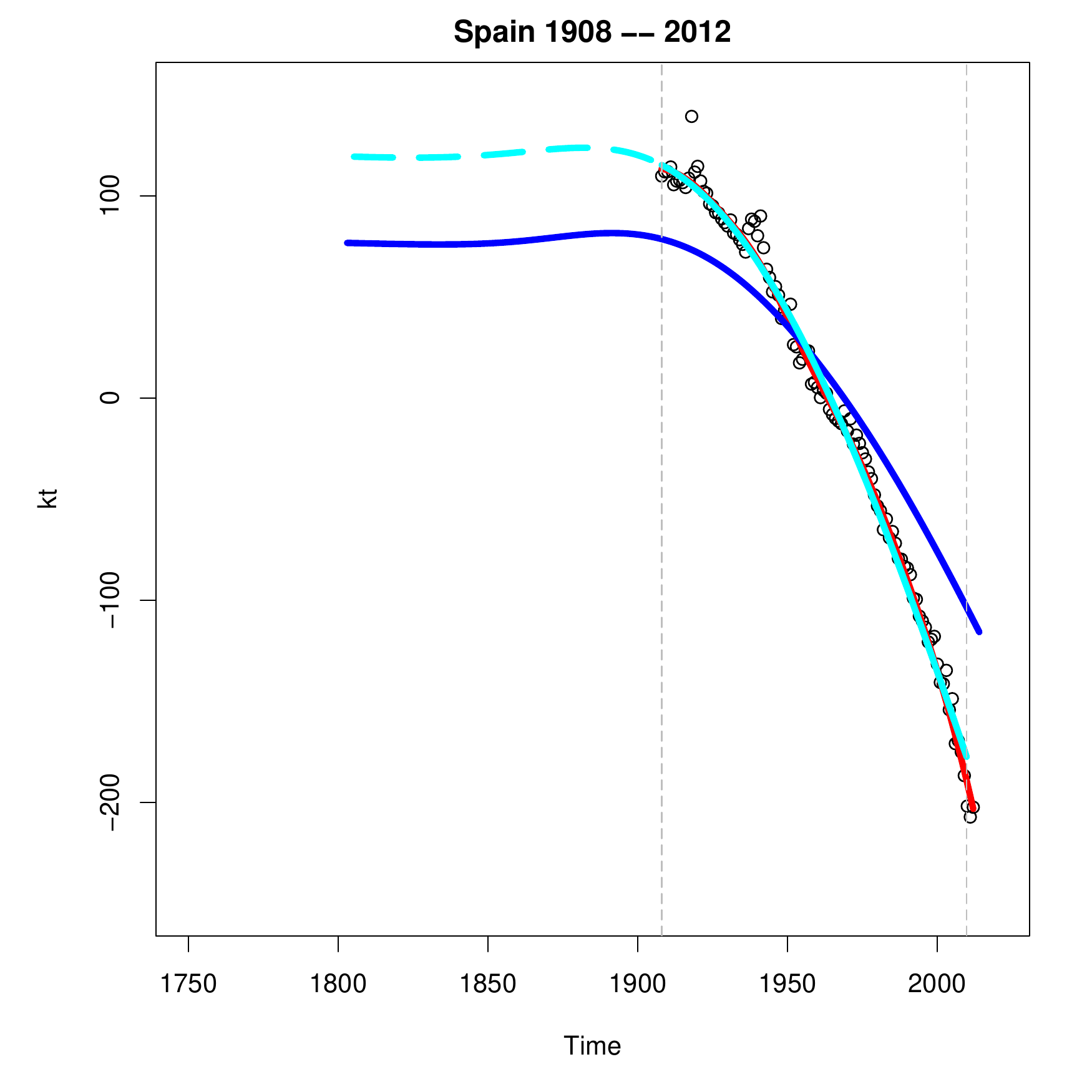}}\hfill
\subfloat[]
  {\includegraphics[width=0.5\textwidth]{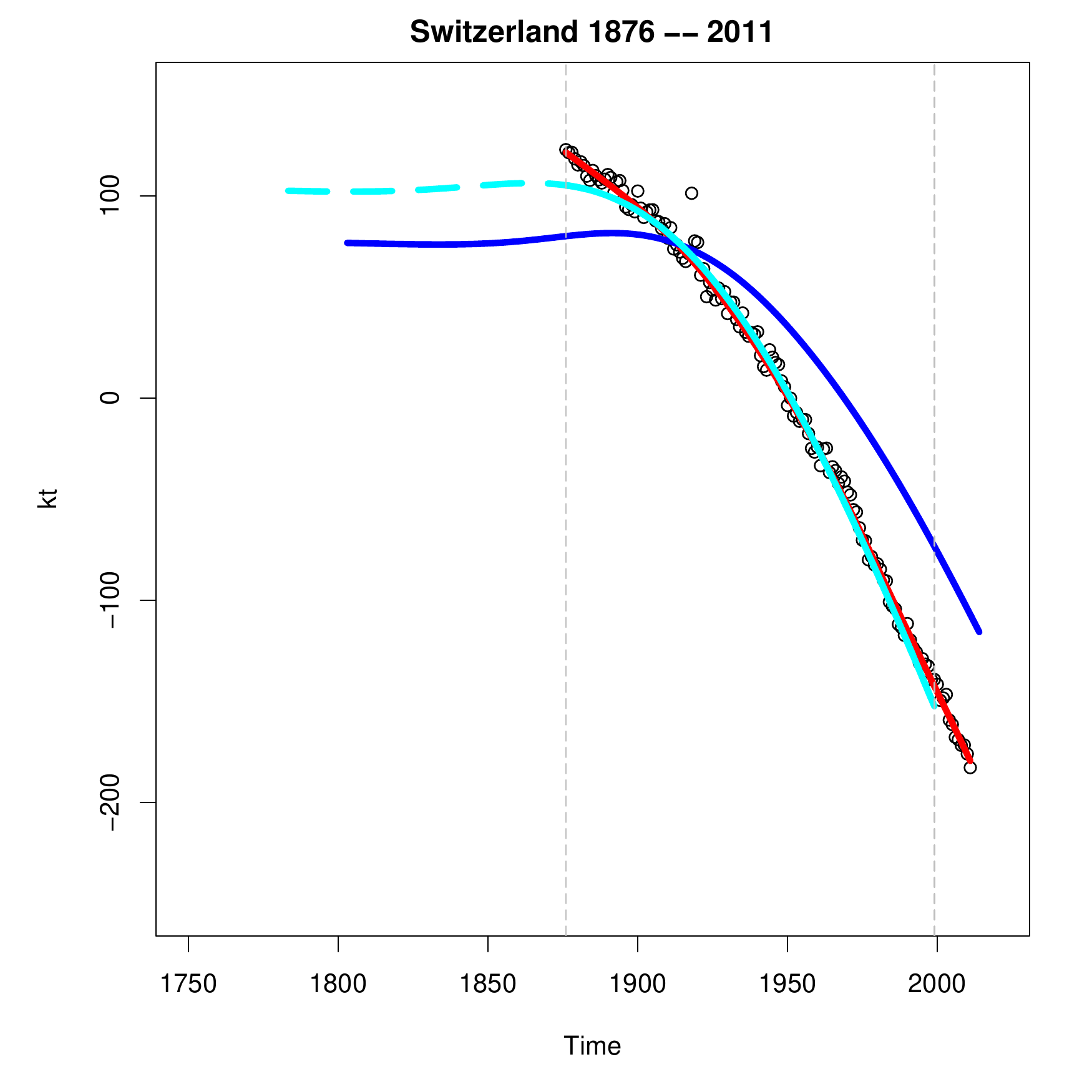}}\\
\vspace{-1.2 cm}
\subfloat[]
  {\includegraphics[width=0.5\textwidth]{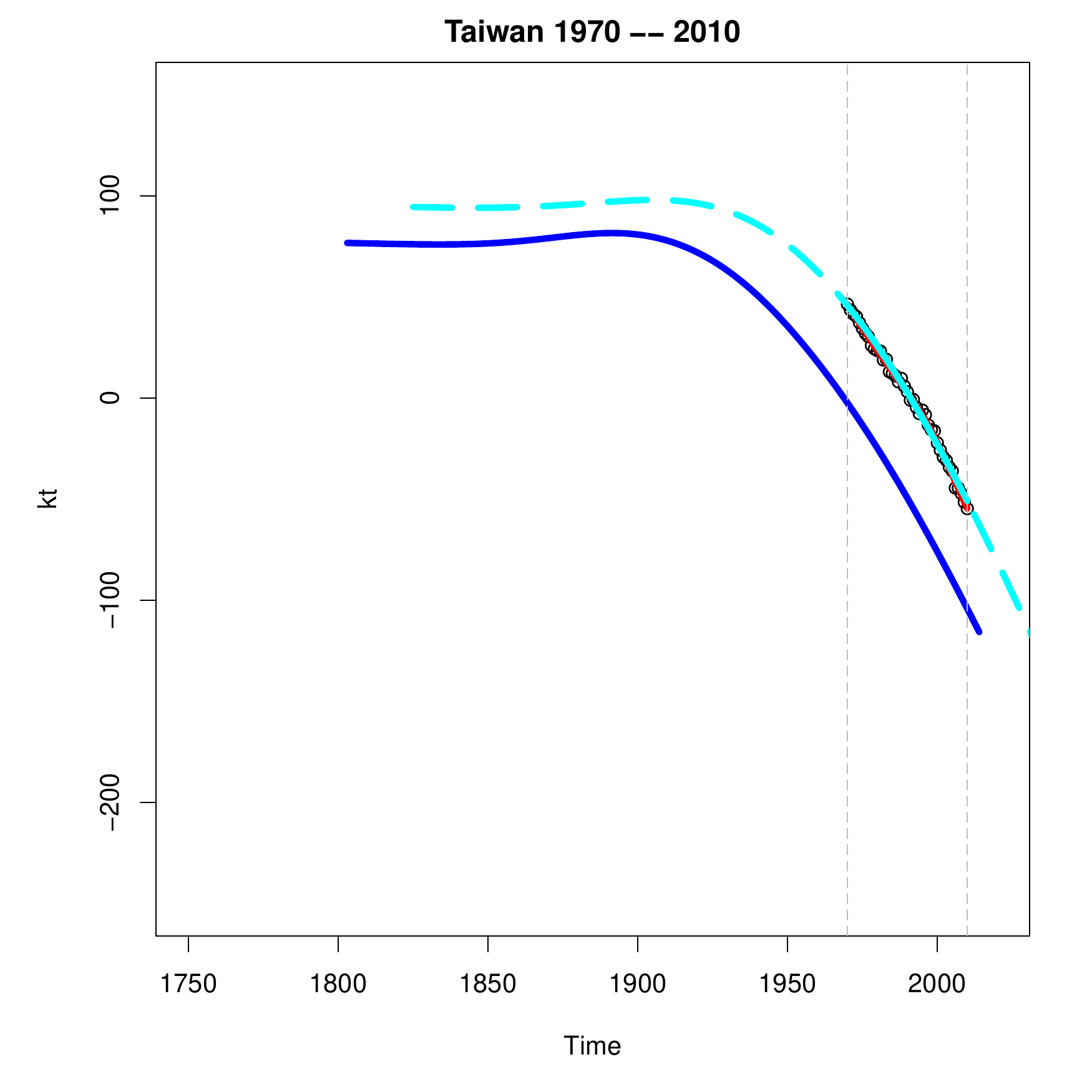}}\hfill
\subfloat[]
  {\includegraphics[width=0.5\textwidth]{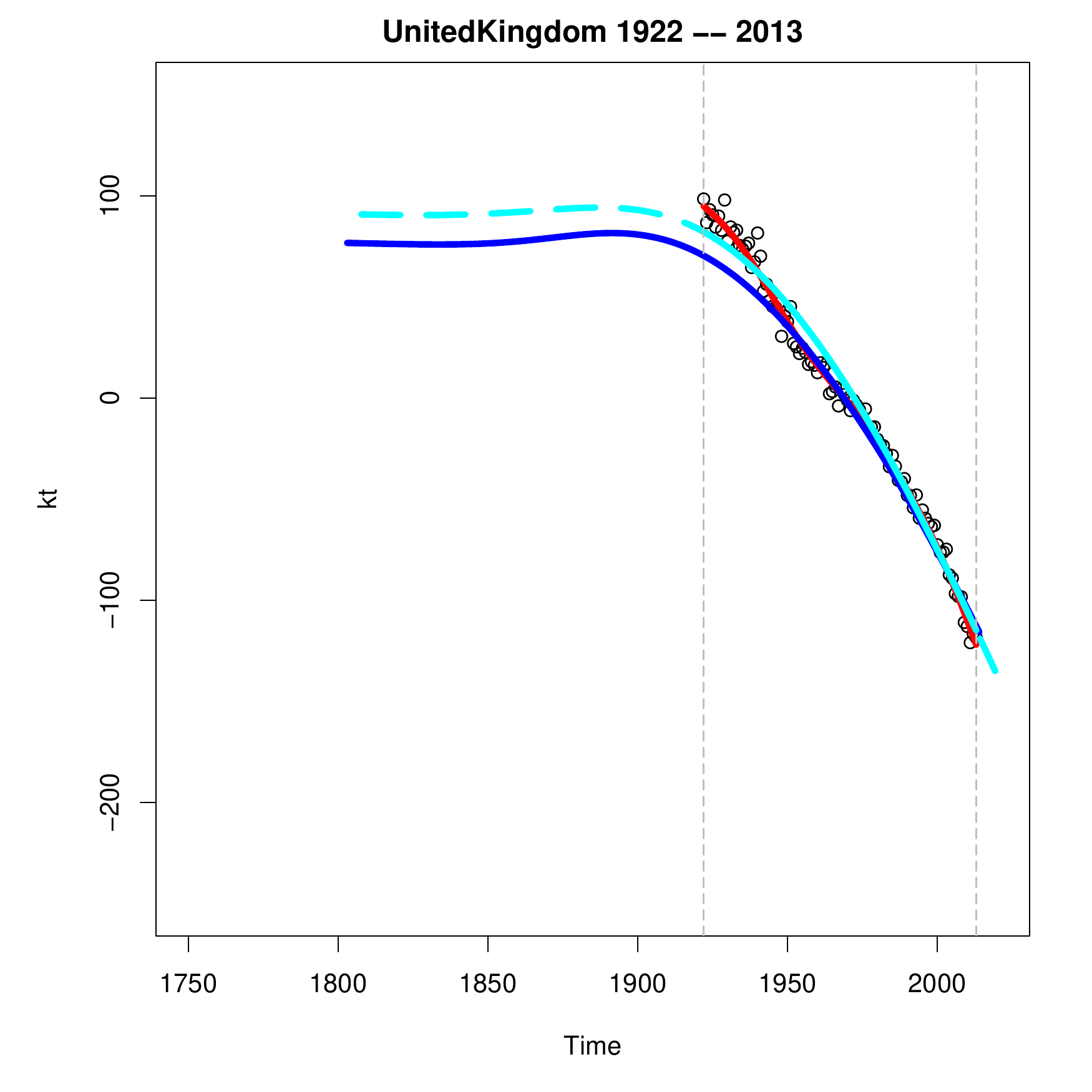}}\\
\vspace{-1.2 cm}
  \subfloat[]
  {\includegraphics[width=0.5\textwidth]{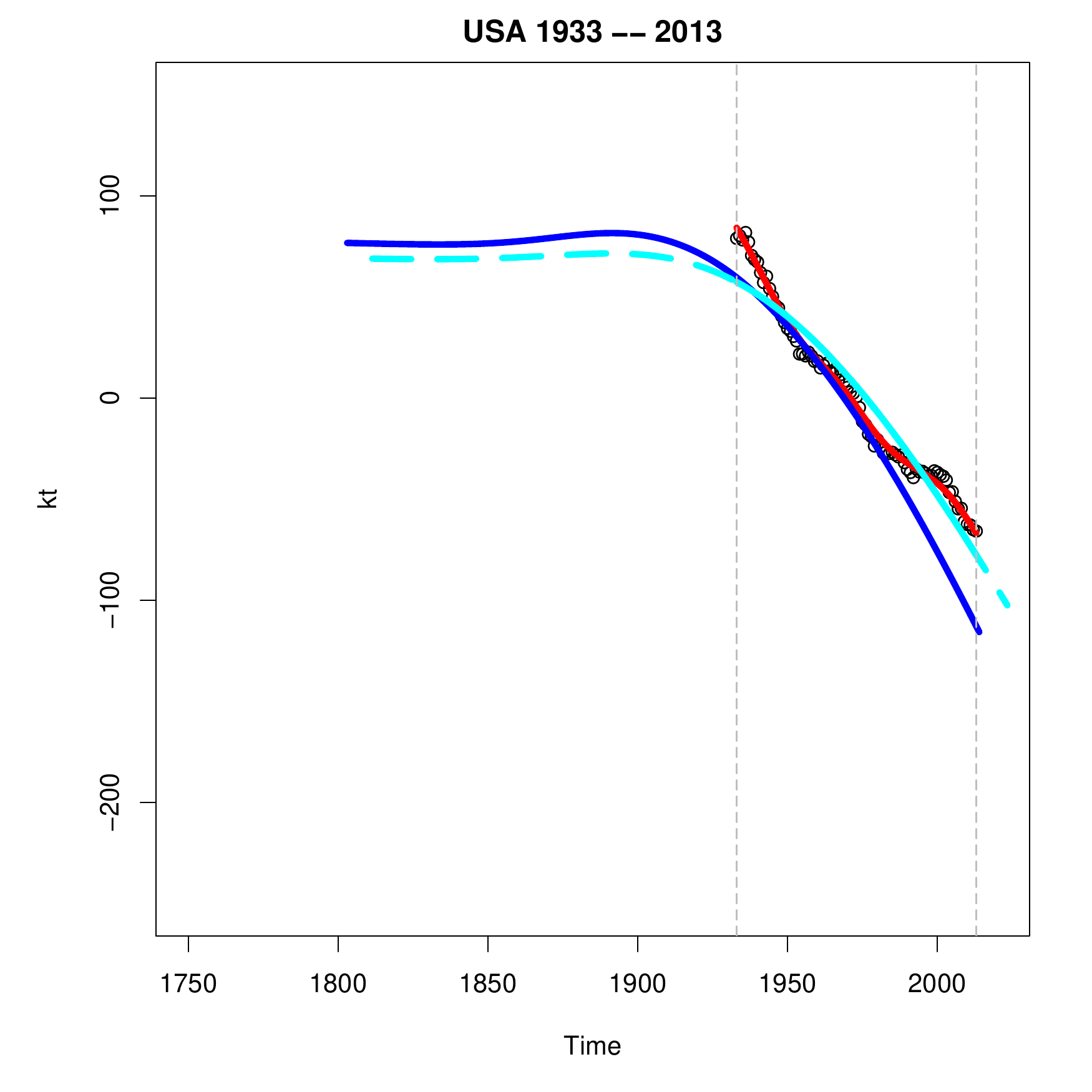}}\hfill
\subfloat[]
  {\includegraphics[width=0.5\textwidth]{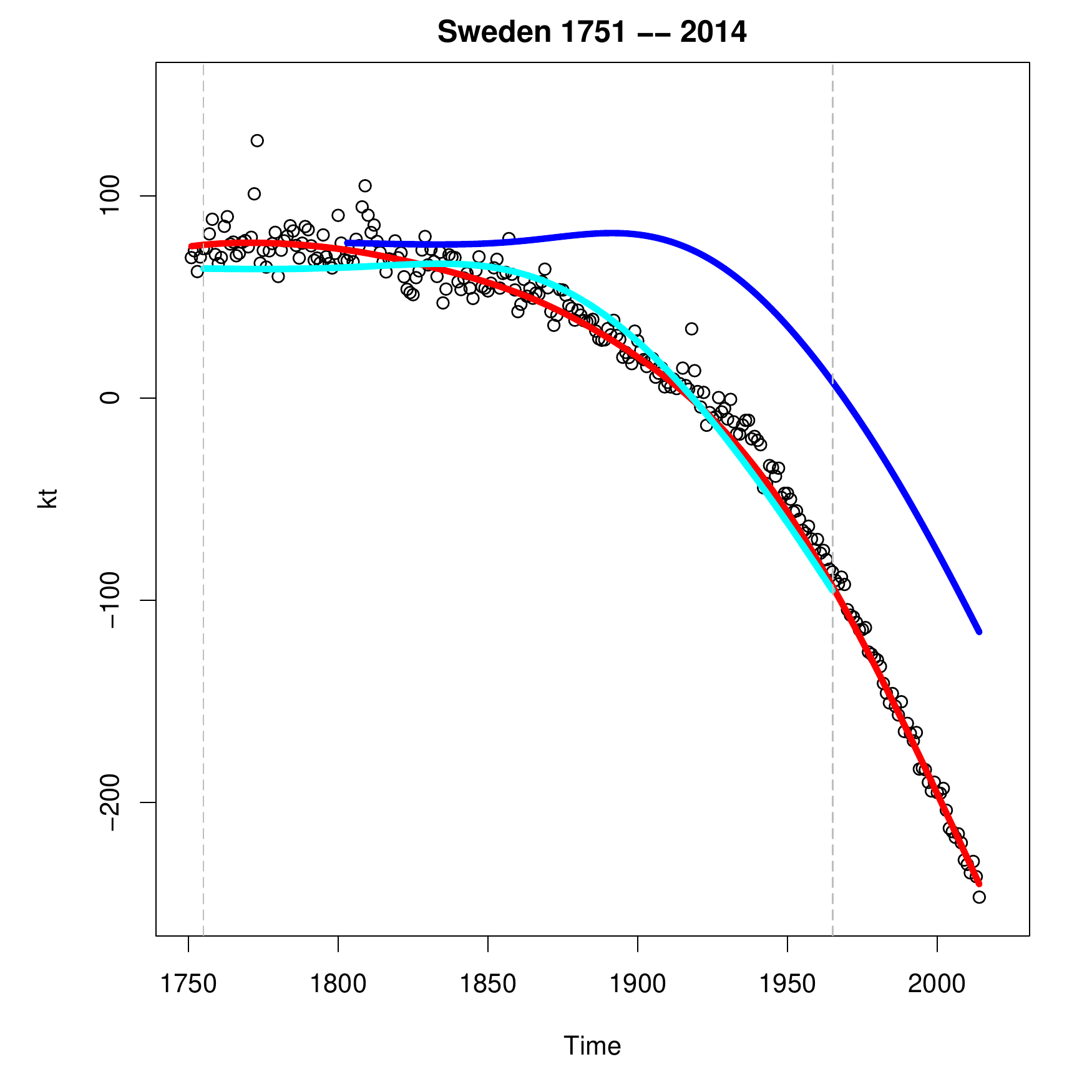}}\\
\vspace{-1.2 cm}
\href{https://github.com/QuantLet/MuPoMo}{\quantnet MuPoMo}
\end{figure}

\end{document}